\documentclass[a4paper,11pt]{article}
\pdfoutput=1
\usepackage{jcappub}
\usepackage{ulem}
\usepackage{bm}
\usepackage[dvipsnames]{xcolor}
\usepackage{float}
\usepackage{arydshln}
\usepackage{multirow}

\usepackage{cancel}

\allowdisplaybreaks

\renewcommand{\thesection}{\arabic{section}}

\def\laq{~\raise 0.4ex\hbox{$<$}\kern -0.8em\lower 0.62ex\hbox{$\sim$}~}
\def\gaq{~\raise 0.4ex\hbox{$>$}\kern -0.7em\lower 0.62ex\hbox{$\sim$}~}

\def\beq{\begin{equation}}
\def\eeq{\end{equation}}
\def\bea{\begin{eqnarray}}
\def\eea{\end{eqnarray}}

\def \pa {\partial}

\def \mc {\mathcal}
\def \ms {\mathscr}
\def \pr {\prime}
\def \t {\text}

\def \ka {\epsilon}

\def \ga {\gamma}

\def \Hcal {\mathcal{H}}

\def \U {\Upsilon}

\newcommand{\quotes}[1]{``#1''} 
\def \o {_\text{o}}
\def \oz {|^\text{o}_z}
\def \t {\text}
\def \nex {\notag \\[1ex]}

\def \psiI {\Psi^\text{I}}
\def \psiA {\Psi^\text{A}}

\def \vo {v_{||\text{o}}}
\def \vz {v_{||z}}

\usepackage{amsmath, amssymb, amscd}
\usepackage{amsthm}
\usepackage[scr=boondox]{mathalfa}
\usepackage{tikz,tikz-cd}
\usepackage{booktabs}
\usepackage{subfig}

\title{Nonlinear Relativistic Effects on Cosmological Redshift Drift}

\author[a,b]{P. B\'echaz,}
\author[c]{G. Fanizza,}
\author[a,b]{G. Marozzi}
\author[d,b]{and M. R. Medeiros Silva}

\affiliation[a]{Dipartimento di Fisica, Universit\`a di Pisa, Largo B. Pontecorvo 3, 56127 Pisa, 
Italy}
\affiliation[b]{Istituto Nazionale di Fisica Nucleare, Sezione di Pisa, Largo B. Pontecorvo 3, 56127 Pisa, Italy}
\affiliation[c]{Dipartimento di Ingegneria, Universit\`a LUM, S.S. 100 km 18 - 70010 Casamassima (BA), Italy}
\affiliation[d]{Departamento de Física, Universidade Estadual de Londrina, Rod. Celso Garcia Cid, Km 380, 86057-970, Londrina, Paraná, Brazil}

\emailAdd{pierre.bechaz@phd.unipi.it}
\emailAdd{fanizza@lum.it}
\emailAdd{giovanni.marozzi@unipi.it}
\emailAdd{matheusrmsilva@uel.br}

\abstract{Using  a fully gauge-invariant approach, we compute for the first time in the literature  relativistic effects on the redshift drift up to second order in cosmological perturbation theory. This is achieved by employing a set of light-cone coordinates that simplify the description of light propagation in an inhomogeneous and anisotropic universe. We show that redshift-space distortion occurs only as a second-order effect whereas, as known, it is not present among the linear perturbations.
We then derive analytical expressions of the bispectrum for the leading-order perturbative contributions on sub-Hubble scales, providing some numerical evaluations.  Our finding is that, at low redshift and for large momenta, the non-linearities in the bispectrum are  enhanced more than   the squared power spectrum.}

\keywords{bispectrum, cosmological perturbation theory, 
geodesic light-cone gauge, 
redshift drift, relativistic cosmology
 
\vskip13pt plus8pt minus11pt

\noindent{\bfseries\large\sffamily{Preprints:}}}

\begin{document}

\maketitle


\section{Introduction}
Observational cosmology has entered an era of high-precision measurements. Thanks to recent technological and instrumental advances, Large-Scale Structure (LSS) surveys can probe our Universe on smaller and smaller scales. Indeed, both ground-based surveys such as the Dark Energy Spectroscopic Instrument (DESI) \cite{DESI:2016fyo}, the Vera Rubin Observatory's LSST \cite{Abate:2012za} and the Square Kilometre Array (SKA) \cite{Dewdney:2009tmd}, and space-based ones such as Euclid \cite{Amendola:2012ys} and the forthcoming Nancy Grace Roman Space Telescope \cite{Spergel:2015sza} can provide us with full three-dimensional images of the universe. This means that, ranging the Hubble size down to scales $\sim h^{-1} \, \t{Mpc}$, there are about $10^{10}$ orders of magnitude in $k$-space that can be tested. 

This provides a significant improvement when compared to observations based on the Cosmic Microwave Background (CMB) radiation \cite{Ade:2013zuv,Planck:2018vyg}, since the latter can only  intrinsically  map the two-dimensional surface\footnote{For the sake of completeness, we recall that secondary effects on CMB can also provide  information about the 3-dimensional LSS. We redirect the  interested reader to, for example, \cite{Lewis:2006fu,Durrer:2020fza} and \cite{Bohm:2016gzt,Marozzi:2016uob,Pratten:2016dsm,
Lewis:2016tuj,Marozzi:2016qxl,Marozzi:2016und,Bohm:2018omn} for linear and non-linear treatments respectively.} of the last scattering, thus giving access to approximately $10^7$ orders of magnitude. However, a correct theoretical interpretation of data coming from LSS surveys is more difficult than for CMB. Indeed, CMB physics is almost entirely captured by linear and Gaussian cosmological perturbations because, at the time of recombination, fluctuations were still in the linear regime. This means that two-point correlation functions encode nearly all the physical information carried  by the baryons and photons fluid. Conversely, the processes driving structure formation and gravitational collapse are inherently non-linear, being dominated by the non-linear growth of the dark matter density field. As a consequence, late-time LSS observables develop a significant level of non-Gaussianity, even when the initial conditions are Gaussian, as predicted by inflation (see \textit{e.g} \cite{Bartolo:2004if} for a review).

Therefore, since the power spectrum is not able to capture all the non-linear behavior of structure formation, it is necessary to consider higher-order statistics. In particular, the first next-to-leading order  statistic which can measure the non-Gaussian signature of gravity is the three-point correlation function, namely the bispectrum. To properly compute the latter, the study of the cosmological perturbation theory beyond the linear approximation is required. 

Currently, in the literature several authors   developed  a cosmological perturbation theory to second order (see \textit{e.g.} \cite{Bruni:1996im,Matarrese:1997ay} for some pioneering works), and applied it to compute general relativistic effects on cosmological observables (see, for instance, second-order corrections to the luminosity distance–redshift relation in \cite{Umeh:2012pn,BenDayan:2012wi,Umeh:2014ana,Marozzi:2014kua,Scaccabarozzi:2017ncm,Fanizza:2019pfp,Fanizza:2021tuh,Magi:2022nfy,Schiavone:2023olz,Bechaz:2025ojy}, galaxy number count in \cite{Bertacca:2014dra,Bertacca:2014wga,Yoo:2014sfa,DiDio:2014lka,DiDio:2015bua}, and CMB lensing in \cite{Pratten:2016dsm,Marozzi:2016uob,Lewis:2016tuj,Marozzi:2016qxl,Marozzi:2016und}). These effects can be  also tested against numerical simulations done for LSS observables \cite{Adamek:2016zes} and CMB \cite{Fabbian:2017wfp,Beck:2018wud}. 

In this paper, we consider another cosmological observable, known as the redshift drift, whose first theoretical prediction dates back to the 1960s \cite{Sandage:1962,McVittie:1962}, while some observational evidences have been discussed only recently \cite{Trost:2025kro, Trost:2026zmt}. Defined as the real-time variation of the redshift emitted by distant sources, the measurement of the redshift drift would allow for the determination of the evolution of the Hubble factor, $H=H(z)$. Therefore, while measuring the redshift provides kinematic evidence for the expansion of the universe, measuring the redshift drift can provide  a new probe for the acceleration of the universe throughout the various phases of its evolution.

Consequently, the redshift drift can be used to further constrain the parameters of the $\Lambda$CDM model, and to test dark energy and modified gravity theories \cite{Jain:2009bm,Heinesen:2021nrc,Melia:2024qha,Alfano:2026sxg}. More generally, being sensitive to the local cosmic expansion rate, the redshift drift paves the way for the so-called real-time cosmology \cite{Quercellini:2010zr}, which aims to extract cosmological information from the redshift evolution of observables measured on the past light-cone of an observer.

The redshift drift has already been computed in an arbitrary space-time, as well as in an FLRW universe, for example in \cite{Kim:2014uha,Piattella:2017uat,Marcori:2018cwn,Heinesen:2020pms,Bessa:2023qrr}. In particular, the authors of  \cite{Bessa:2023qrr}  provided a fully gauge-invariant formula for the linear relativistic effects on the redshift drift including scalar fluctuations, and computed its power spectrum both analytically and numerically. Furthermore, Refs.~\cite{Koksbang:2023tun,Bessa:2024beh,Oestreicher:2025qcs} investigated the redshift drift through numerical simulations.

On the other hand, second-order perturbations to the redshift drift have not yet been addressed  in the literature. Therefore, the aim of this paper is to fill this gap, by providing a full derivation of the second-order perturbative effects and the analytical expression of  the bispectrum induced by the leading terms on sub-Hubble scales.  To achieve this aim, both light-like and time-like geodesic equations must be perturbed up to second order in the cosmological perturbations, and integrated along perturbed photon trajectories connecting a source to an observer. This leads to the so-called post-Born lensing effects. Furthermore, Redshift-Space Distortion (RSD) originates from light-rays propagating along the past light-cone of an observer, and  other relativistic effects induced by second-order perturbations come into play (light-cone distortions, frame-dragging effects, and coupling of linear modes). Various approaches have been proposed in the literature  to properly compute these effects at non-linear order, such as using the cosmic rulers \cite{Jeong:2011as,Schmidt:2012ne,Jeong:2014ufa,Villey:2025xfz} or the Geodesic Light-Cone (GLC) gauge \cite{Gasperini:2011usf,Fanizza:2013doa,Fanizza:2018tzp}.

In particular, a strong motivation for the use of the GLC gauge, first introduced in \cite{Gasperini:2011usf}, is that observations are made on our past light-cone and  can then be  better interpreted by adopting a light-like foliation of  space-time. Indeed, by foliating the space-time into light-cones, one can define a physically motivated gauge specifically developed to obtain fully non-linear expressions for light-like cosmological observables. This is because, since  the angles  are kept constant along photon trajectories in this gauge, observables can be computed at any order in perturbation theory directly in terms of the observed light-cone and observed angles. This can be achieved through purely geometrical (or kinematical) computations, because  both time-like and light-like geodesics have simple, non-perturbative solutions in the GLC gauge.

In the past, the GLC coordinates were successfully used to compute perturbations to cosmological observables such as the redshift, the luminosity distance–redshift relation, and the number count \cite{Gasperini:2011usf,BenDayan:2012wi,Fanizza:2013doa,Marozzi:2014kua, DiDio:2014lka,DiDio:2015bua,Scaccabarozzi:2017ncm}. More recently, a  perturbation theory directly on the past light-cone has been introduced in \cite{Fanizza:2020xtv, Fanizza:2023ixk} at first order  and in \cite{Bechaz:2025ojy} at second order (see also \cite{Fanizza:2015gdn,Mitsou:2020czr,Frob:2021ore,Fanizza:2023fus} for other applications of the GLC gauge). In particular, in \cite{Bechaz:2025ojy}, we established a novel framework for evaluating cosmological observables to second order in a fully gauge-invariant way. In that work, we proved that relativistic effects can be more easily handled by working on the past light-cone. We also described a complete gauge-fixing procedure around the observer position. This  makes the results devoid of any divergence and gauge invariant not only at the source but also at the observer position, in a rigorous and consistent way. Moreover, in \cite{Bechaz:2025ojy} these procedures were used to compute the angular distance–redshift relation at second order as seen by a free-falling observer, and the matching of the results with those of the literature \cite{BenDayan:2012wi,Fanizza:2013doa,Marozzi:2014kua,Fanizza:2015swa} was shown, thus providing a validation of the methodology.

In the present manuscript, inspired by the motivations outlined before and relying on the analytical techniques developed in \cite{Bechaz:2025ojy}, we derive fully gauge-invariant expressions for the first- and second-order perturbative effects on the redshift drift. We properly account for all the perturbations evaluated at the source position and along the world-line of a free-falling observer, by also including non-vanishing anisotropic stress. We then use these formulae to compute the bispectrum of the leading terms on sub-Hubble scales, obtaining that the amplitude of the three-point function is comparatively enhanced w.r.t. that of the two-point function.

More in detail, the manuscript is structured as follows. In Sect.~\ref{sec:RD-GLC} we introduce the background redshift drift and, recalling how the GLC gauge is defined, we provide the fully non-linear formula for the redshift drift in this gauge. In Sect.~\ref{sec:LC-pert-theory} we summarize from \cite{Bechaz:2025ojy} how to formulate a cosmological perturbation theory on top of a background light-cone geometry, working up to second order. Then, we provide first- and second-order gauge-invariant variables corresponding to the values that the light-cone perturbations assume in the perturbative realizations of the GLC gauge.  Thereafter, in Sect.~\ref{sec:RD-perturbations} we proceed with an in-depth evaluation of the first- and second-order  corrections to the redshift drift in a fully gauge-invariant way, by also comparing our first-order results with \cite{Bessa:2023qrr}. In Sect.~\ref{eq:bispectrum} we provide analytical expressions for the leading contributions on sub-Hubble scales to the relativistic bispectrum of the redshift drift, and show some numerical results, discussing the overall amplitude of the non-linear effects. Finally, our main findings are summarized in Sect.~\ref{eq:summary}. In Appendix~\ref{app:gauge-inv-formulae}, we report some mathematical formulae that define gauge-invariant variables on the light-cone.  

\section{Redshift Drift: Background and Fully Non-linear Expressions}
\label{sec:RD-GLC}
The redshift drift is defined as the ratio between the variation of the redshift $\Delta z$ of a given source and the time interval  $\Delta \tau\o$ measured in the observer's rest frame needed for the measurement. In a flat Friedmann–Lema\^itre–Robertson–Walker (FLRW)  universe, it has the simple expression \cite{Quercellini:2010zr}
\begin{align}
    \frac{\Delta z}{\Delta \tau_\t{o}}
    =H\o (1+z) - H_\t{s} \, ,  
    \label{eq:bkg-RD}
\end{align}
where  the subscripts \quotes{s} and \quotes{o} stand for evaluation at the source and observer position. 

To model the redshift drift in a generic inhomogeneous and anisotropic universe, the use of the so-called \textit{Geodesic Light-Cone} (GLC) coordinates is particularly convenient \cite{Gasperini:2011usf}. Thus, in the following  we will first define the  GLC gauge and then provide the fully non-linear formula the redshift drift in this gauge. 

\paragraph{Fully non-linear Geodesic Light-Cone gauge.} Since light signals  propagate along the past light-cone of an   observer,   their physical properties can be easily described  using the GLC coordinates $x^\mu = (\tau, w, \tilde{\theta}^a)$, first introduced in \cite{Gasperini:2011usf}. These coordinates are intrinsically defined by the line element\footnote{See \cite{Fleury:2016htl} for a \quotes{bottom-up} derivation of this metric.}
\begin{equation}
\text{d}s^2 =  - 2 \Upsilon \text{d} \tau \text{d}w + \Upsilon^2 \text{d}w^2 + g_{ab} (\text{d}\tilde{\theta}^a-\mathcal{U}^a \text{d}w)(\text{d}\tilde{\theta}^b - \mathcal{U}^b \text{d}w) \, ,
\label{eq:GLCmetric-start}
\end{equation}
and are such that $\tau$ is the proper time of a free-falling observer, while the level sets of $w$ identify its past light-cone. Then, the intersection between  the $\tau=\t{const.}$ and  $w=\t{const.}$ hyper-surfaces defines a  surface diffeomorphic to a 2-sphere $\mathbb{S}^2$, parameterized by the angles $\tilde{\theta}^a$ and thus generated by the vectors $\partial_{\tilde{\theta}^a}$.

In Eq.~\eqref{eq:GLCmetric-start}, we have introduced the six arbitrary functions $\Upsilon$, $\mathcal{U}^a$ and $g_{ab} = g_{ba}$, where the symmetric 2-tensor $g_{ab}$ is the metric induced on $\mathbb{S}^2$. In matrix form, the metric and its inverse are
\begin{equation}
g_{\mu \nu} = 
\begin{pmatrix}
0 & -\Upsilon & \mathbf{0}\\
-\Upsilon & \Upsilon^2 + \mathcal{U}^2 & -\mathcal{U}_b\\
\mathbf{0}^\t{T} & - \mathcal{U}^{\text{T}}_a & g_{ab}
\end{pmatrix} \qquad , \qquad 
g^{\mu \nu} = 
\begin{pmatrix}
- 1 & -1/\Upsilon & -\mathcal{U}^b/\Upsilon\\
-1/\Upsilon & 0  & \mathbf{0}\\
- (\mathcal{U}^a)^\text{T} / \Upsilon & \mathbf{0}^\t{T}  & g^{ab}
\end{pmatrix}\,, \label{eq:GLC-matrix}
\end{equation}
with $\mathcal{U}^a \equiv g^{ab}\mathcal{U}_b$, $\mathcal{U}^2 \equiv \mathcal{U}^a \mathcal{U}_a$ and $g^{ab} \equiv \big (g_{ab} \big )^{-1}$. Then, by using Eqs.~\eqref{eq:GLCmetric-start} and \eqref{eq:GLC-matrix}, it can be proved that
\begin{equation}
     \partial^\mu w \partial_\mu w =0 \qquad , \qquad \partial^\nu \tau \nabla_\nu \big (\partial_\mu \tau \big )=0 \, . 
\end{equation}
In particular, the second equality shows that $\pa_\mu \tau$ defines a geodesic flow, which can be seen as the 4-velocity of a free-falling observer. Hence, the solution to time-like geodesic equations has the form
\begin{equation}
u_\mu = - \partial_\mu \tau = -\delta^\tau_\mu \qquad , \qquad  
    u^\mu = \bigg ( 1, \frac{1}{\Upsilon}, \frac{\mc{U}^a}{\Upsilon} \bigg ) \, .
    \label{eq:4-velocity-GLC}
\end{equation}
Furthermore, the normal vector to the $w = \text{const.}$ hyper-surfaces can be identified with 
\begin{equation}
    k^\mu \equiv -\omega_{\text{phys}} g^{\mu \nu} \partial_\nu w = -\omega_{\text{phys}} g^{\mu w} = \frac{\omega_{\text{phys}}}{\Upsilon} \delta^\mu_\tau \, ,
    \label{eq:normal-vector-glc-light-cone}
\end{equation}
and, by using the metric in Eq.~\eqref{eq:GLCmetric-start}, we see that this vector is null ($k^\mu k_\mu = 0$), geodesic  ($k^\nu \nabla_\nu k^\mu=0$) and orthogonal to $\mathbb{S}^2$ ($k^\mu \partial_\mu \tilde{\theta}^a \propto g^{w a}=0$). Thus, it can be correctly interpreted as the wave-vector of photons incoming to an observer, being  the solution to the light-like geodesics $\tilde{\theta}^a = \text{const.}$ Finally, $\omega_{\text{phys}}$ is the physical frequency of the signal. The fact that, in these coordinates, photons propagate with constant values of $w$ and $\tilde{\theta}^a$ makes it possible to  obtain
fully non-linear expressions for light-like cosmological observables, as shown in \cite{Gasperini:2011usf, Fanizza:2013doa}.

From a geometrical point of view,   the background FLRW metric is recovered by starting from the GLC one and by setting
\begin{equation}
\begin{split}
\tau = t \quad  , \quad w = r+\eta (\tau) \quad  ,  \quad  \Upsilon = a(t) \quad , \quad \mathcal{U}^a = 0 \quad , \quad g_{ab} \text{d}\tilde{\theta}^a \text{d}  \tilde{\theta}^b = a^2(t) r^2 \text{d} \Omega^2 \, ,
\end{split}
\label{eq:GLC-background-FLRW}
\end{equation}
where here $t$  stands for the background proper time and $\t{d}\Omega^2 \equiv \t{d}\theta^2 +\sin^2 \theta \t{d}\varphi^2$.
In particular,  $\mc{U}^a$ vanishes in the FLRW limit. 

To conclude, the metric of Eq.~\eqref{eq:GLCmetric-start} does not completely specify the gauge, because the   residual gauge transformations given by 
\begin{equation}
    w \rightarrow w^\prime = w^\prime (w) \qquad  , \qquad
      \tilde{\theta}^a \rightarrow \tilde{\theta}^{a \prime} = \tilde{\theta}^{a \prime} (w, \tilde{\theta}^a)  
     \label{eq:glc-residual-freedom}
\end{equation}
preserve the form of the metric. As  emphasized in \cite{Fanizza:2013doa,Fleury:2016htl},  the first above transformation amounts to re-labelling light-cones, while the second one to re-labelling light-rays with the angles $\tilde{\theta}^a$ and to  how
such a  labelling is transferred from one light-cone to another.

\paragraph{Fully non-linear redshift drift in the GLC gauge.} 
\begin{figure}
\centering
\includegraphics[scale=0.4]
{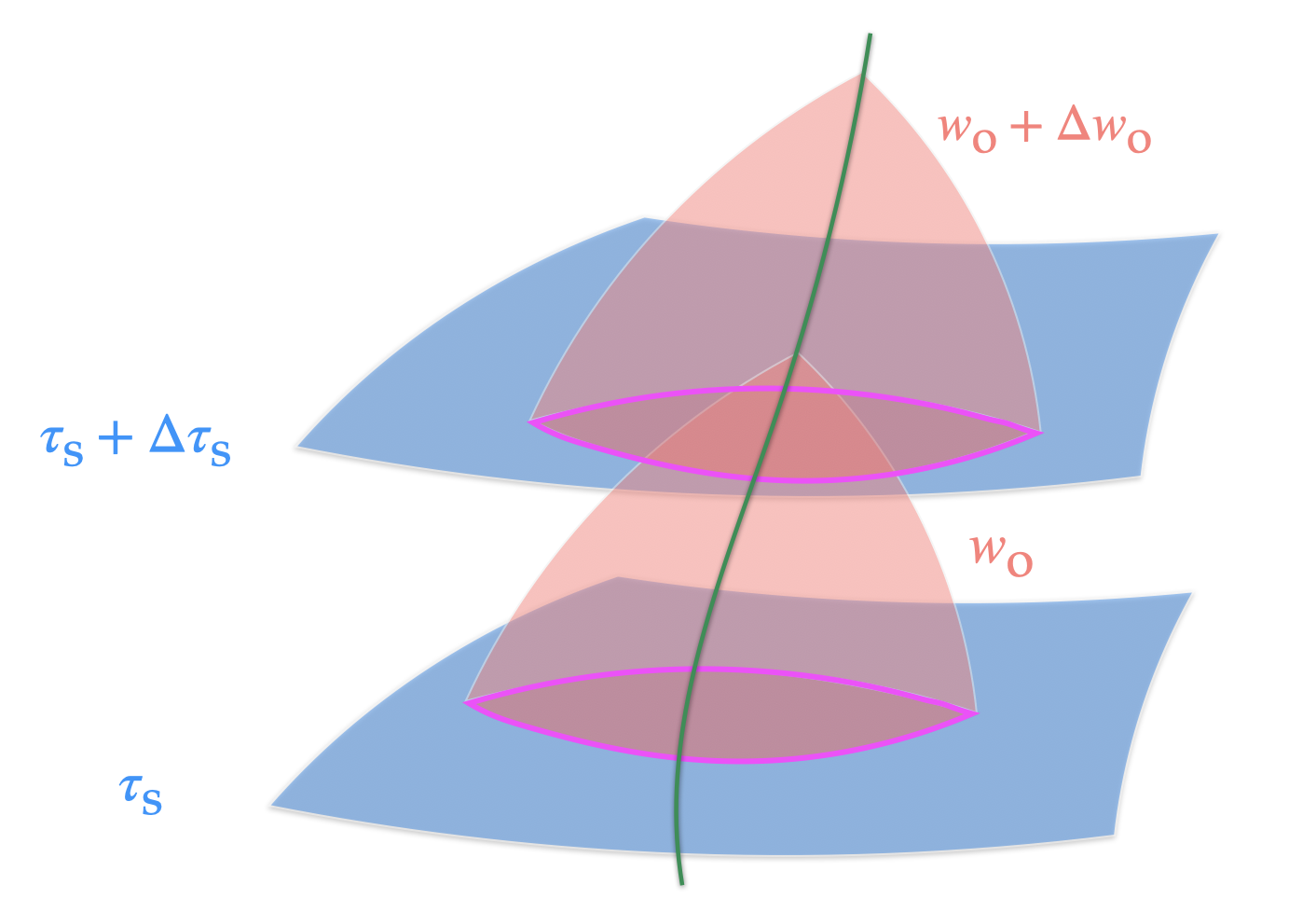}
\caption{Graphical illustration of the redshift drift effect, where in green the world-line of a given observer is represented.}
\label{graf:RD} 
\end{figure}
Using Eqs.~\eqref{eq:4-velocity-GLC} and \eqref{eq:normal-vector-glc-light-cone}, in \cite{Gasperini:2011usf} it was shown that the  redshift  in GLC coordinates is exactly given by
\begin{equation}
    1+z = \frac{(k^\mu u_\mu)_\text{s}}{(k^\mu u_\mu)_\text{o}} = \frac{\Upsilon (\tau_\text{o}, w, \tilde{\theta}^a_\text{o})}{\Upsilon (\tau_\text{s}, w, \tilde{\theta}^a_\text{s})} \, , 
    \label{eq:z-fully-non-linear-glc}
\end{equation}
where we have accounted for the fact that the source and the observer belong to the same light-cone, namely $w_\text{s}=w_\text{o}\equiv w$.  Therefore, we can compute the variation of the redshift as  (see  \cite{Gasperini:2011usf} and Fig.~\ref{graf:RD})
\begin{align}
    \Delta (1+z) = \frac{\pa (1+z)}{\pa w_\t{o}}\Delta w\o +\frac{\pa (1+z)}{\pa \tau_\t{o}}\Delta \tau_\t{o}+\frac{\pa (1+z)}{\pa \tau_\t{s}}\Delta \tau_\t{s}+\frac{\pa (1+z)}{\pa \tilde{\theta}^a_\t{s}}\Delta \tilde{\theta}^a_\t{s} \, . 
\end{align}
Then, to write  a proper expression of $\Delta z / \Delta \tau_\t{o}$, we need to relate $\Delta \tau_\t{o}$ to the variation of the other coordinates. Using that, in the GLC gauge, the velocity of a free-falling observer has the expression \eqref{eq:4-velocity-GLC},  the parameterization of his/her geodesic curve $\tau \to x^\mu(\tau)$ satisfies
\begin{align}
    \frac{\t{d}x^\mu}{\t{d}\tau} = u^\mu = \bigg ( 1, \frac{1}{\Upsilon}, \frac{\mc{U}^a}{\Upsilon} \bigg ) \, . 
\end{align}
Hence, we have 
\begin{equation}
    \Delta \tau = \Upsilon \Delta w \qquad , \qquad \Delta \tilde{\theta}^a = \mathcal{U}^a  \Delta w  \, , 
\end{equation}
and thus $\Delta w \o = \Delta \tau \o / \Upsilon \o = \Delta \tau_\t{s} / \Upsilon_\t{s}$. We can finally obtain\footnote{Note that, all along the observer world-line,  the redshift can not depend on the angles  $\tilde{\theta}^a\o$. This point will be more rigorously addressed  in the final paragraph of Subsect.~\ref{subsec:gaue-inv-LC}.} the fully non-linear formula of the redshift drift in the GLC gauge as \cite{Gasperini:2011usf}
\begin{align}
\frac{\Delta z}{\Delta \tau_\text{o}} &= (1+z) \frac{1}{\Upsilon_\text{o}} \frac{\partial \Upsilon_\text{o}}{\partial \tau_\text{o}} + \frac{1}{\Upsilon_\text{o}} \frac{\partial (1+z)}{\partial w_\text{o}} + \frac{\partial \log (1+z)}{\partial \tau_\text{s}} + \frac{\mathcal{U}^a_\text{s}}{\Upsilon_\text{s}} \frac{\partial (1+z)}{\partial \tilde{\theta}^a_\text{s}}  \, .
\label{eq:rsd-glc-fully-nonlinear}
\end{align}
In order to properly compute perturbative expressions up to second order of the above Eqs.~\eqref{eq:z-fully-non-linear-glc} and \eqref{eq:rsd-glc-fully-nonlinear},  we will make use of a second-order cosmological perturbation theory directly built on the past light-cone, which we are going to recall in the next section.

\section{Second-Order Perturbation Theory on the Light-Cone}
\label{sec:LC-pert-theory}
In this section,  following the  works \cite{Fanizza:2020xtv, Fanizza:2023ixk,Bechaz:2025ojy}, we will introduce the formalism necessary to compute the first- and second-order relativistic corrections  to the redshift drift. In particular, by building a perturbation theory on top of a background light-cone geometry, we will provide a set of  gauge-invariant variables up to  second order corresponding to the values that perturbations acquire in the perturbative counterparts of the fully non-linear GLC gauge.

\subsection{Standard and light-cone metric perturbations}
\paragraph{Standard perturbations.} In standard perturbation theory, using the conformal time and spherical spatial coordinates, $y^\mu = (\eta, r, \theta^a)$, the perturbed FLRW metric is 
\begin{align}
\text{d}s^2 & \equiv g_{\mu \nu} \text{d}y^\mu \text{d}y^\nu = (\bar{g}_{\mu \nu} +  g^{(1)}_{\mu \nu} +g^{(2)}_{\mu \nu} )\text{d}y^\mu \text{d}y^\nu \notag \\[1ex]
& = a^2(\eta) \Big [-\big (1+2(\phi^{(1)}+\phi^{(2)})\big )\text{d}\eta^2 -2(\mathcal{B}^{(1)}_r+\mathcal{B}^{(2)}_r) \text{d}\eta \text{d}r -2 (\mathcal{B}^{(1)}_a+ \mathcal{B}^{(2)}_a) \text{d}\eta \text{d} \theta^a  \notag \\[1ex]
& \quad + (1+ \mathcal{C}^{(1)}_{rr}+\mathcal{C}^{(2)}_{rr}) \text{d}r^2  + 2 (\mathcal{C}^{(1)}_{ra} + \mathcal{C}^{(2)}_{ra})\text{d} r \text{d} \theta^a+ (\bar{\gamma}_{ab} + \mathcal{C}^{(1)}_{ab}+ \mathcal{C}^{(2)}_{ab}) \text{d}\theta^a \text{d}\theta^b \Big ] \, ,
\label{eq:FLRW-metric-perturb}
\end{align}
where $\bar{g}_{\mu \nu}$ is the homogeneous and isotropic background metric, the 2-dimensional angular metric is $\bar{\gamma}_{ab} = \text{diag} \left(r^2, r^2 \sin^2 \theta\right)$, while $g^{(1)}_{\mu \nu}$ and $g^{(2)}_{\mu \nu}$ respectively stand for first- and second-order fluctuations. 
Using an index $i=(r, \theta, \varphi)$ to label spatial coordinates and adopting from now on the superscript $n = 1,2$ to indicate first- and second-order perturbations, one can split the fluctuations according to the Scalar-Vector-Tensor (SVT) decomposition, \textit{i.e.}
\begin{align}
\mathcal{B}^{(n)}_i &= \partial_i B^{(n)} + B^{(n)}_i  \qquad , \qquad \mathcal{C}^{(n)}_{ij} = -2\psi^{(n)} \bar{\gamma}_{ij} +2 D_{ij}E^{(n)} +2 \nabla_{(i}F^{(n)}_{j)}+2h^{(n)}_{ij}   \, , 
\end{align}
where 
\begin{align}
\bar{\gamma}_{ij} = \text{diag}\left(1, \bar{\gamma}_{ab}\right) \qquad , \qquad D_{ij} \equiv \nabla_{(i} \nabla_{j)} -\frac{1}{3} \bar{\gamma}_{ij} \Delta_3 \, , 
\end{align}
and the symmetrization between the two lower indices of two vectors is defined as $X_{(\mu}Y_{\nu)} \equiv (1/2)(X_\mu Y_\nu + X_\nu Y_\mu)$. We recall that, in the SVT decomposition, the vectors $B^{(n)}_i$ and $F^{(n)}_i$ are divergenceless, and the tensor $h^{(n)}_{ij}$ is traceless and divergenceless, namely
\begin{equation}
\nabla^i B^{(n)}_i =0 \qquad  , \qquad \nabla^i F^{(n)}_i =0 \qquad , \qquad h^{(n)}_{ii}=0 \qquad ,  \qquad \nabla^i h^{(n)}_{ij}=0 \, .
\end{equation}

\paragraph{Light-cone perturbations.} We can now move to the set of background light-cone coordinates indicated as $x^\mu = (\tau, w, \tilde{\theta}^a)$, where $\tau$ is the proper time of a time-like particle, $w$ defines the background light-cone of an observer and the angles $\tilde{\theta}^a$ coincide with the two polar angles  we observe in the sky. 
These coordinates are related to $y^\mu$ through the relations
\begin{equation}
\eta (\tau)-\eta_{\t{in}} = \int_{\tau_{\text{in}}}^\tau \frac{\text{d}\tau^\prime}{a(\tau^\prime)} \qquad  , \qquad r = w - \eta(\tau) \qquad  ,  \qquad \theta^a = \tilde{\theta}^a \, ,
\label{eq:spherical-to-glc}
\end{equation}
where  the initial condition $\eta_{\text{in}}$ is chosen to cancel with the integral on the r.h.s. evaluated at $\tau_{\text{in}}$. Using the  transformation rule under  diffeomorphisms, \textit{i.e.} 
\begin{equation}
f_{\mu \nu} = \frac{\partial y^\rho}{\partial x^\mu} \frac{\partial y^\sigma}{\partial x^\nu} \, g_{\rho \sigma}  \, ,
\label{eq:diff}
\end{equation}
we can compute the background light-cone geometry $\bar{f}_{\mu \nu}$ and then add first- and second-order perturbations on top of it:
\begin{align}
\text{d}s^2 & \equiv f_{\mu \nu} \text{d}x^\mu \text{d}x^\nu = (\bar{f}_{\mu \nu} +  f^{(1)}_{\mu \nu}+ f^{(2)}_{\mu \nu}) \text{d}x^\mu \text{d}x^\nu  \notag \\[1ex]
& = a^2(\tau) \bigg [(L^{(1)}+L^{(2)}) \text{d}\tau^2 -\frac{2}{a}\big (1-a(M^{(1)}+M^{(2)}) \big ) \text{d}\tau \text{d}w\notag \\[1ex]
& \quad + 2 (V^{(1)}_a+V^{(2)}_a)  \text{d}\tau \text{d}\tilde{\theta}^a +(1+N^{(1)}+N^{(2)})\text{d}w^2  \notag \\[1ex]
& \quad + 2  (U^{(1)}_a+U^{(2)}_a) \text{d}w \text{d}\tilde{\theta}^a   +(\bar{\gamma}_{ab}+ \gamma^{(1)}_{ab}+\gamma^{(2)}_{ab}) \text{d}\tilde{\theta}^a \text{d}\tilde{\theta}^b \bigg ]  \, .
\label{eq:metricGLC}
\end{align}
Here, we are using 
\begin{equation}
\bar{\gamma}_{ab} =  \big [ w- \eta(\tau)\big ]^2\text{diag} (1, \sin^2 \tilde{\theta}^1) \equiv \big [ w- \eta(\tau)\big ]^2 q_{ab} \, , 
\end{equation}
where $q_{ab}$ is the metric induced on the unit 2-dimensional sphere $\mathbb{S}^2$. Then, we indicate with $D_a$ the covariant derivative on $\mathbb{S}^2$ and with $\tilde{D}_a \equiv \epsilon^b_a D_b$ its dual, where $\epsilon^b_a$ is the covariant volume form on $\mathbb{S}^2$ defined as
\begin{equation}
    \ka_{ab} \equiv \sqrt{\t{det}[q_{cd}]}  \, \varepsilon_{ab}  = \sin \tilde{\theta}^1 \, \varepsilon_{ab} \, , 
    \label{eq:volume-form}
\end{equation}
$\varepsilon_{ab}$ being the totally anti-symmetric Levi-Civita symbol.

The perturbations in Eq.~\eqref{eq:metricGLC} can be split into scalars and pseudo-scalars (indicated with a \quotes{hat}) as
\begin{equation}
\begin{split}
V^{(n)}_a & =r^2 \big [ D_a v^{(n)} + \tilde{D}_a \hat{v}^{(n)} \big ]  \, ,  \\[1ex]
U^{(n)}_a &=r^2 \big [ D_a u^{(n)}+ \tilde{D}_a \hat{u}^{(n)}  \big ]\, , \\[1ex]
 \gamma^{(n)}_{ab} &= 2r^2 \big [ q_{ab} \nu^{(n)} + D_{ab} \mu^{(n)} + \tilde{D}_{ab} \hat{\mu}^{(n)}  \big ] \, , 
 \label{eq:SPS-vec-tensor}
\end{split}
\end{equation}
where we have defined the traceless derivative operators
\begin{equation}
D_{ab} \equiv D_{(a}D_{b)} - \frac{1}{2}q_{ab}D^2 \qquad  , \qquad  \tilde{D}_{ab} \equiv \tilde{D}_{(a}D_{b)}= D_{(a} \tilde{D}_{b)} \, 
\label{eq:D-ab-def}
\end{equation}
with $D^2$  the angular Laplacian. This is the so-called \textit{Scalar-PseudoScalar} (SPS) decomposition, because the perturbations $L^{(n)}$, $M^{(n)}$, $N^{(n)}$, $v^{(n)}$, $u^{(n)}$, $\nu^{(n)}$ and $\mu^{(n)}$ are scalars under rotations on $\mathbb{S}^2$, while $\hat{v}^{(n)}$, $\hat{u}^{(n)}$ and $\hat{\mu}^{(n)}$ behave as pseudo-scalars.

\paragraph{Map between standard and light-cone perturbations.}
For future use, here we recall from \cite{Fanizza:2020xtv,Bechaz:2025ojy} the dictionary between the standard metric perturbations $f^{(n)}_{\mu \nu}$ and the light-cone ones $g^{(n)}_{\mu \nu}$. In order to do so, exploiting the diffeomorphism law of Eq.~\eqref{eq:diff}, 
we obtain the following relations and their inverse expressions \cite{Fanizza:2020xtv}:
\begin{align}
\begin{cases}
a^2L = -2 \big ( \phi - \frac{1}{2} \mathcal{C}_{rr} - \mathcal{B}_r\big )\\
aM = - \mathcal{B}_{r}- \mathcal{C}_{rr}\\
N = \mathcal{C}_{rr}\\
aV_a = -\mathcal{B}_a - \mathcal{C}_{ra}\\
U_a = \mathcal{C}_{ra}\\
\delta \gamma_{ab} = \mathcal{C}_{ab} 
\end{cases}
\qquad \Rightarrow \qquad 
\begin{cases}
\phi =-\frac{1}{2} ( a^2 L + N + 2aM )\\
\mathcal{B}_r =  -N-aM\\
\mathcal{C}_{rr} = N\\
\mathcal{B}_a = -U_a - aV_a\\
\mathcal{C}_{ra} = U_a\\
\mathcal{C}_{ab} =\delta \gamma_{ab}
\end{cases} \,\,\,\,\,\,\,\, , 
\label{eq:dictionary}
\end{align}
where we have omitted the superscript $n = 1,2$ in the above equations  because they have a fully non-linear behavior, meaning that they are valid at any perturbative order. 

As a consequence, using Eqs.~\eqref{eq:dictionary}, we can express the SPS variables in terms of the SVT ones at any order in perturbation theory, with the following results \cite{Bechaz:2025ojy}: 
\begin{align}
    L &= -\frac{2}{a^2} \bigg \{\phi + \psi - \bigg (\nabla_r \nabla_r - \frac{1}{3}\Delta_3 \bigg ) E - \partial_r B - B_r - \nabla_r F_r - h_{rr}\bigg \} \, , \notag \\[1ex]
    M &= -\frac{1}{a} \bigg \{-2\psi + 2 \bigg (\nabla_r \nabla_r - \frac{1}{3}\Delta_3 \bigg ) E + \partial_r B + B_r + 2 \nabla_r F_r + 2 h_{rr}  \bigg \} \, , \notag \\[1ex]
    N &= -2\psi + 2 \bigg (\nabla_r \nabla_r - \frac{1}{3}\Delta_3  \bigg ) E + 2 \nabla_r F_r + 2 h_{rr} \, , \notag \\[1ex]
    v &=-\frac{1}{a}\frac{1}{D^2}\bigg \{ \bar{\ga}^{ab} D_{a} \big [ 2 \nabla_{(r}\nabla_{b)}E + \partial_{b}B + B_{b} + 2 \nabla_{(r} F_{b)} + 2h_{rb}\big ] \bigg \}\, \, , \notag \\[1ex]
    \hat{v} &=-\frac{1}{a}\frac{1}{D^2}\bigg \{ \bar{\ga}^{ab} \tilde{D}_{a} \big [2 \nabla_{(r}\nabla_{b)}E + \partial_{b}B + B_{b} + 2 \nabla_{(r} F_{b)} + 2h_{rb}\big ] \bigg \}\, , \notag \\[1ex]
    u &= \frac{2}{D^2}\bigg \{ \bar{\ga}^{ab} D_{a} \big [\nabla_{(r}\nabla_{b)}E + \nabla_{(r}F_{b)}+ h_{rb}\big ] \bigg \}\, , \notag \\[1ex] 
    \hat{u} &= \frac{2}{D^2}\bigg \{ \bar{\ga}^{ab} \tilde{D}_{a} \big [\nabla_{(r}\nabla_{b)}E + \nabla_{(r}F_{b)}+ h_{rb}\big ] \bigg \}\, , \notag \\[1ex]
    \nu &= - \bigg ( \psi + \frac{1}{3}\Delta_3 E \bigg ) + \frac{\bar{\ga}^{ab}}{2} \bigg \{\nabla_{a} \big (\nabla_{b}E+ F_{b} \big )+ h_{ab} \bigg \}\, , \notag \\[1ex]
    \mu &= \frac{2}{r^2} \frac{1}{D^2 (D^2+2)} \bigg \{D^{ab} \bigg [\nabla_{a} \big (\nabla_{b}E + F_{b} \big ) + h_{ab} \bigg ] \bigg \}\, , \notag \\[1ex]
    \hat{\mu} &= \frac{2}{r^2} \frac{1}{D^2 (D^2+2)} \bigg \{\tilde{D}^{ab} \bigg [\nabla_{a} \big (\nabla_{b}E + F_{b} \big ) + h_{ab} \bigg ] \bigg \}\, .
    \label{eq:SPS-in-terms-SVT}
\end{align}
As stressed in \cite{Bechaz:2025ojy},  relativistic effects on any light-like observable can be more easily computed directly on the past light-cone of an observer. Then,  the  relations \eqref{eq:SPS-in-terms-SVT} can be exploited to express the results in terms of standard perturbations, in particular using  gauge-invariant gravitational potentials, as we did in \cite{Bechaz:2025ojy} for the angular distance–redshift relation and will be doing for the redshift drift in Sect.~\ref{sec:RD-perturbations}.

\subsection{Gauge invariance on the light-cone} 
\label{subsec:gaue-inv-LC}
In this subsection, following \cite{Bechaz:2025ojy}, we will recall how  the non-linear gauge-invariant light-cone perturbation theory is built, and we will show, as done in \cite{Fanizza:2020xtv,Bechaz:2025ojy}, how the residual gauge freedom around the observer position can be completely removed both at first and second order.

\paragraph{Gauge-invariant variables.} In the perturbed light-cone metric of Eq.~\eqref{eq:metricGLC} there are 10 d.o.f.s because the gauge has not yet been specified. To fix the GLC gauge and thus eliminate 4 unphysical d.o.f.s, we can compare the entries of the fully non-linear GLC metric of Eq.~\eqref{eq:GLCmetric-start} with those of  Eq.~\eqref{eq:metricGLC}. 

The  calculations were fully worked out in \cite{Fanizza:2020xtv} at first-order and in \cite{Bechaz:2025ojy} at second-order. We redirect the interested reader to these references for all the technical details. Here we limit ourselves to reporting the set of first- and second-order gauge-invariant variables amounting to the values that perturbations acquire in the perturbative counterparts of the GLC gauge.  To do so, we start from the second-order gauge transformation on the coordinates $x^\mu = (\tau, w, \tilde{\theta}^a)$, \textit{i.e.}
\begin{equation}
x^\mu \rightarrow \tilde{x}^\mu \simeq x^\mu + \xi^\mu_{(1)}+ \frac{1}{2}  \left (\xi^\nu_{(1)} \partial_\nu \xi^\mu_{(1)}+ \xi^\mu_{(2)} \right )  \, , 
\label{eq:gauge-trans-coord}
\end{equation}
and further decompose the angular components of the gauge modes $\xi^\mu_{(n)}$ (for $n=1,2$) as
\begin{equation}
\xi^a_{(n)} = q^{ab} \big ( D_b \chi_{(n)}+ \tilde{D}_b \hat{\chi}_{(n)}\big ) \, , 
\end{equation}
 where $\chi_{(n)}$ are pure scalar gauge d.o.f.s while $\hat{\chi}_{(n)}$ are pure pseudo-scalar ones.
Then, we compute the gauge transformations of all the metric perturbations  of Eq.~\eqref{eq:metricGLC} using 
\begin{align}
    \tilde{f}^{(1)}_{\mu \nu} &= f^{(1)}_{\mu \nu} - \mathcal{L}_{\xi_{(1)}}\bar{f}_{\mu \nu} \, , \nex
    \tilde{f}^{(2)}_{\mu \nu} &= f^{(2)}_{\mu \nu}-\mathcal{L}_{\xi_{(1)}}f^{(1)}_{\mu \nu}+ \frac{1}{2} \left (\mathcal{L}^2_{\xi_{(1)}}\bar{f}_{\mu \nu}-\mathcal{L}_{\xi_{(2)}}\bar{f}_{\mu \nu}\right ) \, , 
\label{eq:gauge-2nd-rule}
\end{align}
where $\mathcal{L}_{\xi_{(1)}}$ and $\mathcal{L}_{\xi_{(2)}}$ are the Lie derivatives along the first- and second-order gauge modes.

In particular, the explicit forms of $\xi^\mu_{(n)}$ are the ones needed to move to the linearized and quadratic counterparts of the GLC gauge within the light-cone perturbation theory. These  conditions are \cite{Fanizza:2020xtv, Bechaz:2025ojy}
\begin{align}
& \tilde{L}^{(1)}=0 \, , \quad\quad\quad\quad\quad\quad\quad\quad \quad\quad\quad\quad \tilde{L}^{(2)}=0\,,\notag \\[1ex]
& \tilde{v}^{(1)} =0 \, ,\quad\quad\quad\quad\quad\quad\quad\quad \quad\quad\quad\quad \tilde{v}^{(2)} =0\, , \notag \\[1ex]
& \tilde{\hat{v}}^{(1)} =0 \, ,\quad\quad\quad\quad\quad\quad\quad\quad \quad\quad\quad\quad  \tilde{\hat{v}}^{(2)} =0\, , \notag \\[1ex]
& \tilde{N}^{(1)}+2a\tilde{M}^{(1)} =0 \, ,\quad\quad \quad\quad\quad\quad\quad \tilde{N}^{(2)}- \frac{1}{4}
(\tilde{N}^{(1)})^2 +2a\tilde{M}^{(2)}- \tilde{U}^2_{(1)} =0\, . 
\label{eq:fixing-GLC-gauge-orde1-2}
\end{align}
It follows that the solutions for the first-order gauge modes are \cite{Fanizza:2020xtv}
\begin{align}
\xi^\tau_{(1)} &= -\frac{1}{2}\int_{\tau_{\text{in}}}^\tau \text{d}\tau^\prime \, \big ( a^2L^{(1)} + N^{(1)}+2aM^{(1)}\big )\big (\tau^\prime, w-\eta (\tau)+ \eta (\tau^\prime) \big ) \ ,  \notag \\[1ex]
\xi^w_{(1)} &= \frac{1}{2} \int_\tau^{\tau_\t{o}} \text{d}\tau^\prime aL^{(1)} + w^{(1)}\o(w) \, , \notag\\[1ex]
\chi_{(1)} &= - \int_\tau^{\tau_\t{o}} \text{d}\tau^\prime \bigg (v^{(1)}+\frac{1}{2ar^2} \int_{\tau^\prime}^{\tau_\t{o}} \text{d}\tau^{\prime \prime} aL^{(1)} + \frac{w^{(1)}_0}{ar^2} \bigg )+ \chi^{(1)}\o(w, \tilde{\theta}^a) \, , \notag \\[1ex]
\hat{\chi}_{(1)} &= -\int_\tau^{\tau_\t{o}} \text{d}\tau^\prime \, \hat{v}^{(1)} + \hat{\chi}^{(1)}\o(w, \tilde{\theta}^a) \, , 
\label{eq:xi-1st-order}
\end{align}
where $\tau\o$ is the present time, and the  free  functions $w\o^{(1)}$, $\chi\o^{(1)}$ and $\hat{\chi}\o^{(1)}$ account for the residual gauge transformations of Eqs.~\eqref{eq:glc-residual-freedom}. 

The solutions for the second-order gauge modes are \cite{Bechaz:2025ojy}
\begin{align}
\xi^\tau_{(2)} & =  \int_{\tau_{\text{in}}}^\tau \text{d}\tau^{\prime} \bigg (-(a^2L^{(2)}+2aM^{(2)}+N^{(2)})+\frac{1}{4}(N^{(1)})^2+F^{(2)} \bigg )\big (\tau^{\prime}, w-\eta (\tau) + \eta (\tau^{\prime}) \big ) \, , \nex
\xi^w_{(2)} & = \int_\tau^{\tau_\t{o}} \text{d}\tau^\prime \, a \bigg [ L^{(2)} -\xi^i_{(1)} \partial_i L^{(1)}+ aL^{(1)} \bigg (M^{(1)}+ \frac{a}{4}L^{(1)} \bigg )-r^2 q^{ab} \mathscr{L}^{(2)}_{ab} \bigg ] \notag\\[1ex]
& \quad + a L^{(1)}\xi^\tau_{(1)}+\xi^\mu_{(1)}\pa_\mu \xi^w_{(1)} + w^{(2)}\o (w) \, , \nex
\chi_{(2)} 
& = - \int_\tau^{\tau_\t{o}} \text{d}\tau^\prime \bigg \{
2v^{(2)}+\frac{1}{ar^2} \bigg [\int_{\tau^\prime}^{\tau_\t{o}} \text{d}\tau^{\prime \prime} \, a \bigg (L^{(2)}-\xi^i_{(1)}\partial_i L^{(1)}+ aL^{(1)}\bigg (M^{(1)}+\frac{a}{4}L^{(1)} \bigg ) \notag \\[1ex]
& \quad - r^2 q^{ab} \mathscr{L}^{(2)}_{ab}
\bigg ) +aL^{(1)}\xi^\tau_{(1)}+w^{(2)}\o (w, \tilde{\theta}^a)\bigg ] + \frac{2}{r^2D^2}\mc{V}^{(2)}
\bigg \}  + \chi^{(2)}\o (w, \tilde{\theta}^a)\, , \notag \\[1ex]
\hat{\chi}_{(2)} &=-2 \int_\tau^{\tau_\t{o}} \text{d}\tau^\prime \, \bigg [\hat{v}^{(2)}+ \frac{1}{r^2D^2}\hat{\mc{V}}^{(2)} \bigg ] + \hat{\chi}^{(2)}\o(w, \tilde{\theta}^a) \, , 
\label{eq:xi-2nd-order}
\end{align}
where the quantities $F^{(2)}$, $\mathscr{L}^{(2)}_{ab}$, $\mathcal{V}^{(2)}$ and $\hat{\mathcal{V}}^{(2)}$ are reported in Appendix~\ref{app:gauge-inv-formulae}. Analogously to the first-order gauge modes of Eqs.~\eqref{eq:xi-1st-order}, the presence of the functions $w^{(2)}\o$, $\chi\o^{(2)}$ and $\hat{\chi}\o^{(2)}$ is a consequence of the residual gauge freedom of the GLC gauge described in Eqs.~\eqref{eq:glc-residual-freedom}.

Then, by using Eqs.~\eqref{eq:xi-1st-order} and \eqref{eq:xi-2nd-order}, we compute a set of first- and second-order gauge-invariant variables, which are identified as the values that perturbations acquire in the linearized and quadratic versions of the perturbed GLC gauge. At first order we have \cite{Fanizza:2020xtv}
\begin{align}
\mathscr{V}^{(1)} &\equiv \nu^{(1)} - \frac{1}{2}D^2 \chi_{(1)} - \xi^\tau_{(1)} \bigg (H-\frac{1}{ar} \bigg )- \frac{\xi^w_{(1)}}{r} \, , \notag \\[1ex]
\mathscr{N}^{(1)} & \equiv N^{(1)}-2H\xi^\tau_{(1)} +\frac{2}{a}\partial_w \xi^\tau_{(1)} - 2\partial_w \xi^w_{(1)} \, , \notag \\[1ex]
\mathscr{M}^{(1)} &\equiv  \mu^{(1)} - \chi_{(1)} \, , \notag \\[1ex]
\mathscr{\hat{M}}^{(1)} &\equiv \hat{\mu}^{(1)} -\hat{\chi}_{(1)} \, , \notag \\[1ex]
\mathscr{U}^{(1)} &\equiv u^{(1)} + \frac{\xi^\tau_{(1)}}{ar^2} - \frac{\xi^w_{(1)}}{r^2}-\partial_w \chi_{(1)} \, , \notag \\[1ex]
\mathscr{\hat{U}}^{(1)} &\equiv \hat{u}^{(1)} - \partial_w \hat{\chi}_{(1)} \, , 
\label{eq:gauge-inv-quantities-1st-order}
\end{align}
whereas at second order we obtain \cite{Bechaz:2025ojy}
\begin{align}
\mathscr{V}^{(2)} & \equiv \nu^{(2)} -\frac{1}{4}D^2 \chi_{(2)}- \frac{1}{2}\xi^\tau_{(2)} \bigg (H-\frac{1}{ra} \bigg )-\frac{1}{2}\frac{\xi^w_{(2)}}{r}+\mathbb{V}^{(2)} \, , \notag \\[1ex]
\mathscr{N}^{(2)} & \equiv N^{(2)}-H\xi^\tau_{(2)}+\frac{1}{a}\partial_w \xi^\tau_{(2)}-\partial_w \xi^w_{(2)}+\mc{N}^{(2)} \, , \notag \\[1ex]
\mathscr{M}^{(2)} & \equiv \mu^{(2)}-\frac{1}{2}\chi_{(2)} +\frac{1}{r^2} \mathcal{D}^{-1} D^{ab} \mc{M}^{(2)}_{ab} \notag \, , \\[1ex]
\hat{\mathscr{M}}^{(2)} & \equiv \hat{\mu}^{(2)}-\frac{1}{2}\hat{\chi}_{(2)} +\frac{1}{r^2} \mathcal{D}^{-1} \tilde{D}^{ab} \mc{M}^{(2)}_{ab} \, , 
\notag \\[1ex]
\mathscr{U}^{(2)} & \equiv u^{(2)}+ \frac{1}{2ar^2}\xi^\tau_{(2)}-\frac{1}{2r^2} \xi^w_{(2)}-\frac{1}{2} \partial_w \chi_{(2)}-\frac{1}{2ar^2}\xi^\mu_{(1)} \partial_\mu \xi^\tau_{(1)}+ \frac{1}{r^2D^2}\mc{U}^{(2)} \, , 
\notag \\[1ex]
\hat{\mathscr{U}}^{(2)} & \equiv \hat{u}^{(2)}- \frac{1}{2} \partial_w \hat{\chi}_{(2)}+ \frac{1}{r^2D^2}\hat{\mc{U}}^{(2)} \, ,
\label{eq:glc-gauge-inv-2nd-order}
\end{align}
where 
\begin{align}
\mc{N}^{(2)} & \equiv  r^2 \big [-2 (D_a u^{(1)}+\tilde{D}_a \hat{u}^{(1)})+ \partial_w (D_a \chi_{(1)}+\tilde{D}_a \hat{\chi}_{(1)}) \big ] \big [q^{ab}\partial_w (D_b \chi_{(1)}+\tilde{D}_b \hat{\chi}_{(1)})] \notag \\[1ex]
& \quad + \frac{\dot{a}}{a^2}\xi^\tau_{(1)} \big (\dot{a} \xi^\tau_{(1)}-2\partial_w  \xi^\tau_{(1)} \big ) -2 \big (M^{(1)}\partial_w \xi^\tau_{(1)}+  N^{(1)} \partial_w \xi^w_{(1)} \big ) + \big (\partial_w \xi^w_{(1)} \big )^2  \notag \\[1ex]
& \quad + \partial_w \big (\xi^\mu_{(1)}\partial_\mu \xi^w_{(1)} \big ) - \xi^\mu_{(1)}\partial_\mu N^{(1)}-\frac{1}{a} \bigg \{-\ddot{a} (\xi^\tau_{(1)})^2-\dot{a} \xi^\mu_{(1)}\partial_\mu \xi^\tau_{(1)}+2 \partial_w \xi^\tau_{(1)} \partial_w \xi^w_{(1)}\notag \\[1ex]
& \quad + \partial_w \big (\xi^\mu_{(1)} \partial_\mu \xi^\tau_{(1)} \big )+ \dot{a}\xi^\tau_{(1)} \big (2N^{(1)}-4 \partial_w \xi^w_{(1)} \big ) \bigg \} \, , 
\label{eq:definitions-glc-gauge-inv-2nd}
\end{align}
and $\mathcal{D}$ is the operator defined as $\mathcal{D} \equiv \left ((D^2)^2 + 2D^2 \right )$, while the \quotes{dot} indicates the derivative w.r.t. the time $\tau$.
The quantities $\mathbb{V}^{(2)}$,  $\mc{M}^{(2)}_{ab}$, $\mc{U}^{(2)}$ and $\hat{\mc{U}}^{(2)}$ are not needed for the purposes of this paper. However, for the sake of completeness, we report their expressions in Appendix~\ref{app:gauge-inv-formulae}.

\paragraph{Evaluation in terms of Bardeen potentials.}  Since  the variables written in Eqs.~\eqref{eq:gauge-inv-quantities-1st-order} and \eqref{eq:glc-gauge-inv-2nd-order} are gauge invariant, they can be used to compute gauge-invariant observables which can be re-expressed in any gauge. We can then rewrite these observables in terms of the gravitational (or Bardeen) potentials $\Phi$ and $\Psi$, which correspond to the value that the standard perturbations $\phi$ and $\psi$ appearing in Eq.~\eqref{eq:FLRW-metric-perturb} acquire in the so-called Poisson Gauge (PG). The latter is defined by the conditions $E^{(n)}=0$ and $B^{(n)}=0$, at any perturbative order $n$, by neglecting vector and tensor fluctuations. Therefore,  we   fix the gauge corresponding to PG within the light-cone  perturbation theory,  by imposing the two aforementioned relations in Eqs.~\eqref{eq:SPS-in-terms-SVT}, valid at any order, and obtaining the following non-perturbative definition of the PG:
\begin{align}
    L &= -\frac{2}{a^2} (\Phi + \Psi) \quad  , \quad  M = \frac{2}{a}\Psi \quad  , \quad N = -2\Psi \quad  , \quad \nu = -\Psi \, ,  \notag \\[1ex]
    v &=0 \quad  , \quad \hat{v} =0 \quad 
    ,\quad u = 0
    \quad ,\quad \hat{u} = 0
    \quad ,\quad \mu = 0
    \quad ,\quad \hat{\mu} =0 
    \, .
\label{eq:SPS-NG}
\end{align}
From now on, following the notation of \cite{Fanizza:2015swa,Marozzi:2014kua,BenDayan:2012wi}, we will use the definitions
\begin{equation}
  P \equiv \frac{1}{a(\tau)}\int_{\tau_{\t{in}}}^\tau \t{d}\tau^\pr \, \Phi^{(1)}\big ( w-\eta(\tau)+\eta(\tau^\pr)\big ) \qquad 
   , \qquad Q \equiv \int_{\tau}^{\tau_\t{o}} \frac{\t{d}\tau^\pr}{a} (\Psi^{(1)}+\Phi^{(1)} ) \, ,
  \label{eq:P-Q-defined}
\end{equation}
 and, for brevity, we will omit the superscript \quotes{1} for the first-order perturbations $\Phi$ and $\Psi$. Moreover, we introduce the isotropic and anisotropic parts of the Bardeen potentials as
\begin{align}
    \Psi^\t{I} \equiv \frac{\Psi +\Phi}{2} \qquad  ,  \qquad  \Psi^\t{A} \equiv \frac{\Psi - \Phi}{2} \, .
    \label{eq:IA-potentials-defined}
\end{align}
Keeping $\Phi \neq \Psi$ already at first order allows us to consider also the case of non-vanishing anisotropic stress in the Einstein equations, which would be relevant to dark energy or modified gravity models (see, for example, \cite{Sawicki:2012re, Amendola:2012ky, Amendola:2013qna, Sobral-Blanco:2021cks, Castello:2022uuu}). 

Since we will extensively use them in Sect.~\ref{sec:RD-perturbations}, here we also report from \cite{Bechaz:2025ojy} the explicit expressions of the first- and second-order gauge modes needed to fix the GLC gauge rewritten in terms of the gravitational potentials. At first order we have
\begin{align}
    \xi^\tau_{(1)} &= aP \, , \notag \\[1ex]
    \xi^w_{(1)} &= -Q + w^{(1)}\o (w, \tilde{\theta}^a) \, , \notag \\[1ex]
    \xi^a_{(1)} &= q^{ab}\int_\tau^{\tau_\t{o}} \frac{\text{d}\tau^\prime}{ar^2}\, \pa_b \big [Q-w^{(1)}\o (w)\big ] + \xi^{a}_{(1)\t{o}} (w, \tilde{\theta}^a) \, , 
     \label{eq:gauge-modes-NG-1st}
\end{align}
and at second order \begin{align}
    \xi^\tau_{(2)}  &= \int_{\tau_{\t{in}}}^\tau \t{d}\tau^\pr \,  \bigg [2\Phi^{(2)}-\Phi^2 + (\pa_w P)^2 + \bar{\ga}^{ab}\pa_a P \pa_b P \bigg ]   - \pa_w P \o  \int_{\tau_{\t{in}}}^\tau \t{d}\tau^\pr \, \pa_w P \nex
    & \quad -a(\Psi^\t{I}-\Psi^\t{A}) P + aP\pa_w P - a (P\o - Q) \pa_w P - a\xi^a_{(1)}\pa_a P \, , \nex
    \xi^w_{(2)}&= \int_{\tau}^{\tau_\t{o}}\, \frac{\t{d}\tau^\pr}{a} \bigg [-4\Psi^\t{I}_{(2)}-\bar{\ga}^{ab} \pa_a Q \pa_b Q +4 \Psi^\t{I}\pa_w Q  - 8 \Psi^\t{I}\Psi^\t{A}-4 (\Psi^\t{I})^2 \bigg ]+2\Psi^\t{I}_\t{o} P\o -2  \Psi^\t{I}P\nex
    & \quad  +(P\o -2Q)\pa_w Q - Q (\pa_w P\o -\pa_w Q) + \xi^a_{(1)}\pa_a Q+w^{(2)}\o(w, \tilde{\theta}^a)\,  ,\nex
    \xi^a_{(2)} &= q^{ab}\int_{\tau}^{\tau_\t{o}}\frac{\t{d}\tau^\pr}{ar^2} \, \bigg \{\int_{\tau^\pr}^{\tau_\t{o}}\frac{\t{d}\tau^{\pr \pr}}{a} \, \pa_b \Big [4\Psi^\t{I}_{(2)} + \bar{\ga}^{cd}\pa_c Q \pa_d Q - 4\Psi^\t{I}\pa_w Q +8\Psi^\t{I}\Psi^\t{A}+4(\Psi^\t{I})^2 \Big ]\nex
    & \quad -2 \pa_b (\xi^c\pa_c Q)- \pa_b \big [(P\o -2Q)\pa_w Q \big ]+\frac{2}{r}(P\o -2Q)\pa_b Q + 2 \pa_c Q (\pa_b \xi^c + \bar{\ga}^{dc}\xi^e\pa_e \bar{\ga}_{bd})\nex
    & \quad +4\Psi^\t{I}\pa_b Q \bigg \} + \bar{\ga}^{ab}P \pa_b Q +\xi^w_{(1)}\pa_w \xi^a_{(1)}+\xi^b_{(1)}\pa_b \xi^a_{(1)}+ \xi^a_{(2)\t{o}}(w, \tilde{\theta}^a)  \, . 
    \label{eq:gauge-modes-NG-2nd}
\end{align}
Such Eqs.~\eqref{eq:gauge-modes-NG-1st} and \eqref{eq:gauge-modes-NG-2nd} are derived by plugging the conditions \eqref{eq:SPS-NG} directly into the gauge modes written in Eqs.~\eqref{eq:xi-1st-order} and \eqref{eq:xi-2nd-order}. This is because,  adopting a gauge-invariant procedure, we can already fix the PG at the level of
the gauge modes.

\paragraph{Gauge fixing at the observer position.} As we have already pointed out, the functions $w^{(n)}\o(w, \tilde{\theta}^a)$ and $\xi^a_{(n)\t{o}}(w, \tilde{\theta}^a)$  appearing in Eqs.~\eqref{eq:gauge-modes-NG-1st} and \eqref{eq:gauge-modes-NG-2nd} reflect the fact that the GLC gauge is defined up to the residual transformations of Eqs.~\eqref{eq:glc-residual-freedom}. This residual gauge freedom around the observer position  is removed once a  choice of the free functions $w^{(n)}\o(w, \tilde{\theta}^a)$ and $\xi^a_{(n)\t{o}}(w, \tilde{\theta}^a)$ is specified.

We choose the observer to be set at the center of the polar frame, \textit{i.e.} $r\o = w\o-\eta(\tau \o)=0$. Then, in order to preserve the condition $w\o=\eta(\tau \o)$ in any other gauge, by taking the first-order gauge transformations
\begin{equation}
    \tau \rightarrow \tilde{\tau} = \tau + \xi^\tau_{(1)} \qquad , \qquad w \rightarrow \tilde{w} = w + \xi^w_{(1)}  \,  , 
\end{equation}
we have to demand that, locally around the observer, it holds
\begin{equation}
    \left(\xi^w_{(1)}\right)\o \equiv w^{(1)}\o =\frac{\left(\xi^\tau_{(1)}\right)\o}{a\o} \, , \label{eq:monopole-firstorder}
\end{equation}
so that $w^{(1)}\o$ is a pure monopole, namely $w^{(1)}\o=w^{(1)}\o(w)$.

Furthermore,  the residual angular gauge mode $\xi^a_{(1)\t{o}}(w, \tilde{\theta}^a)$ can be fixed by requiring that  the angles $\tilde{\theta}^a$ coincide with the \textit{observed} angular directions of incoming photons in the sky, as in the so-called Fermi Normal Coordinates  \cite{Fanizza:2018tzp}. Let us consider the first-order gauge transformation mapping the angles as $\tilde{\theta}^a_\t{o} \rightarrow \tilde{\theta}^a_\t{o} + \xi^a_{(1)\t{o}}$. In the GLC gauge, the angles are precisely the background ones of the polar frame, namely $\tilde{\theta}^a = \theta^a_\t{o}$ (see also \cite{Fanizza:2015swa}), hence 
\begin{align}
   \xi^a_{(1)}(w\o,\tilde{\theta}^a\o)=0  \, . \label{eq:angles-residual}
\end{align}
More in general, in this \quotes{observational gauge}, since no preferable angular  direction at the center of the polar frame is allowed, any quantity $X$ evaluated at the observer position must satisfy $D_a X \o \equiv 0$.

As we have discussed in \cite{Bechaz:2025ojy}, the first-order gauge conditions \eqref{eq:monopole-firstorder} and \eqref{eq:angles-residual} can be consistently  extended also to second order. Thus, the gauge at the observer position can be fixed by the geometrical conditions
\begin{align}
    w^{(1,2)}\o(w)= \frac{\xi^\tau_{(1,2)}(w\o)}{a\o} \qquad , \qquad \xi^{a}_{(1,2)\t{o}}(w,\tilde{\theta}^a) = 0\, .
    \label{eq:obs-gauge-fixing}
\end{align}
We refer to \cite{Bechaz:2025ojy} for a more in-depth discussion about such a gauge fixing.

\section{Relativistic Effects on the Redshift Drift}
\label{sec:RD-perturbations}
As highlighted in the Introduction, the robustness of  the second-order light-cone perturbation theory fully developed in  \cite{Bechaz:2025ojy} and summarized in Sect.~\ref{sec:LC-pert-theory} has already been validated in \cite{Bechaz:2025ojy}. In this section we will derive the extended formulae for the first- and second-order relativistic corrections to the redshift drift, by starting from the fully non-linear expression \eqref{eq:rsd-glc-fully-nonlinear} and thoroughly singling out perturbations evaluated at the source and observer position. For the sake of clearness, here we outline the logical steps on which our computations are based:
\begin{enumerate}
    \item  We expand  Eq.~\eqref{eq:rsd-glc-fully-nonlinear}, valid in the GLC gauge,  up to second order, and replace each perturbation with its gauge-invariant counterpart provided in Eqs.~\eqref{eq:gauge-inv-quantities-1st-order} and \eqref{eq:glc-gauge-inv-2nd-order}. This automatically gives a formula for the redshift drift written in terms of the observed light-cone and observed angles;
    \item Then, by Taylor expanding each quantity around the time of the source corresponding to the observed redshift, we obtain the fully gauge-invariant formula for the redshift drift  in terms of the observed redshift;
    \item Finally, we express the  relativistic effects in terms of the gravitational potentials $\Phi$ and $\Psi$ as pointed out in Subsect.~\ref{subsec:gaue-inv-LC}.
\end{enumerate}

\subsection{Redshift perturbations and redshift parameterization}
\label{subsec:redshift-pert}
The perturbations to the redshift can be obtained  by expanding the fully non-linear formula of Eq.~\eqref{eq:z-fully-non-linear-glc} up to second order as
\begin{align}
    1+z & = (1+\bar{z})(1+\delta^{(1)}z+\delta^{(2)}z)\nex
    &=\frac{a_\text{o}}{a_\text{s}} \bigg \{ 1+\Upsilon^{(1)}|^\t{o}_\t{s} +\Upsilon^{(2)}|^\t{o}_\t{s} + (\Upsilon^{(1)}_\t{s})^2 - \Upsilon^{(1)}_\t{o}\Upsilon^{(1)}_\t{s} \bigg \} \nex
    & = \frac{a_\text{o}}{a_\text{s}} \bigg \{ 1+\frac{N^{(1)}|^\text{o}_\text{s}}{2}+\frac{N^{(2)}|^\text{o}_\text{s}}{2}-\frac{(N^{(1)})^2|^\text{o}_\text{s}}{8}+ \bigg (\frac{N^{(1)}_\text{s}}{2}\bigg )^2- \frac{N^{(1)}_\text{o}N^{(1)}_\text{s}}{4}-\frac{1}{2}\left[U_{(1)}^2\right]^\text{o}_\text{s} \bigg \} \, ,
    \label{eq:perturbed-z}
\end{align}
where, to move from the second to the third line,  we  have expressed the functions $\Upsilon$ and $\mathcal{U}^a$ in terms of the GLC perturbations appearing in the metric of Eq.~\eqref{eq:metricGLC}, by writing\footnote{We account for the fact that  $\mathcal{U}^a$ has zero background value, as we have specified after Eqs.~\eqref{eq:GLC-background-FLRW}.} 
\begin{equation}
    \Upsilon = \bar{\Upsilon} (1+ \Upsilon^{(1)} + \Upsilon^{(2)})    \qquad  , \qquad 
    \mathcal{U}^a = \mathcal{U}^a_{(1)}+ \mathcal{U}^a_{(2)} \, ,
\label{eq:decomposition-LC-functions}
\end{equation}
with (see \cite{Bechaz:2025ojy})\footnote{Eqs.~\eqref{eq:Upsilon_pertubations} and \eqref{eq:U-perturbations-map} are obtained by comparing the $ww$ and $wa$ entries of the fully non-linear GLC metric in Eq.~\eqref{eq:GLCmetric-start} with its perturbative counterpart in Eq.~\eqref{eq:metricGLC}.} 
\begin{equation}
\bar{\Upsilon} = a\qquad , \qquad 
\Upsilon^{(1)} = \frac{1}{2}N^{(1)} \qquad , \qquad 
\Upsilon^{(2)} = \frac{1}{2}N^{(2)}- \frac{1}{8} \big (N^{(1)} \big )^2 - \frac{1}{2}U^2_{(1)} \,  ,
\label{eq:Upsilon_pertubations}
\end{equation}
and 
\begin{align}
    \mathcal{U}^a_{(1)} =  \bar{\gamma}^{ab} U^{(1)}_b \qquad , \qquad \mathcal{U}^a_{(2)} =\bar{\gamma}^{ab}  (U^{(2)}_b - \gamma^{(1)}_{bc} \bar{\gamma}^{cd} U^{(1)}_d )\, . \label{eq:U-perturbations-map}
\end{align}

Typically, using cosmological surveys catalogs, for each value of the redshift $z$ of some source, we have  the correspondent value of an observable. Hence, we can perturbatively redefine the proper time at the source as
\begin{equation}
    \tau_\t{s} = \tau_z + \tau^{(1)}_z + \tau^{(2)}_z \, , 
    \label{eq:time-source-expanded}
\end{equation}
where $\tau_z$ is the proper time of the source evaluated at the observed redshift $z$, and $\tau^{(1,2)}_z$ are the distortions at first and second order due to inhomogeneities and anisotropies  along the line of sight (see, for example,  \cite{BenDayan:2012wi,Marozzi:2014kua,Scaccabarozzi:2017ncm, Fanizza:2018qux}). We do not have to perform analogous expansions  also for the coordinates $w$ and $\tilde{\theta}^a$ because, since we are working in the GLC gauge, observations are made on the past light-cone defined by $w = \text{const.}$, and the angles coincide with those of the background. Therefore, by using Eq.~\eqref{eq:time-source-expanded}, we first Taylor expand the scale factor as
\begin{equation}
    a(\tau) = a_z \bigg [1+H_z \tau^{(1)}_z + H_z \tau^{(2)}_z -\frac{1}{2}q_z (\tau^{(1)}_z)^2\bigg ] \, , 
    \label{eq:a-s-z-expanded}
\end{equation}
with
\begin{equation}
    a_z \equiv a(\tau_z) \qquad , \qquad H_z \equiv H(\tau_z) \qquad , \qquad q_z \equiv -\frac{\ddot{a}_z}{a_z} \, .
\end{equation}

More generally, any  function $X$ can be expressed in terms of the observed redshift working up to second order in perturbation theory. In the  simplest case in which its background counterpart only depends on the time $\tau$, namely $\bar{X}(x^\mu) \equiv \bar{X} (\tau)$, as it is the case for the redshift drift of Eq.~\eqref{eq:bkg-RD}, we have that
\begin{equation}
    X(x^\mu) = \bar{X} (\tau) + X^{(1)}(x^\mu) + X^{(2)}(x^\mu) \, . 
\end{equation}
Then, by using Eq.~\eqref{eq:time-source-expanded}, we can perform the following Taylor expansions, \textit{i.e.}
\begin{align}
\bar{X}(\tau) &= \bar{X}(\tau_z) + \dot{\bar{X}}(\tau_z) \tau^{(1)}_z + \dot{\bar{X}}(\tau_z) \tau^{(2)}_z + \frac{1}{2}\ddot{\bar{X}}(\tau_z) (\tau^{(1)}_z)^2 \, ,  \nex
X^{(1)} (x^\mu) &= X^{(1)}(\tau_z) + \dot{X}^{(1)} (\tau_z) \tau^{(1)}_z \, ,  \nex
X^{(2)} (x^\mu) &= X^{(2)} (\tau_z) \, ,
\end{align}
where, for  brevity, we have omitted to  report the dependence of the  perturbations on the other three coordinates $(w, \tilde{\theta}^a)$.
Therefore, we are left with
\begin{equation}
    X(x^\mu) = \bar{X}_z + X^{(1)}_z + X^{(2)}_z \, , 
    \label{eq:expansion-obs-redshift}
\end{equation}
where
\begin{align}
    \bar{X}_z &\equiv \bar{X}(\tau_z) \, ,  \nex
    X^{(1)}_z &\equiv  \dot{\bar{X}}(\tau_z) \tau^{(1)}_z + X^{(1)}(\tau_z) \, , \nex
     X^{(2)}_z &\equiv \dot{\bar{X}}(\tau_z) \tau^{(2)}_z + \frac{1}{2}\ddot{\bar{X}}(\tau_z) (\tau^{(1)}_z)^2 + \dot{X}^{(1)}(\tau_z) \tau^{(1)}_z \, .
     \label{eq:perturbations-with-z}
\end{align}
Then, Eqs.~\eqref{eq:a-s-z-expanded} and \eqref{eq:perturbations-with-z} can be inserted into Eq.~\eqref{eq:perturbed-z} to obtain
\begin{align}
    1+z &=\frac{a_\text{o}}{a_z} \bigg \{1+\frac{N^{(1)}|^\text{o}_z}{2} -\frac{\dot{N}^{(1)}_z}{2}\tau^{(1)}_z + \frac{N^{(2)}|^\text{o}_z}{2} + \frac{3}{8}(N^{(1)}_z)^2 - \frac{(N^{(1)}_\text{o})^2}{8}-\frac{1}{2}[U^2_{(1)}]^\text{o}_z-\frac{N^{(1)}_\text{o}N^{(1)}_z}{4}\notag \\[1ex]
    & \quad -H_z \tau^{(1)}_z -H_z \tau^{(2)}_z +  \bigg (\frac{1}{2}q_z + H^2_z \bigg ) (\tau^{(1)}_z)^2 -\frac{1}{2}H_z \tau^{(1)}_z N^{(1)}|^\text{o}_z 
    \bigg \}   \, ,  
    \label{eq:z-z-expansion}
\end{align}
where, from now on, the subscript \quotes{$z$} means evaluation at the time $\tau_z$.
Thus, following \cite{Fanizza:2020xtv,Bechaz:2025ojy}, by demanding that the redshift is the \textit{observed} one, \textit{i.e.} $1+z = a_\text{o}/a_z$, we obtain  that the first- and second-order fluctuations $\tau^{(1)}_z$ and $\tau^{(2)}_z$ have the expressions
\begin{align}
    \tau^{(1)}_z &= \frac{1}{2H_z} N^{(1)}|^\text{o}_z \, , \notag \\[1ex]
    \tau^{(2)}_z &= \frac{1}{2H_z} \Bigg \{ N^{(2)}|^\text{o}_z-\frac{1}{4}\left(N^{(1)}_\text{o}\right)^2+ \frac{3}{4}\left(N^{(1)}_z\right)^2 -\frac{1}{2}N^{(1)}_\text{o}N^{(1)}_z  \nex
    &  \quad -\left[U^2_{(1)} \right]^\text{o}_z + \frac{1}{4}\frac{q_z}{H^2_z} \left(N^{(1)}|^\text{o}_z\right)^2 - \frac{N^{(1)}\oz \dot{N}^{(1)}_z}{2H_z}\Bigg \} \, .
    \label{eq:time-z-perturbations}
\end{align}

\subsection{Perturbative expansion of the redshift drift}
We now expand Eq.~\eqref{eq:rsd-glc-fully-nonlinear} up to second order, using the second line of Eq.~\eqref{eq:perturbed-z} for the perturbations of the redshift:
\begin{align}
    \frac{\Delta z}{\Delta \tau_\t{o}} &= \frac{a\o}{a_\t{s}} \bigg [ H\o \bigg (1+\U^{(1)}|^\t{o}_\t{s} +\U^{(2)}|^\t{o}_\t{s}   - \U_\t{s}^{(1)}\U \o^{(1)} +\left (\U^{(1)}_\t{s}\right )^2  \bigg )+ \dot{\U}\o^{(1)} + \dot{\U}^{(2)}\o - \dot{\U}\o^{(1)} \U_\t{s}^{(1)}\bigg ]\nex
    & \quad  +\frac{1}{a_\t{s}}\bigg [ \pa_w \U^{(1)}|^\t{o}_\t{s} +\frac{1}{a_\t{s}} \pa_w \U^{(2)}|^\t{o}_\t{s} +2\U_\t{s}^{(1)}\pa_w \U_\t{s}^{(1)}-\U_\t{s}^{(1)}\pa_w \U_\t{o}^{(1)} -\U_\t{o}^{(1)}\pa_w \U_\t{o}^{(1)} \bigg ]\nex 
    & \quad - H_\t{s}-\dot{\U}^{(1)}_\t{s}-\dot{\U}^{(2)}_\t{s}+\U^{(1)}_\t{s}\dot{\U}^{(1)}_\t{s}-\frac{1}{a_\t{s}}U^{a(1)}_{\t{s}}\pa_a \U_\t{s}^{(1)}\, . 
    \label{eq:rsd-2nd-start}
\end{align}

Next, we Taylor-expand the scale factor $a_\t{s}$ and $\U_\t{s}$  according to the redshift parameterization  detailed in Subsect.~\ref{subsec:redshift-pert}.  Thus, omitting for brevity the superscript \quotes{1} for the first-order perturbation $N$, we obtain:
\begin{align}
   \frac{1}{a_\t{s}(\tau_z + \tau^{(1)}_z+\tau^{(2)}_z)}&= \frac{1}{a_z}\bigg [ 1-\frac{N\oz}{2} -\frac{N^{(2)}\oz}{2}+\frac{3N^2\o}{8}-\frac{N^2_z}{8}-\frac{N\o N_z}{4}-\frac{U^a_zU_{az}}{2}+\frac{N\oz \dot{N}_z}{4H_z}\bigg] \, , \nex
    \U_\t{s}^{(1)}(\tau_z + \tau^{(1)}_z) &=\U_z^{(1)} + \dot{\U}_z^{(1)}\tau^{(1)}_z = \frac{N_z}{2}+\frac{N\oz \dot{N}_z}{4H_z} \, , \nex
    \pa_w \U_\t{s}^{(1)}(\tau_z+\tau^{(1)}_z) &= \pa_w \U_z^{(1)} + \tau^{(1)}_z\pa_\tau \left ( \pa_w \U_z^{(1)}\right ) = \frac{\pa_w N_z}{2} +\frac{N\oz \pa_w \dot{N}_z}{4H_z}\,. 
    \label{eq:z-expansions}
\end{align}
It is worth  highlighting that the derivative $\pa_w$ must act \textit{before} expanding $\U_\t{s}$ around the observed redshift. This is due to the fact that partial derivatives  w.r.t. any of $(\tau, w, \tilde{\theta}^a)$ keep constant the remaining three.

Furthermore, we also have to expand the derivative w.r.t. the time of the source, \textit{i.e.}
\begin{align}
  \frac{\pa}{\pa \tau_\t{s}} = \frac{\pa \tau_z}{\pa \tau_\t{s}} \frac{\pa}{\pa \tau_z}= \left (1-\dot{\tau}^{(1)}_z-\dot{\tau}^{(2)}_z\right ) \frac{\pa}{\pa \tau_z}  \, , 
  \label{eq:exp-der-time}
\end{align}
from which it follows that  
\begin{align}  
\dot{\U}_\t{s}&=\pa_{\tau_\t{s}}\U_\t{s}^{(1)}(\tau_z + \tau^{(1)}_z)=\dot{\U}_z^{(1)} +\ddot{\U}_z^{(1)} \tau^{(1)}_z = \frac{\dot{N}_z}{2}+\frac{\ddot{N}_z N\oz}{4H_z}\, , \nex
    H_\t{s}(\tau_z + \tau^{(1)}_z +\tau^{(2)}_z) &= H_z +\dot{H}_z\tau^{(1)}_z +\dot{H}_z\tau^{(2)}_z +\frac{\ddot{H}_z}{2}\left ( \tau^{(1)}_z\right)^2 -H_z \left ( \dot{\tau}^{(1)}_z\right)^2 \nex
    &=H_z +\frac{\dot{H}_z}{2H_z}N\oz +\frac{\dot{H}_z}{2H_z} \left ( N^{(2)}\oz + U^a_z U_{az}\right )+\frac{\dot{H}_z}{4H_z}\left ( N^2_z - N^2\o \right )\nex
    & \quad +\frac{1}{H_z}\left ( -\frac{3\dot{H}^2_z}{8H^2_z}+\frac{\ddot{H}_z}{8H_z}\right )\left ( N\oz \right)^2 -\frac{3\dot{H}_z}{4H^2_z} \dot{N}_z N\oz -\frac{\dot{N}^2_z}{4H_z} \, . \label{eq:exp-hubble}
\end{align}

Finally, we redshift expand Eq.~\eqref{eq:rsd-2nd-start} by using  Eqs.~\eqref{eq:z-expansions} and \eqref{eq:exp-hubble}, and we replace the perturbations with their gauge-invariant counterparts of Eqs.~\eqref{eq:gauge-inv-quantities-1st-order} and \eqref{eq:glc-gauge-inv-2nd-order}, obtaining the following results
\begin{align}
    \frac{\Delta z}{\Delta \tau_\t{o}} = \frac{\Delta \bar{z} }{\Delta \tau_\t{o}} +\left ( \frac{\Delta z}{\Delta \tau_\t{o}}\right )^{(1)}+\left ( \frac{\Delta z}{\Delta \tau_\t{o}}\right )^{(2)}
\end{align}
with 
\begin{align}
    \frac{\Delta \bar{z} }{\Delta \tau_\t{o}} & = \frac{a\o H\o}{a_z}-H_z \, , \nex
    \left ( \frac{\Delta z}{\Delta \tau_\t{o}}\right )^{(1)}&=\frac{a_\t{o}}{a_z}\frac{\dot{\ms{N}}_\t{o}}{2}
    +\frac{1}{2a_z}\pa_w \ms{N}|^\t{o}_z -\frac{\dot{H}_z}{2H_z}\ms{N}|^\t{o}_z  -\frac{\dot{\ms{N}}_z}{2} \, , \nex  
    \left ( \frac{\Delta z}{\Delta \tau_\t{o}}\right )^{(2)} &=\frac{a\o}{a_z}\frac{\dot{\ms{N}}^{(2)}\o}{2}+\frac{1}{2a_z}\pa_w \ms{N}^{(2)}\oz-\frac{\dot{H}_z}{2H_z}\ms{N}^{(2)}\oz-\frac{\dot{\ms{N}}^{(2)}_z}{2} -\frac{a\o}{a_z}\frac{\ms{N}\o \dot{\ms{N}}\o}{2}\nex
    & \quad   +\frac{1}{a_z}\Bigg [ -\frac{\dot{\ms{N}}_z \pa_w \ms{N}\oz}{4H_z}-\frac{\ms{N}\oz \pa_w \dot{\ms{N}}_z}{4H_z}+\ms{U}^a_z \pa_w \ms{U}_{az}+\frac{\ms{N}_z \pa_w \ms{N}_z}{2} \nex
    & \quad
    -\frac{3\ms{N}\o \pa_w \ms{N}\o}{4}+\frac{\ms{N}\o \pa_w \ms{N}_z}{4}\Bigg ] 
    -\frac{\dot{H}_z}{2H_z}    \ms{U}^a_z \ms{U}_{az}\nex
    & \quad -\frac{\dot{H}_z}{4H_z}\left ( \ms{N}^2_z - \ms{N}^2\o \right ) -\frac{1}{H_z}\left ( -\frac{3\dot{H}^2_z}{8H^2_z}+\frac{\ddot{H}_z}{8H_z}\right )\left ( \ms{N}\oz \right)^2\nex
    & \quad +\frac{3\dot{H}_z}{4H^2_z}\dot{\ms{N}}_z \ms{N}\oz +\frac{\dot{\ms{N}}^2_z}{4H_z}-\frac{\ddot{\ms{N}}_z \ms{N}\oz}{4H_z}+\frac{\ms{N}_z\dot{\ms{N}}_z}{4}+\ms{U}^a_z \dot{\ms{U}}_{az}\nex
    & \quad +\frac{\ms{N}_z \dot{\ms{N}}_z}{4}-\frac{\ms{U}^a_z \pa_a \ms{N}_z}{2a_z}\, . 
\label{eq:rsd-2-notreplaced} 
\end{align}
These are the fully gauge-invariant formulae for the first- and second-order redshift drift written as function of the observed redshift, the observed past light-cone and the observed angles.

The next step is to replace $\ms{N}$, $\ms{N}^{(2)}$ and $\ms{U}_a$ with their forms \eqref{eq:gauge-inv-quantities-1st-order} and \eqref{eq:glc-gauge-inv-2nd-order}, and then to fix the GLC gauge rewritten in terms of gravitational potentials  using Eqs.~\eqref{eq:gauge-modes-NG-1st} and \eqref{eq:gauge-modes-NG-2nd}. Since the final results will be expressed in standard perturbations, let us recall that the derivatives w.r.t. light-cone and standard coordinates are connected via (see Eqs.~\eqref{eq:spherical-to-glc})
\begin{equation}
    \pa_\tau = \frac{1}{a}(\pa_\eta - \pa_r) \qquad , \qquad \pa_w = \pa_r \qquad , \qquad \pa_{\tilde{\theta}^a} = \pa_{\theta^a}  \, . 
    \label{eq:map-der}
\end{equation}
We will also use that, for a generic quantity $X$, it holds \cite{Fanizza:2020xtv,Scaccabarozzi:2017ncm}
\begin{align}
    \pa_w X \big (\tau^\prime, w-\eta(\tau)+\eta(\tau^\prime) \big ) &= \pa_r X(\eta^\prime, \eta_\t{o}-\eta) \,   \, ,   \label{eq:map-deriv} 
    \end{align}
 and 
 \begin{align}
    \pa_w X(\tau^\prime, w) &= \pa_r X(\eta^\prime, \eta_\t{o}-\eta^\prime)  = \pa_{\eta^\prime} X(\eta^\prime, \eta_\t{o}-\eta^\prime) -\frac{\t{d}}{\t{d}\eta^\prime}X(\eta^\prime, \eta_\t{o}-\eta^\prime)  \, .
    \label{eq:map-deriv-2}
\end{align}
Moreover, let us define the first-order radial peculiar velocity
\begin{align}
    v_{|| \alpha} \equiv \pa_r P_\alpha = \frac{1}{a_\alpha} \int_{\eta_{\t{in}}}^{\eta_\alpha} \t{d}\eta \, a \pa_r \Phi \qquad  , \qquad \alpha = \t{o}, z \, .
    \label{eq:pecvel-1st-defined}
\end{align}
Then, by using Eqs.~\eqref{eq:map-deriv} and \eqref{eq:map-deriv-2} and the definition \eqref{eq:pecvel-1st-defined} in the first-order gauge modes of Eq.~\eqref{eq:gauge-modes-NG-1st}, we see that each gauge mode corresponds to a GR effect \cite{Bechaz:2025ojy}. In particular, $\xi^\tau_{(1)}$ is the velocity potential, the term 
\begin{align}
    \pa_w \xi^w_{(1)z} = -2\int_{\eta_z}^{\eta_\t{o}} \t{d}\eta \, \pa_\eta \Psi^\t{I} + 2 \Psi^\t{I}\o - 2\Psi^\t{I}_z \,  
\end{align}
gives the local and integrated Sachs-Wolfe effects, and finally 
\begin{align}
D_a \xi^a_{(1)z} = \frac{2}{r_z} \int_{\eta_z}^{\eta_\t{o}} \t{d}\eta \, \frac{\eta - \eta_z}{\eta_\t{o}-\eta} D^2 \Psi^\t{I}
\end{align}
amounts to lensing.  As also remarked in \cite{Bechaz:2025ojy}, the advantage of using our  methodology to compute observables is then twofold: by directly perturbing the light-cone, we have compact expressions for relativistic effects -- which are simply encoded in gauge modes and their derivatives -- and we also gain a better understanding of the perturbed structure of the observed past light-cone.

\subsection{First-order relativistic effects}
\label{subsec:fitst-order-RD}
We are now ready to compute all the source and observer perturbative effects on the redshift drift, starting from the first-order ones. Our result will be compared with \cite{Bessa:2023qrr}, where the corresponding expression of source effects with standard techniques was derived. As a development of that work, perturbations at the observer position will be properly accounted for, using the method developed in \cite{Fanizza:2020xtv,Bechaz:2025ojy}, in the case of a free-falling observer.

We then report the expressions of the gauge-invariant variable $\ms{N}$ at the source and observer positions, in terms of gravitational potentials (from now on, we will always omit the superscript \quotes{1} for first-order perturbations):
\begin{align}
    \ms{N}_z &=2 \left [ -\Psi  -H\xi^\tau +v_{||}\right]_z+2\int_{\eta_z}^{\eta_\t{o}}\t{d}\eta \,  \pa_\eta  \left (\Phi + \Psi \right ) -2 \left [\Phi + \Psi \right ]^\t{o}_z -2v_{||\t{o}}\, , \nex
    \ms{N}\o &= -2\left [ \Psi  +H\xi^\tau \right]\o \, . 
    \label{eq:N-gi-firstorder}
\end{align}
Note that, only in this subsection, we directly work with the gravitational potentials $\Phi$ and $\Psi$ rather than their isotropic and anisotropic counterparts defined in Eqs.~\eqref{eq:IA-potentials-defined}, in order to have a direct matching with \cite{Bessa:2023qrr}.

Considering Eqs.~\eqref{eq:rsd-2-notreplaced}, as an intermediate step we compute
\begin{align}
    -\frac{1}{2a_z}\pa_w \ms{N}_z-\frac{\dot{\ms{N}}_z}{2}&=H_zv_{||z} -\frac{1}{a_z}\,\int_{\eta_z}^{\eta_\t{o}}\t{d}\eta \, \pa^2_\eta \left ( \Phi + \Psi  \right ) +\frac{1}{a_z}\pa_\eta \left [ \Phi + \Psi  \right ]\o -\frac{1}{a_z}\pa_\eta \Phi_z -\frac{1}{a_z}\pa_r \Phi_z\notag \\[1ex]
    & \quad +\frac{1}{a_z} \pa_r \left [\Phi + \Psi \right ]\o + \dot{H}_z\xi^\tau_z + H_z\Phi_z +\frac{1}{a_z}\pa_r v_{||\t{o}}\, . 
    \label{eq:step-first-order}
\end{align}
We emphasize that the two terms on the l.h.s. of the above equation both contain a contributions proportional to $\pa_r \vz$, which cancel among each other. We will be commenting on this point at the end of the next subsection.

Then, moving to the conformal time $\eta$ and exploiting the usual relations $H=\Hcal/a$ and $\dot{H}/H = (\Hcal^\prime -\Hcal^2 )/a\Hcal$, our final background and first-order results including  source and observer terms are
\begin{align}
    \frac{\Delta \bar{z}}{\Delta \tau_\t{o}} &=   \frac{\Hcal_\t{o} -\Hcal_z}{a_z}  \, , 
    \label{eq:RD-0order-final-complete}
    \\[1ex]
    \left (\frac{\Delta z}{\Delta \tau_\t{o}}\right )^{(1)} &=\frac{1}{a_z}\bigg \{ -\int_{\eta_z}^{\eta_\t{o}}\t{d}\eta \, \pa^2_\eta \left ( \Phi + \Psi \right ) -\pa_\eta \Phi_z -\pa_r \Phi_z + \frac{\Hcal^\prime_z}{\Hcal_z } \left [ v_{||} + \Phi \right ]_z   \nex
    & \quad + \frac{\Hcal_z^\prime-\Hcal_z^2}{\Hcal_z} \int_{\eta_z}^{\eta_\t{o}}\t{d}\eta \,  \pa_\eta  \left (\Phi + \Psi \right ) + \pa_\eta \Phi_\t{o} +\pa_r \Phi_\t{o}\nex
    & \quad -\frac{\Hcal^\pr_\t{o}-\Hcal^2_\t{o}}{a\o}\int_{\eta_{\t{in}}}^{\eta \o} \t{d}\eta \, a \Phi   -\Hcal_\t{o} \Phi \o  \nex
    & \quad +\frac{\Hcal^\pr_z-\Hcal^2_z}{\Hcal_z} \left [ \frac{ \Hcal_\t{o}}{a\o}\int_{\eta_{\t{in}}}^{\eta \o} \t{d}\eta \, a \Phi -\Phi_\t{o} -v_{||\t{o}}\right ]  + \pa_r \Psi \o    + \pa_r v_{||\t{o}}\bigg \} \, . 
    \label{eq:RD-1st-final-complete}
\end{align}
As a first consistency check, we notice that the unphysical gauge effects proportional to  $\xi^\tau_z$  (namely the source velocity potential) cancel. Then, the purely source terms of Eq.~\eqref{eq:RD-1st-final-complete} agree\footnote{Note that their $V$ is the velocity potential that corresponds to $\xi^\tau$ in our formalism, and the definitions of the gravitational potentials $\Phi$ and $\Psi$ are inverted w.r.t. our notation.} with Eq.~(3.28) of \cite{Bessa:2023qrr}. Therefore, this provides a validation of Eq.~\eqref{eq:RD-1st-final-complete}, which  also represents an improvement of \cite{Bessa:2023qrr} because we have properly kept track of  all the observer contributions.

Finally, let us comment on the background redshift drift of Eq.~\eqref{eq:RD-0order-final-complete}.
Unlike what done in \cite{Bessa:2023qrr}, we have properly accounted for the background observer effect $H\o (1+z)$.  As we shall discuss in Subsect.~\ref{subsec:plots-RD}, if we neglected the term $H\o (1+z)$, we  would miss some interesting physical information.

\subsection{Second-order relativistic effects}
\label{subsec:second-order-RD}
We now proceed with the  calculations of the second-order perturbations of the redshift drift. By plugging Eqs.~\eqref{eq:gauge-inv-quantities-1st-order} and \eqref{eq:glc-gauge-inv-2nd-order} into the third of Eqs.~\eqref{eq:rsd-2-notreplaced}, and in terms of Bardeen potentials $\Phi$ and $\Psi$, the full formula is
\begin{align}
\left ( \frac{\Delta z}{\Delta \tau \o}\right)^{(2)}&= \frac{a\o}{a_z} \Bigg \{ \bigg [-\dot{\Psi}^{(2)}-\frac{\dot{H}}{2}\left (\xi^\tau_{(2)}-\xi^\mu \pa_\mu \xi^\tau \right ) - \frac{H}{2} \left ( \pa_\tau + \frac{\pa_w}{a}\right ) \left (\xi^\tau_{(2)}-\xi^\mu \pa_\mu \xi^\tau \right ) \nex
& \quad +\frac{1}{2a}\pa_w \pa_\tau \left (\xi^\tau_{(2)}-\xi^\mu \pa_\mu \xi^\tau \right )-\frac{1}{2}\pa_w \pa_\tau \left (\xi^w_{(2)}-\xi^\mu \pa_\mu \xi^w \right )-\frac{r}{a}\pa_w \xi^a \pa_w \xi_a\nex
& \quad +\frac{r^2}{a}\bar{\ga}^{ab}\pa_w \pa_b \xi^w \pa_w \xi_a + \bigg ( 2H\dot{H}+\frac{\ddot{H}}{2}\bigg ) \left ( \xi^\tau \right )^2 + \xi^\tau \bigg (2H^2 \Phi +\frac{H}{a}\pa_w \Phi \nex
& \quad + \frac{2H}{a}\pa_w \Psi + \ddot{\Psi} + \dot{H}\Phi+2\dot{H}\Psi + 2H \dot{\Psi} \bigg ) - \left ( H^2 + 2 \dot{H}\right) \xi^\tau \frac{\pa_w \xi^\tau}{a}\nex
& \quad -\frac{\pa_w \xi^\tau}{a}\bigg (H\Phi +3\dot{\Psi} +\frac{1}{a}\pa_w (\Phi + \Psi)\bigg )+\frac{H}{a^2}\left ( \pa_w \xi^\tau \right )^2 \nex
& \quad + \frac{\pa^2_w \xi^\tau}{a^2}\left (H\xi^\tau + 2\Psi + \pa_w \xi^w \right ) +\frac{3}{a}\Psi \pa_w \Psi + \dot{\Psi}\Phi +\frac{1}{a}\Phi \pa_w \Psi + 2H\Phi \Psi + 2\dot{\Psi}\pa_w \xi^w\nex
& \quad +\frac{1}{a}\pa_w \Psi \pa_w \xi^w + \xi^w \pa_w \dot{\Psi}+\frac{1}{a}\bar{\ga}^{ab}\pa_a\xi^w \pa_b \Psi  + \xi^a \pa_a \dot{\Psi} - \frac{H}{a}\pa_w \xi^\tau \pa_w \xi^w \nex
& \quad + 2\dot{H}\xi^\tau \pa_w \xi^w +2H\Phi \pa_w \xi^w 
\bigg ]\o  -2\bigg [-\Psi -H\xi^\tau +\frac{1}{a}\partial_w \xi^\tau - \partial_w \xi^w \bigg ] \o  \nex
& \quad \times \bigg [-\dot{\Psi}-\dot{H}\xi^\tau - H  \Phi  +\frac{1}{a}\left ( \pa_w \Phi -\frac{1}{a}\pa^2_w \xi^\tau \right ) -\frac{1}{a}\pa_w (\Phi +\Psi) \bigg ]\o \Bigg \}\nex
& \quad + \frac{1}{a_z}\Bigg \{\bigg [-\pa_w \Psi^{(2)}-\frac{H}{2}\pa_w \left (\xi^\tau_{(2)}-\xi^\mu \pa_\mu \xi^\tau \right ) +\frac{1}{2a}\pa^2_w \left (\xi^\tau_{(2)}-\xi^\mu \pa_\mu \xi^\tau \right ) \nex
& \quad - \frac{1}{2}\pa^2_w \left (\xi^w_{(2)}-\xi^\mu \pa_\mu \xi^w \right ) +\xi^\tau \bigg (2H^2 \pa_w \xi^\tau -\frac{H}{a}\pa^2_w \xi^\tau +\pa_w \dot{\Psi}+\dot{H}\pa_w \xi^\tau\nex
& \quad +2H\pa_w \Psi +2H \pa^2_w\xi^w\bigg ) +r \pa_w \xi^a \pa_w \xi_a +r^2 \pa^2_w \xi^a \pa_w \xi_a -\frac{H}{a}\left ( \pa_w \xi^\tau \right )^2\nex
& \quad  -\frac{2}{a}\Psi \pa^2_w \xi^\tau +3\pa_w \Psi \pa_w \xi^w +\pa^2_w \xi^w \bigg (2\Psi + \pa_w \xi^w -\frac{\pa_w\xi^\tau}{a} \bigg )\nex
& \quad  +\xi^w \pa^2_w \Psi +\pa_w \xi^a \pa_a \Psi + \xi^a \pa_a \pa_w \Psi -\frac{1}{a}\pa^2_w \xi^\tau \pa_w \xi^w \nex
& \quad  +
\pa_w \xi^\tau \bigg (-\frac{2}{a}\pa_w \Psi +\dot{\Psi}+2H \left (\Psi + \pa_w \xi^w \right )\bigg )\bigg]^\t{o}_z \nex
& \quad -\frac{1}{H_z}\bigg [-\Psi-H\xi^\tau +\frac{\partial_w \xi^\tau}{a} - \partial_w \xi^w \bigg ]^\t{o}_z  \nex
& \quad \times \bigg [-\pa_w \dot{\Psi}-\dot{H}\pa_w \xi^\tau - H  \pa_w \Phi  +\frac{1}{a}\left ( \pa^2_w \Phi -\frac{1}{a}\pa^3_w \xi^\tau \right ) -\frac{1}{a}\pa^2_w (\Phi +\Psi) \bigg ]_z\nex
& \quad +\bigg [\frac{1}{r^2}D_a \bigg (\frac{\pa_w \xi^\tau}{a}-\frac{2\xi^\tau}{ar}-\pa_w \xi^w +\frac{2\xi^w}{r} \bigg ) - \pa^2_w\xi_a \bigg ]_z \nex
& \quad \times  \bigg [D^a \bigg (\frac{\xi^\tau}{a}-\xi^w \bigg )-r^2\pa_w \xi^a \bigg ]_z +2\bigg [-\Psi-H\xi^\tau +\frac{\partial_w \xi^\tau}{a} - \partial_w \xi^w  \bigg ]_z  \nex
& \quad \times \bigg [-\pa_w \Psi-H\pa_w \xi^\tau +\frac{\partial^2_w \xi^\tau}{a} - \partial^2_w \xi^w \bigg ]_z -3\bigg [-\Psi-H\xi^\tau +\frac{\partial_w \xi^\tau}{a} - \partial_w \xi^w  \bigg ]\o \nex
& \quad \times \bigg [-\pa_w \Psi-H\pa_w \xi^\tau +\frac{\partial^2_w \xi^\tau}{a} - \partial^2_w \xi^w \bigg ]\o +\bigg [-\Psi-H\xi^\tau +\frac{\partial_w \xi^\tau}{a} - \partial_w \xi^w  \bigg ]\o \nex 
& \quad \times  \bigg [-\pa_w \Psi-H\pa_w \xi^\tau +\frac{\partial^2_w \xi^\tau}{a} - \partial^2_w \xi^w \bigg ]_z 
\Bigg \} \nex
& \quad -\frac{\dot{H}_z}{H_z} \bigg [-\Psi^{(2)}-\frac{H}{2}\left ( \xi^\tau_{(2)}-\xi^\mu \pa_\mu \xi^\tau \right ) +\frac{1}{2a}\pa_w \left ( \xi^\tau_{(2)}-\xi^\mu \pa_\mu \xi^\tau \right ) \nex
& \quad -\frac{1}{2}\pa_w \left ( \xi^w_{(2)}-\xi^\mu \pa_\mu \xi^w \right )+\frac{r^2}{2}\pa_w \xi^a \pa_w \xi_a +\frac{H^2}{2}\left ( \xi^\tau \right )^2 -\frac{H}{a}\xi^\tau \pa_w \xi^\tau \nex
& \quad -\frac{2}{a}\Psi \pa_w \xi^\tau +2\Psi \pa_w \xi^w + \frac{1}{2}\left ( \pa_w \xi^w \right )^2+\xi^\tau \dot{\Psi}+\xi^w\pa_w \Psi + \xi^a \pa_a \Psi \nex
& \quad +\frac{\dot{H}+H^2}{2}\left ( \xi^\tau \right )^2 -\frac{1}{a}\pa_w \xi^\tau \pa_w \xi^w +2H\xi^\tau \left (\Psi +\pa_w \xi^w \right )\bigg ]^\t{o}_z\nex
& \quad -\frac{\dot{H}_z}{2H_z}\bigg [D^a\left (\frac{\xi^\tau}{a} - \xi^w \right ) -r^2\partial_w \xi^a  \bigg ]_z \bigg [\frac{1}{r^2}D_a \left (\frac{\xi^\tau}{a} - \xi^w \right ) -\partial_w \xi_a \bigg ]_z\nex
& \quad -\frac{\dot{H}_z}{H_z}\Bigg \{ \bigg [-\Psi -H\xi^\tau +\frac{\partial_w \xi^\tau}{a} - \partial_w \xi^w \bigg ]^2_z - \bigg [-\Psi -H\xi^\tau +\frac{\partial_w \xi^\tau}{a} - \partial_w \xi^w \bigg ]^2_\t{o}  \Bigg \}\nex
& \quad +\frac{1}{2H_z}\left (\frac{3\dot
{H}_z^2}{H^2_z} -\frac{\ddot{H}_z}{H_z}\right ) \Bigg \{-\Psi\oz - [H\xi^\tau]^\t{o}_z +\pa_w \bigg [\frac{\xi^\tau}{a}-\xi^w \bigg ]^\t{o}_z \Bigg \}^2 \nex
& \quad + \frac{3\dot{H}_z}{H^2_z}\bigg [-\dot{\Psi}-\dot{H}\xi^\tau - H  \Phi  +\frac{1}{a}\left ( \pa_w \Phi -\frac{1}{a}\pa^2_w \xi^\tau \right ) -\frac{1}{a}\pa_w (\Phi +\Psi) \bigg ]_z  \nex
& \quad \times \bigg [-\Psi-H\xi^\tau +\frac{\partial_w \xi^\tau}{a} - \partial_w \xi^w \bigg ]^\t{o}_z\nex
& \quad +\frac{1}{H_z}\bigg [-\dot{\Psi}-\dot{H}\xi^\tau - H  \Phi  +\frac{1}{a}\left ( \pa_w \Phi -\frac{1}{a}\pa^2_w \xi^\tau \right ) -\frac{1}{a}\pa_w (\Phi +\Psi) \bigg ]^2_z\nex
& \quad -\frac{1}{H_z}\bigg [- \ddot{\Psi}-\ddot{H}\xi^\tau -\dot{H}\left ( \Phi - \frac{1}{a}\pa_w \xi^\tau \right ) -\dot{H} \Phi -H \dot{\Phi}\nex
 & \quad +\frac{2H}{a^2}\pa^2_w \xi^\tau -\frac{1}{a^2} \left (\pa^2_w\Phi - \frac{1}{a}\pa^3_w\xi^\tau \right )  +\frac{H}{a}\pa_w  \Psi -\frac{1}{a}\pa_w \dot{\Psi} \bigg ]_z  \nex
 & \quad \times \bigg [-\Psi -H\xi^\tau +\frac{\partial_w \xi^\tau}{a} - \partial_w \xi^w \bigg ]^\t{o}_z\nex
 & \quad + \bigg [\dot{\Psi}^{(2)}+\frac{\dot{H}}{2}\left (\xi^\tau_{(2)}-\xi^\mu \pa_\mu \xi^\tau \right ) + \frac{H}{2} \left ( \pa_\tau + \frac{\pa_w}{a}\right ) \left (\xi^\tau_{(2)}-\xi^\mu \pa_\mu \xi^\tau \right ) \nex
& \quad -\frac{1}{2a}\pa_w \pa_\tau \left (\xi^\tau_{(2)}-\xi^\mu \pa_\mu \xi^\tau \right )+\frac{1}{2}\pa_w \pa_\tau \left (\xi^w_{(2)}-\xi^\mu \pa_\mu \xi^w \right )+\frac{r}{a}\pa_w \xi^a \pa_w \xi_a\nex
& \quad -\frac{r^2}{a}\bar{\ga}^{ab}\pa_w \pa_b \xi^w \pa_w \xi_a - \bigg ( 2H\dot{H}+\frac{\ddot{H}}{2}\bigg ) \left ( \xi^\tau \right )^2 - \xi^\tau \bigg (2H^2 \Phi +\frac{H}{a}\pa_w \Phi \nex
& \quad + \frac{2H}{a}\pa_w \Psi + \ddot{\Psi} + \dot{H}\Phi+2\dot{H}\Psi + 2H \dot{\Psi} \bigg ) + \left ( H^2 + 2 \dot{H}\right) \xi^\tau \frac{\pa_w \xi^\tau}{a}\nex
& \quad +\frac{\pa_w \xi^\tau}{a}\bigg (H\Phi +3\dot{\Psi} +\frac{1}{a}\pa_w (\Phi + \Psi)\bigg )-\frac{H}{a^2}\left ( \pa_w \xi^\tau \right )^2 \nex
& \quad - \frac{\pa^2_w \xi^\tau}{a^2}\left (H\xi^\tau + 2\Psi + \pa_w \xi^w \right ) -\frac{3}{a}\Psi \pa_w \Psi - \dot{\Psi}\Phi -\frac{1}{a}\Phi \pa_w \Psi - 2H\Phi \Psi -2\dot{\Psi}\pa_w \xi^w\nex
& \quad -\frac{1}{a}\pa_w \Psi \pa_w \xi^w - \xi^w \pa_w \dot{\Psi}-\frac{1}{a}\bar{\ga}^{ab}\pa_a\xi^w \pa_b \Psi  - \xi^a \pa_a \dot{\Psi} + \frac{H}{a}\pa_w \xi^\tau \pa_w \xi^w \nex
& \quad - 2\dot{H}\xi^\tau \pa_w \xi^w -2H\Phi \pa_w \xi^w \bigg ]_z + 2\bigg [-\Psi -H\xi^\tau +\frac{\partial_w \xi^\tau}{a} - \partial_w \xi^w \bigg ]_z  \nex
& \quad \times \bigg [ -\dot{\Psi}-\dot{H}\xi^\tau - H  \Phi  +\frac{1}{a}\left ( \pa_w \Phi -\frac{1}{a}\pa^2_w \xi^\tau \right ) -\frac{1}{a}\pa_w (\Phi +\Psi)\bigg ]_z\nex
& \quad + \frac{1}{ar^2}\bigg [D^a\bigg (\frac{\xi^\tau}{a}-\xi^w \bigg ) -r^2\pa_w \xi^a \bigg ]_z D_a \bigg [-2H\xi^\tau +\frac{2\xi^\tau}{ar}-\Phi -\frac{\pa_w \xi^\tau}{a} -\pa_w \xi^w \bigg ]_z \nex
& \quad -\frac{1}{a_z}\bigg [\frac{1}{r^2}D^a \left ( \frac{\xi^\tau}{a} - \xi^w\right )-\partial_w \xi^a \bigg ]_z D_a \bigg [-\Psi-H\xi^\tau +\frac{\partial_w \xi^\tau}{a} - \partial_w \xi^w  \bigg ]_z
\, . 
\label{eq:rsd-2nd-fullformula}
\end{align}
By properly computing the derivatives on the second-order gauge modes using Eqs.~\eqref{eq:gauge-modes-NG-1st} and \eqref{eq:gauge-modes-NG-2nd}, and expanding  the products, all the terms proportional to $\xi^\tau_{(1)z}$ and $\xi^\tau_{(2)z}$ (namely those with the source velocity potential) cancel out. This quite a long calculation is a very important consistency check of our formula for the second-order redshift drift.

Now, we can proceed with a systematic derivation of all the second-order perturbative effects on the redshift drift, written in terms of the gravitational potentials. 
As a technical note, we have to use that
\begin{align}
\pa_w \xi^a &=2q^{ab} \int_{\eta_z}^{\eta_{\t{o}}}\frac{\t{d}\eta}{r^2}\bigg [ -\frac{1}{r}\int_{\eta}^{\eta_{\t{o}}}\t{d}\eta^\pr \, \pa_b \psiI  + \pa_b\psiI + \int_{\eta}^{\eta_{\t{o}}}\t{d}\eta^\pr \,\pa_{\eta^\pr }\pa_b\psiI\bigg ] \, , \nex
\pa^2_w \xi^a &= 2q^{ab}\int_{\eta_z}^{\eta_{\t{o}}} \frac{\t{d}\eta}{r^2}\bigg [\frac{1}{r^2} \int_{\eta}^{\eta_{\t{o}}}\t{d}\eta^\pr \, \pa_b \psiI -\frac{2}{r}\bigg (\pa_b \psiI + \int_{\eta}^{\eta_{\t{o}}}\t{d}\eta^\pr \, \pa_{\eta^\pr}\pa_b \psiI  \bigg ) \nex
& \quad +2 \pa_r \pa_b \psiI + \int_{\eta}^{\eta_{\t{o}}}\t{d}\eta^\pr \, \pa^2_{\eta^\pr}\pa_b \psiI\bigg ]\, .
\end{align}
Moreover, we define the first-order orthogonal velocity and the second-order peculiar velocity as \cite{Marozzi:2014kua}
\begin{align}
    v_{a\perp \alpha} &\equiv \pa_a P_\alpha \, , \nex
    v^{(2)}_{||\alpha} &\equiv \Psi_\alpha v_{|| \alpha} +\frac{1}{2a_\alpha}\int_{\eta_{\t{in}}}^{\eta_\alpha} \t{d}\eta \,  a \pa_r \big [2\Phi^{(2)}-\Phi^2 + (\pa_r P)^2 + \bar{\ga}^{ab}\pa_a P \pa_b P \big ]
\end{align}
for $\alpha = \t{o}, z$.
Then, we decompose
\begin{align}
    \left (\frac{\Delta z}{\Delta \tau \o}\right)^{(2)}  &\equiv  \left (\frac{\Delta z}{\Delta \tau \o}\right)_{\t{pos}}^{(2)}  + \left (\frac{\Delta z}{\Delta \tau \o}\right)_{\t{mixed}}^{(2)}  +\left (\frac{\Delta z}{\Delta \tau \o}\right)_{\t{path}}^{(2)}  \nex
    & \quad +\left (\frac{\Delta z}{\Delta \tau \o}\right)_{P,\t{pos}}^{(2)}  + \left (\frac{\Delta z}{\Delta \tau \o}\right)_{P,\t{mixed}}^{(2)} +\left (\frac{\Delta z}{\Delta \tau \o}\right)_{v,\t{int}}^{(2)} \,  , 
\end{align}
where
\begin{itemize}
    \item \quotes{pos} indicates terms generated by peculiar velocities;
    \item \quotes{mixed} refers to the cross effects \quotes{peculiar velocities $\times$ (SW/ISW/angular deflections)};
    \item \quotes{path} contains all the effects due to SW, ISW and angular deflections;
    \item The subscript \quotes{$P$} indicates all the corrections due to the monopole $P\o$;
    \item Finally, \quotes{$v,\t{int}$} refers to terms with peculiar velocities  integrated between $\eta_{\t{in}}$ and $\eta_{\alpha}$.
\end{itemize}
We further split the contributions \quotes{mixed} and \quotes{path} into two subgroups of terms: the subscript \quotes{I} refers to couplings to the only isotropic potential $\Psi^\t{I}$, while \quotes{A} to couplings involving the anisotropic potential $\Psi^\t{A}$. Thus, our results for the full redshift drift are the following:
\begin{align}
  \left (\frac{\Delta z}{\Delta \tau \o}\right)_{\t{pos}}^{(2)}  &= \frac{1}{a_z}\Bigg \{-\bigg ( \Hcal \o  +\frac{\Hcal^\pr_z-\Hcal^2_z} {\Hcal_z}\bigg ) \vo^{(2)}+\frac{\Hcal^\pr_z}{\Hcal_z}\vz^{(2)} +7 \vo \pa_r \vo +\frac{1}{2}\vz \pa_r \vo\nex
  & \quad +\bigg (\frac{7}{2}-\frac{3\Hcal^\pr_z}{\Hcal^2_z} \bigg ) \vo \pa_r \vz -2 \vz \pa_r \vz +\bigg (\Hcal \o -\frac{1}{2} \frac{\Hcal^\pr_z - \Hcal^2_z}{\Hcal_z} \bigg )(\vo)^2\nex
  & \quad + \frac{3\Hcal_z}{2} \left ((\vz)^2-\vo \vz \right ) -\frac{1}{r^2_z}\bigg (2 v^a_{\perp z}\pa_a \vz+ \frac{\Hcal^\pr_z+2\Hcal^2_z}{2\Hcal_z}v^a_{\perp z} v_{a \perp z}\bigg ) \nex
  & \quad +\frac{1}{\Hcal_z} (\pa_r \vz)^2 +\frac{1}{2\Hcal^2_z}\bigg (\frac{3(\Hcal^\pr_z)^2}{\Hcal_z} + \Hcal^3_z - \Hcal^{\pr \pr}_z - 2 \Hcal_z \Hcal^\pr_z \bigg ) (\vo - \vz)^2  \Bigg \} \label{eq:pos}\, , \\[1ex] 
  \left (\frac{\Delta z}{\Delta \tau \o}\right)_{\t{mixed,I}}^{(2)}  &= \frac{1}{a_z}\Bigg \{ 2\vo \int_{\eta_z}^{\eta \o} \t{d}\eta \, \pa^2_\eta \psiI + 2\bigg ( \int_{\eta_z}^{\eta \o} \t{d}\eta \, \pa_\eta \psiI\bigg )\bigg [\bigg (\frac{3\Hcal^\pr_z}{\Hcal^2_z}-2\bigg )\pa_r \vz +\Hcal_z \vz \nex
  & \quad + \frac{1}{2\Hcal^2_z}\bigg (\frac{3(\Hcal^\pr_z)^2}{\Hcal_z}+\Hcal^3_z - \Hcal^{\pr \pr}_z -2 \Hcal_z \Hcal^2_z \bigg )(\vz-\vo)\bigg ]  \nex
  & \quad +\frac{2}{r^2_z}v^a_{\perp z} \int_{\eta_z}^{\eta \o}\t{d}\eta \, \pa_\eta \pa_a \psiI +\frac{2\Hcal^\pr_z}{\Hcal_z}\pa_r \vz \int_{\eta_z}^{\eta \o} \t{d}\eta \,  \psiI \nex
  & \quad + 2\bigg (\frac{5\Hcal_z}{2r^2_z} -\frac{1}{r^3_z}-\frac{\Hcal^\pr_z}{\Hcal_z r^2_z} \bigg ) v^a_{\perp z}   \int_{\eta_z}^{\eta \o} \t{d}\eta \,  \pa_a \psiI\nex
  & \quad -\frac{4}{r^2_z}\pa_r v^a_{\perp z} \int_{\eta_z}^{\eta \o} \t{d}\eta \,  \pa_a \psiI  -\frac{2\Hcal^\pr_z}{\Hcal_z}q^{ab}\pa_r v_{a\perp z} \int_{\eta_z}^{\eta_{\t{o}}}\frac{\t{d}\eta}{r^{2}}\int_{\eta}^{\eta_{\t{o}}}\t{d}\eta^\pr\, \pa_b \psiI
  \nex
  & \quad + 2q^{ab} (2\pa_r v_{a \perp z} - \Hcal_z v_{a \perp z})\int_{\eta_z}^{\eta_{\t{o}}}\frac{\t{d}\eta}{r^2}\bigg [ -\frac{1}{r}\int_{\eta}^{\eta_{\t{o}}}\t{d}\eta^\pr \, \pa_b \psiI  + \pa_b\psiI  \nex
  & \quad + \int_{\eta}^{\eta_{\t{o}}}\t{d}\eta^\pr \,\pa_{\eta^\pr }\pa_b\psiI\bigg ] + \bigg (\frac{\Hcal^\pr_z}{\Hcal_z}-\Hcal_z \bigg )  (\psiI_z \vz - 2 \psiI \o \vo  )  \nex
  & \quad +\frac{3}{2}\frac{\Hcal^\pr_z - \Hcal^2_z}{\Hcal_z}\psiI \o \vo+ \psiI \o (\Hcal \o \vo - \Hcal_z \vz) - \bigg (\frac{\Hcal^\pr_z}{\Hcal_z}+\Hcal_z \bigg ) \psiI_z \vo\nex
 & \quad +\frac{1}{\Hcal^2_z}\bigg (\frac{3(\Hcal^\pr_z)^2}{\Hcal_z} + \Hcal^3_z - \Hcal^{\pr \pr}_z - 2 \Hcal_z \Hcal^\pr_z \bigg ) (\vo - \vz)(\psiI \o - \psiI_z)      \nex
 & \quad + \frac{3\Hcal^\pr_z}{\Hcal^2_z}(\psiI_z-\psiI \o)+   (2 \psiI -3 \psiI_z )\pa_r \vz +\frac{5}{2}\psiI \o \pa_r \vo + \frac{2}{\Hcal_z} \pa_\eta \psiI_z \pa_r \vz\nex
 & \quad   +\frac{2}{r^2_z}v^a_{\perp z}\pa_a \psiI_z +\vo \bigg [\pa_\eta \psiI \o +\frac{7}{2}\pa_r \psiI \o  -2 \pa_r \psiI_z +\bigg ( 1-\frac{3\Hcal^\pr_z}{\Hcal^2_z} \bigg )\pa_\eta \psiI_z\bigg ]
 \nex 
 & \quad + \bigg (\frac{3\Hcal^\pr_z}{\Hcal^2_z}-4 \bigg ) \vz \pa_\eta \psiI_z -\frac{1}{\Hcal_z}(\vz - \vo) (\pa^2_\eta \psiI_z +\pa^2_r \psiI_z )\Bigg \} \label{eq:mixedI}\, , \\[1ex] 
  \left (\frac{\Delta z}{\Delta \tau \o}\right)_{\t{mixed,A}}^{(2)}  &=\frac{1}{a_z}\Bigg \{\vo (\pa_r \psiA \o + \pa_\eta \psiA \o - \Hcal \o \psiA \o ) -2 \psiA \o \pa_r \vo  - \frac{\Hcal^\pr_z - \Hcal^2_z}{\Hcal_z}\vz \psiA_z \nex
  & \quad +\vo \bigg [\bigg ( 1-\frac{3\Hcal^\pr_z}{\Hcal^2_z} \bigg ) \pa_\eta \psiA_z - \pa_r \psiA_z +\frac{1}{\Hcal_z} (\pa^2_\eta \psiA_z -\pa^2_r \psiA_z ) \bigg ] -\frac{1}{\Hcal_z}\psiA_z \pa^2_r \vz\nex
  & \quad -\frac{3\Hcal^\pr_z}{\Hcal^2_z}\psiA_z \pa_r \vz + \bigg (\frac{3\Hcal^\pr_z}{\Hcal^2_z}-2 \bigg )\vz \pa_\eta \psiA_z +\frac{2}{\Hcal_z}\pa_r \vz \pa_\eta \psiA_z\nex
  &\quad +\frac{1}{\Hcal_z}\vz (\pa^2_r \psiA_z - \pa^2_\eta \psiA_z) +\psiA \o \bigg (\Hcal_z \vz -4 \pa_r \vz +\frac{3\Hcal^\pr_z}{\Hcal^2_z}\pa_r \vz \bigg )\nex
  & \quad +\frac{1}{\Hcal^2_z}\bigg (\frac{3(\Hcal^\pr_z)^2}{\Hcal_z} +\Hcal^3_z - \Hcal^{\pr \pr}_z - 2 \Hcal_z \Hcal^\pr_z \bigg ) (\vo - \vz)(\psiA_z - \psiA \o) \nex
  & \quad +\frac{\Hcal^\pr_z -\Hcal^2_z}{\Hcal_z} \psiA_z \vo +\frac{6}{r^2_z}v^a_{\perp z} \pa_a \psiA_z \Bigg \} \label{eq:mixedA}\, , \\[1ex]
  \left (\frac{\Delta z}{\Delta \tau \o}\right)_{\t{path,I}}^{(2)}  &=\frac{1}{a_z}\Bigg \{ \bigg (\Hcal \o +\frac{\Hcal^\pr_z-\Hcal^2_z}{\Hcal_z} \bigg )(\Psi^{\t{I}(2)}_z - \Psi^{\t{I}(2)}\o)+ \pa_\eta (\Psi^{\t{I}(2)}\o  -  \Psi^{\t{I}(2)}_z )\nex
  & \quad + \pa_r( 3\Psi^{\t{I}(2)}\o  - \Psi^{\t{I}(2)}_z) + \frac{1}{2} \int_{\eta_z}^{\eta \o} \t{d}\eta \, \pa^2_\eta \bigg [-2 \Psi^\t{I(2)}+2(\psiI)^2 \nex
  & \quad -2 \bar{\ga}^{ab}\bigg (\int_{\eta}^{\eta \o} \t{d}\eta \, \pa_a \psiI \bigg )\bigg (\int_{\eta}^{\eta \o} \t{d}\eta \, \pa_b \psiI \bigg ) \nex
  & \quad +4 \psiI \bigg (\int_{\eta}^{\eta \o} \t{d}\eta \, \psiI - \psiI \o \bigg ) \bigg ]-\frac{\Hcal^\pr_z - \Hcal^2_z}{2\Hcal_z} \int_{\eta_z}^{\eta \o} \t{d}\eta \, \pa_\eta \bigg [-2 \Psi^\t{I(2)}+2(\psiI)^2\nex
  & \quad -2 \bar{\ga}^{ab}\bigg (\int_{\eta}^{\eta \o} \t{d}\eta \, \pa_a \psiI \bigg )\bigg (\int_{\eta}^{\eta \o} \t{d}\eta \, \pa_b \psiI \bigg ) +4 \psiI \bigg (\int_{\eta}^{\eta \o} \t{d}\eta \, \psiI - \psiI \o \bigg ) \bigg ] \nex
  & \quad -4\bigg (\int_{\eta_z}^{\eta \o}\t{d}\eta \, \psiI \bigg )  \bigg (\int_{\eta_z}^{\eta \o}\t{d}\eta \, \pa^3_\eta \psiI \bigg ) \nex
  & \quad +2 \bigg (2\int_{\eta_z}^{\eta \o}\t{d}\eta \, \psiI - \psiI_z\bigg )  \bigg (\int_{\eta_z}^{\eta \o}\t{d}\eta \, \pa^2_\eta \psiI \bigg )  \nex
  & \quad -8\bar{\ga}^{ab}\bigg (\int_{\eta_z}^{\eta \o}\t{d}\eta \, \pa_a \psiI \bigg ) \bigg (\int_{\eta_z}^{\eta \o}\t{d}\eta \, \pa_\eta \pa_b \psiI \bigg )   + \bigg (2\int_{\eta_z}^{\eta \o}\t{d}\eta \, \pa_\eta \psiI \bigg ) \bigg [3\pa_r \psiI_z \nex 
  & \quad + \bigg (\frac{3\Hcal^\pr_z}{\Hcal^2_z}-1 \bigg ) \pa_\eta \psiI_z -\frac{1}{\Hcal_z} (\pa^2_\eta \psiI_z + \pa^2_r \psiI_z)  + \Hcal_z (2\psiI_z - 3 \psiI \o)\nex
  & \quad +\frac{1}{\Hcal^2_z}\bigg (\frac{3(\Hcal^\pr_z)^2}{\Hcal_z}-\Hcal^{\pr \pr } \bigg ) \bigg (\psiI_z - \psiI \o + 2\int_{\eta_z}^{\eta \o}\t{d}\eta \, \pa_\eta \psiI  \bigg )\bigg ]  \nex
  & \quad +\frac{\Hcal^\pr_z}{\Hcal_z}\bigg (4 \psiI \o -3 \psiI_z -\int_{\eta_z}^{\eta \o}\t{d}\eta \, \pa_\eta \psiI\bigg )     \nex
  & \quad +2q^{ab} \bigg [\int_{\eta_z}^{\eta_{\t{o}}}\frac{\t{d}\eta}{r^2}\bigg ( -\frac{1}{r}\int_{\eta}^{\eta_{\t{o}}}\t{d}\eta^\pr \, \pa_b \psiI  + \pa_b\psiI + \int_{\eta}^{\eta_{\t{o}}}\t{d}\eta^\pr \,\pa_{\eta^\pr }\pa_b\psiI\bigg ) \bigg ]  \nex
  & \quad \times \bigg [3\int_{\eta_z}^{\eta_{\t{o}}} \t{d}\eta \, \pa_\eta \pa_b \psiI  + 2\bigg ( \frac{2}{r_z}-\frac{\Hcal^\pr_z - \Hcal^2_z}{2\Hcal_z}\bigg ) \int_{\eta_z}^{\eta_{\t{o}}} \t{d}\eta \,  \pa_b \psiI + \pa_b \psiI_z \nex
  & \quad -r_z \int_{\eta_z}^{\eta_{\t{o}}}\frac{\t{d}\eta}{r^2}\bigg ( -\frac{1}{r}\int_{\eta}^{\eta_{\t{o}}}\t{d}\eta^\pr \, \pa_b \psiI  + \pa_b\psiI + \int_{\eta}^{\eta_{\t{o}}}\t{d}\eta^\pr \,\pa_{\eta^\pr }\pa_b\psiI\bigg )    \bigg ]\nex
  & \quad + 2q^{ab} \bigg ( \int_{\eta_z}^{\eta_{\t{o}}}\frac{\t{d}\eta}{r^{2}}\int_{\eta}^{\eta_{\t{o}}}\t{d}\eta^\pr\, \pa_b \psiI\bigg ) \bigg [2\int_{\eta_z}^{\eta \o}\t{d}\eta \, \pa^2_\eta \psiI  + \pa_\eta \psiI_z + \pa_r \psiI_z -\frac{\Hcal^\pr_z}{\Hcal_z}\psiI_z \nex
  & \quad -2 \frac{\Hcal^\pr_z - \Hcal^2_z}{\Hcal_z}\int_{\eta_z}^{\eta \o}\t{d}\eta \, \pa_\eta \psiI \bigg ] +2\bigg (\int_{\eta_z}^{\eta \o}\t{d}\eta \, \psiI \bigg ) \bigg [2\pa^2_\eta \psiI \o -2 \pa^2_\eta \psiI_z  \nex
  & \quad + 2\pa_\eta \pa_r \psiI \o- \pa_\eta \pa_r \psiI_z + 2 \pa^2_r \psiI \o -  \pa^2_r \psiI_z   \nex
  & \quad + 2\frac{\Hcal^\pr_z - \Hcal^2_z}{\Hcal_z}(\pa_\eta \psiI_z - \pa_\eta \psiI \o - \pa_r \psiI \o ) +\frac{\Hcal^\pr_z}{\Hcal_z}\pa_r \psiI_z \bigg ]\nex
  & \quad -2\bar{\ga}^{ab}\bigg ( \int_{\eta_z}^{\eta \o}\t{d}\eta \, \pa_a \psiI \bigg )\bigg (\pa_b \psiI_z + \frac{4}{r_z}\int_{\eta_z}^{\eta \o}\t{d}\eta \, \pa_b \psiI \bigg ) \nex
  & \quad + \frac{1}{2\Hcal^2_z}\bigg (\frac{3(\Hcal^\pr_z)^2}{\Hcal_z}-\Hcal^{\pr \pr}_z \bigg ) (\psiI_z - \psiI \o)^2 + \frac{3\Hcal \o-\Hcal_z }{2} (\Psi^\t{I}\o)^2 \nex
  & \quad +\bigg (\frac{6\Hcal^\pr_z}{\Hcal_z}-7\Hcal_z \bigg ) (\Psi^\t{I}_z)^2 +\bigg (4\Hcal_z -\frac{3\Hcal^\pr_z}{\Hcal_z} \bigg )\psiI \o \psiI_z \nex
  & \quad -3 \psiI \o (\pa_\eta \psiI \o+\pa_r \psiI_z) +2 \psiI_z (2\pa_r \psiI \o+\pa_r \psiI_z) + 5 \psiI \o (\pa_\eta \psiI_z-\pa_r \psiI_z) \nex
  & \quad + \frac{3\Hcal^\pr_z}{\Hcal^2_z}(\psiI_z - \psiI \o)\pa_\eta \psiI_z +\frac{1}{\Hcal_z}\left [(\pa_\eta \psiI_z)^2 +(\psiI \o - \psiI_z) (\pa^2_\eta \psiI_z +\pa^2_r \psiI_z) \right ]\Bigg \} \label{eq:pathI}\, , \\[1ex]
  \left (\frac{\Delta z}{\Delta \tau \o}\right)_{\t{path,A}}^{(2)}  &=\frac{1}{a_z}\Bigg \{\bigg ( \Hcal \o +\frac{\Hcal^\pr_z - \Hcal^2_z}{\Hcal_z}\bigg )\Psi^\t{A(2)}\o + \pa_\eta (\Psi^\t{A(2)}_z- \Psi^\t{A(2)}\o) +\pa_r (\Psi^\t{A(2)}_z- \Psi^\t{A(2)}\o)  \nex
  & \quad -\frac{\Hcal^\pr_z}{\Hcal_z}\Psi^\t{A(2)}_z -4 \int_{\eta_z}^{\eta \o} \t{d}\eta \, \pa^2_\eta (\psiI \psiA) + 4\frac{\Hcal^\pr_z - \Hcal^2_z}{\Hcal_z}\int_{\eta_z}^{\eta \o} \t{d}\eta \, \pa_\eta (\psiI \psiA) \nex
  & \quad -2 (4\psiA_z + \psiA \o) \int_{\eta_z}^{\eta \o} \t{d}\eta \, \pa^2_\eta \psiI +  \frac{2}{\Hcal_z}\bigg [\bigg (\frac{\Hcal^{\pr \pr}_z}{\Hcal_z} +\Hcal^\pr_z-\frac{3(\Hcal^\pr_z)^2}{\Hcal^2_z}\bigg )(\psiA_z -\psiA \o)  \nex
  & \quad + \Hcal^2_z \psiA \o +\frac{3\Hcal^\pr_z}{\Hcal_z}\pa_\eta \psiA_z -\pa^2_\eta \psiA_z+\pa^2_r \psiA_z -\Hcal_z \pa_\eta \psiA_z \bigg ]\int_{\eta_z}^{\eta \o} \t{d}\eta \, \pa_\eta \psiI  \nex
  & \quad +2\bigg (  \pa_\eta \pa_r \psiA_z+\pa^2_r \psiA_z -\frac{\Hcal^\pr_z}{\Hcal_z}\pa_r \psiA_z \bigg ) \int_{\eta_z}^{\eta \o} \t{d}\eta \,  \psiI +4\bar{\ga}^{ab}\pa_a \psiA_z \int_{\eta_z}^{\eta \o}\t{d}\eta \, \pa_b \psiI \nex
    & \quad -8q^{ab} \pa_a \psiA_z \int_{\eta_z}^{\eta_{\t{o}}}\frac{\t{d}\eta}{r^2}\bigg [ -\frac{1}{r}\int_{\eta}^{\eta_{\t{o}}}\t{d}\eta^\pr \, \pa_b \psiI  + \pa_b\psiI + \int_{\eta}^{\eta_{\t{o}}}\t{d}\eta^\pr \,\pa_{\eta^\pr }\pa_b\psiI\bigg ] \nex
    & \quad -2q^{ab} \bigg ( \pa_\eta \pa_a \psiA_z+ \pa_r \pa_a \psiA_z -\frac{\Hcal^\pr_z}{\Hcal_z}\pa_a \psiA_z\bigg ) \int_{\eta_z}^{\eta_{\t{o}}}\frac{\t{d}\eta}{r^2} \int_{\eta}^{\eta_{\t{o}}}\t{d}\eta^\pr \, \pa_b \psiI \nex
    & \quad + \bigg (\frac{3\Hcal \o}{2}+\frac{\Hcal^\pr_z-\Hcal^2_z}{\Hcal_z} \bigg )(\psiA \o)^2 +\frac{3\Hcal_z}{2}(\psiA_z)^2-\bigg (3\Hcal \o +4 \frac{\Hcal^\pr_z-\Hcal^2_z}{\Hcal_z} \bigg )\psiI \o \psiA \o\nex
    & \quad +\Hcal_z  \psiI_z \psiA_z +  \frac{\Hcal^\pr_z-\Hcal^2_z}{\Hcal_z} (\psiI_z \psiA \o - \psiA \o \psiA_z + \psiI \o \psiA_z)\nex
    & \quad +\frac{1}{2\Hcal^2_z}\left (\frac{3(\Hcal^\pr_z)^2}{\Hcal_z} +\Hcal^3_z - \Hcal^{\pr \pr}_z -2 \Hcal_z \Hcal^\pr_z\right )  \nex
    & \quad \times \left [ (\psiA \o - \psiA_z)^2  +2 \psiI \o (\psiA_z- \psiA \o)\right ] +3 (\psiI \o - \psiA \o)\pa_\eta \psiA \o + 3\psiA \o \pa_\eta \psiI \o\nex
    & \quad +\psiI \o \pa_r \psiA \o +  \psiA \o (3\pa_r \psiI \o + \pa_r \psiA  - 3 \pa_r \psiA \o ) +\psiI_z (2\pa_r \psiA_z - 3 \pa_\eta \psiA_z)\nex
    & \quad 
    +8\psiA_z (\pa_\eta \psiI \o -\pa_r \psiI_z + \pa_r \psiI \o) + \psiI \o (2\pa_\eta \psiA_z -3 \pa_r \psiA_z) \nex
    & \quad - \psiA \o (2 \pa_\eta \psiI_z + \pa_r \psiI_z -2 \pa_\eta \psiA_z ) + \psiA_z (2 \pa_r \psiA_z - \pa_\eta \psiA_z)\nex
    & \quad -3 \bigg ( 1-\frac{\Hcal^\pr_z}{\Hcal^2_z}\bigg ) \left [(\psiI_z - \psiA_z - \psiI \o  + \psiA \o)\pa_\eta \psiA_z  + \psiA \o \pa_\eta \psiI_z \right ]\nex
    & \quad +\frac{1}{\Hcal_z}\left [(\pa_\eta \psiA_z)^2 +2 \pa_\eta \psiI_z \pa_\eta \psiA_z + (\psiA_z - \psiA \o) (\pa^2_\eta \psiI_z + \pa^2_r \psiI_z) \right. \nex
    & \quad \left. + (\psiI \o - \psiA \o - \psiI_z + \psiA \o)(\pa^2_\eta \psiA_z - \pa^2_r \psiA_z) \right ] \label{eq:pathA}
  \Bigg \}\, . 
\end{align}

Finally, all the terms containing observer monopoles  are 
\begin{align}
    \left (\frac{\Delta z}{\Delta \tau \o}\right)_{P,\t{pos}}^{(2)}  &=\frac{1}{a_z}\Bigg \{P\o^2 \bigg [\frac{\Hcal^{\pr \pr}\o}{2}-\Hcal^3 \o -\frac{\Hcal^\pr_z - \Hcal^2_z}{2\Hcal_z} (\Hcal^\pr \o - \Hcal^2\o) \bigg ] \nex
    & \quad + \frac{1}{2}\bigg (\frac{\Hcal \o}{\Hcal_z} (\Hcal^\pr_z - \Hcal^2_z) - \Hcal^\pr \o + \Hcal^2 \o \bigg )\int_{\eta_{\t{in}}}^{\eta \o} \t{d}\eta \, a \left [2\Phi^{(2)}-\Phi^2 + v^2_{||} +v^{a}_\perp v_{a \perp} \right ] \nex
    & \quad 
    + P\o \bigg [\frac{5}{2}\pa^2_r \vo - \frac{\Hcal^\pr_z - \Hcal^2_z}{\Hcal_z }  \bigg ( 1-\frac{3\Hcal \o}{\Hcal_z} \bigg )\pa_r \vz  + (\Hcal \o - \Hcal_z)\pa_r \vz  \nex
    &  \quad + \Hcal \o \Hcal_z (\vo +\vz ) + \bigg (\Hcal^\pr \o -\frac{\Hcal \o \Hcal^\pr_z}{\Hcal_z}  -2\Hcal^2 \o\bigg )\vo\bigg ] \Bigg \} \label{eq:Ppos}\, , \\[1ex] 
    \left (\frac{\Delta z}{\Delta \tau \o}\right)_{P,\t{mixed,I}}^{(2)}  &=\frac{P\o}{a_z}\Bigg \{2\int_{\eta_z}^{\eta \o}\t{d}\eta \, \pa^3_\eta \psiI -2\bigg ( \Hcal \o +\frac{\Hcal^\pr_z-\Hcal^2_z}{\Hcal_z} \bigg ) \int_{\eta_z}^{\eta \o}\t{d}\eta \, \pa^2_\eta \psiI -\pa^2_\eta \psiI \o   \nex
    & \quad + 9\pa^2_r \psiI \o  - \pa_\eta \pa_r \psiI \o + \pa_\eta \pa_r \psiI_z - \bigg ( 2 +\frac{\Hcal \o}{\Hcal_z}\bigg ) \pa^2_r \psiI_z + \bigg ( \Hcal \o -\frac{\Hcal^\pr_z}{\Hcal_z}\bigg ) \pa_r \psiI_z\nex
    & \quad +\frac{\Hcal^\pr_z-\Hcal^2_z}{\Hcal_z}(\Hcal \o \psiI \o + \pa_\eta \psiI \o +\frac{1}{2} \pa_r \psiI \o -2 \pa_\eta \psiI_z + \Hcal \o \psiI_z)  \nex
    & \quad +2 (\Hcal^\pr \o - \Hcal^\pr_z )\psiI \o +\Hcal \o \bigg [2\pa_\eta \psiI \o +2 \pa_r \psiI \o -2\Hcal \o \psiI \o \nex
    & \quad + \bigg (\frac{3\Hcal^\pr_z}{\Hcal^2_z}-5 \bigg ) \pa_\eta \psiI_z -\frac{1}{\Hcal_z}\pa^2_\eta \psiI_z \bigg ]\Bigg \} \label{eq:PmixedI}\, , \\[1ex] 
    \left (\frac{\Delta z}{\Delta \tau \o}\right)_{P,\t{mixed,A}}^{(2)}  &=\frac{P\o}{a_z}\Bigg \{\frac{\Hcal \o}{\Hcal_z}  (\pa^2_r \psiA_z - \pa^2_\eta \psiA_z) +2\pa^2_r \psiA_z  + \pa_\eta \pa_r \psiA \o - \pa_\eta \pa_r \psiA_z +\pa^2_\eta \psiA \o \nex
    & \quad + \bigg ( \Hcal \o +\frac{\Hcal^\pr_z}{\Hcal_z}\bigg ) (\pa_r \psiA_z - \pa_\eta \psiA \o) +\frac{\Hcal^\pr_z -\Hcal^2_z}{\Hcal_z}(\Hcal \o \psiA \o - \pa_r \psiA \o - \Hcal \o \psiA_z)  \nex
    & \quad -2\Hcal^\pr \o \psiA \o+\Hcal \o \bigg [2\Hcal \o \psiA \o -\pa_\eta \psiA \o-2 \pa_r \psiA \o + \bigg (\frac{3\Hcal^\pr_z}{\Hcal^2_z}-1 \bigg )\pa_\eta \psiA_z \bigg ]\Bigg \} \label{eq:PmixedA}\, , \\[1ex] 
    \left (\frac{\Delta z}{\Delta \tau \o}\right)_{v,\t{int}}^{(2)}  &=\frac{1}{a_z}\Bigg \{  \frac{1}{2a\o}\bigg (\Hcal^\pr \o - \Hcal^2 \o -\Hcal \o \frac{\Hcal^\pr_z - \Hcal^2_z}{\Hcal_z} \bigg ) \vo \int_{\eta_{\t{in}}}^{\eta_\t{o}} \t{d}\eta \, a v_{||} \nex
    & \quad +\vo \bigg [\frac{1}{2a\o} \bigg (3\Hcal \o +\frac{\Hcal^\pr_z -\Hcal^2_z}{\Hcal_z} \bigg )  \int_{\eta_{\t{in}}}^{\eta_\t{o}} \t{d}\eta \, a \pa_r v_{||} \nex
    & \quad - \frac{1}{a_z}\bigg (\frac{3\Hcal_z}{2}+\frac{\Hcal^\pr_z - \Hcal^2_z}{\Hcal_z} \bigg )\int_{\eta_{\t{in}}}^{\eta_z} \t{d}\eta \, a \pa_r v_{||}\bigg ]\nex
    & \quad + \pa_r \vo \bigg [\frac{1}{a\o} \bigg (\Hcal \o+\frac{\Hcal^\pr_z - \Hcal^2_z}{2\Hcal_z} \bigg )  \int_{\eta_{\t{in}}}^{\eta_\t{o}} \t{d}\eta \, a v_{||} -\frac{1}{a_z}\frac{\Hcal^\pr_z}{\Hcal_z} \int_{\eta_{\t{in}}}^{\eta_z} \t{d}\eta \, a v_{||}\bigg ]\nex
    & \quad +  \frac{3}{2}\pa_r \vo \bigg (\frac{1}{a_z} \int_{\eta_{\t{in}}}^{\eta_z} \t{d}\eta \, a \pa_r v_{||} -\frac{1}{a\o}\int_{\eta_{\t{in}}}^{\eta_\t{o}} \t{d}\eta \, a \pa_r v_{||}\bigg ) \nex
    & \quad + \vo \bigg ( \frac{1}{a_z}\int_{\eta_{\t{in}}}^{\eta_z} \t{d}\eta \, a \pa^2_r v_{||} -\frac{1}{a\o}\int_{\eta_{\t{in}}}^{\eta_\t{o}} \t{d}\eta \, a \pa^2_r v_{||} \bigg )\nex
    & \quad +\frac{1}{2}\pa^2_r \vo \bigg (\frac{1}{a_z}\int_{\eta_{\t{in}}}^{\eta_z} \t{d}\eta \, a v_{||}-\frac{1}{a\o}\int_{\eta_{\t{in}}}^{\eta_\t{o}} \t{d}\eta \, a v_{||} \bigg )\Bigg \} \label{eq:vint}\, . 
\end{align}
On sub-Hubble scales ($k\gg \Hcal$), the leading contributions are those with the highest number of radial and angular derivatives. Considering Eq.~\eqref{eq:RD-1st-final-complete} and Eqs.~\eqref{eq:pos}-\eqref{eq:pathA}, we are simply left with
\begin{align}
 \left (\frac{\Delta z}{\Delta \tau \o}\right)_{\t{leading}}^{(1)}  &=\frac{1}{a_z}\Bigg \{\frac{\Hcal^\pr_z}{\Hcal_z}\vz -\pa_r \Phi_z\Bigg \} \, ,  \nex
\left (\frac{\Delta z}{\Delta \tau \o}\right)_{\t{leading}}^{(2)}  &=\frac{1}{a_z}\Bigg \{\frac{1}{\Hcal_z}(\pa_r\vz)^2 +\frac{\Hcal^\pr_z}{\Hcal_z}v^{(2)}_{||z}- \pa_r \Phi^{(2)}_z + 
\begin{pmatrix}
    \t{Squared terms with 3}\\
    \t{spatial derivatives}
\end{pmatrix}
\Bigg \}\, ,\label{eq:leadingterms}
\end{align}
where, for the second order, we have also reported the genuine second-order velocity and gravitational potentials to have a more neat comparison with the corresponding leading terms at first order, even though at second order they are sub-leading w.r.t. the RSD $(\pa_r v_{||z})^2$.

As a side note, similar leading terms 
are also present in the contributions to Eq.~\eqref{eq:RD-1st-final-complete} and \eqref{eq:Ppos}-\eqref{eq:vint} at the observer position. However, we neglect them because in the next section we will focus just on the leading terms purely at the source position. 

\paragraph{Discussion about the term $(\pa_r \vz)^2$.} The presence of the term $(\pa_r \vz)^2$ in the second of Eqs.~\eqref{eq:leadingterms} deserves further clarifications.

In the first-order redshift drift, the terms involving $\pa_r\vz$ cancel out. On the contrary, this is not
the case for similar terms in the second-order expression, where the term 
$(\pa_r\vz)^2$ is present.  Indeed, at first order, considering Eqs.~\eqref{eq:rsd-glc-fully-nonlinear} and \eqref{eq:N-gi-firstorder}-\eqref{eq:step-first-order}, we see that one term of the form $\pa_r \vz$ stems from the light-cone derivative of the redshift, $\pa_{w}(1+z)$, and another one from its time derivative,  $\pa_{\tau}(1+z)$. These two terms cancel among each others because they are contained in the combination
\begin{equation}
    \bigg (\pa_\tau + \frac{\pa_w}{a} \bigg ) \ms{N} = \left ( \bar{u}^\mu \pa_\mu \right ) \ms{N}\, , 
\end{equation}
where $\bar{u}^\mu$ is the background counterpart of the 4-velocity of a free-falling observer (see Eqs.~\eqref{eq:4-velocity-GLC}). In general, the quantity  $u^\mu \pa_\mu$ is the evolution operator along a given geodesic connecting source and observer belonging to the same light-cone.

Thus, this cancellation is physically due to the fact that the observer measuring the redshift drift of some given source always lies on the same light-cone of the source, and this is valid at any perturbative order. This point was not emphasized in \cite{Bessa:2023qrr}, and the use of GLC coordinates to compute the redshift drift helps in highlighting this subtlety in a natural and clear way. At the same time, there is a second source of this kind of contribution. Indeed,  terms proportional to $\pa_r \vz$ can occur in the redshift expansion of $H_\t{s}$. However, we can easily see, from the second of Eqs.~\eqref{eq:exp-hubble}, that $\pa_r v_{||z}$ in $\dot{\tau}^{(1)}_z$ cancels out at first order. 

The same cancellation occurs also at second order for the terms $\pa_r v^{(2)}_{||z}$, while  the term $(\pa_r \vz)^2$ survives in the redshift expansion of $H_\t{s}$, where a term proportional to $(\dot{\tau}^{(1)}_z)^2$ has no counterpart to be canceled with (see Eqs.~\eqref{eq:exp-hubble}). 

This is a very important result, which can be emphasized when compared against what happens \textit{e.g.} in the galaxy number counts \cite{DiDio:2014lka,DiDio:2015bua}. Indeed, the leading terms in the number counts have two radial/angular derivatives at first order and four at second order. Thus, the leading terms in the bispectrum and in the squared spectrum are of the same order in the sub-Hubble regime, since both of them contain eight radial/angular derivatives. On the contrary, for the redshift drift at  leading order the bispectrum has six radial derivatives, while the square spectrum has only four radial derivatives. This seems to suggest that, on sub-Hubble scales, the magnitude of the three-point function for the redshift drift is significantly enhanced compared to  the magnitude of the two-point function. This fact would make it easier to observe the three-point function w.r.t. what one could naively expect from the  two-point function, with real future galaxy surveys or in relativistic $N$-body simulation (see \textit{e.g.} \cite{Koksbang:2023tun,Bessa:2024beh,Oestreicher:2025qcs} for numerical simulations of the power spectrum of the redshift drift). We will further address this point with some numerical estimates in the next section.

\section{The  Leading Terms in the Bispectrum}
\label{eq:bispectrum}
In this section, following  \cite{DiDio:2014lka, DiDio:2015bua,DiDio:2018unb,Schiavone:2023olz}, we will  compute the bispectrum of the redshift drift considering the leading terms written in Eqs.~\eqref{eq:leadingterms}. Since we are interested in the bispectra of the relative fluctuations w.r.t. the background value (see also \cite{Bessa:2023qrr}), we define
\begin{align}
    & \frac{\widetilde{\Delta z}}{\Delta \tau_\t{o}}\equiv \left ( \frac{\Delta \bar{z}}{\Delta \tau_\t{o}}\right )^{-1}\frac{\Delta z}{\Delta \tau_\t{o}}\qquad , \qquad \frac{\Delta \bar{z}}{\Delta \tau_\t{o}} = \frac{\Hcal \o - \Hcal_z}{a_z} \, , 
\end{align}
so  we have  
\begin{align}
 \left (\frac{\widetilde{\Delta z}}{\Delta \tau \o}\right)_{\t{leading}}^{(1)}  &=\frac{1}{\Hcal \o - \Hcal_z}\Bigg \{\frac{\Hcal^\pr_z}{\Hcal_z}\vz -\pa_r \Phi_z\Bigg \} \, , \nex
\left (\frac{\widetilde{\Delta z}}{\Delta \tau \o}\right)_{\t{leading}}^{(2)}  &=\frac{1}{\Hcal_z \left (\Hcal \o - \Hcal_z\right ) }(\pa_r\vz)^2  \, ,  
 \label{eq:leadingterms-normalized}
\end{align}
where, at second order, we limit ourselves to considering only the leading term with 4 radial derivatives.
 As also stressed at the end of Subsect.~\ref{subsec:fitst-order-RD}, unlike what done in \cite{Bessa:2023qrr}, here we do not \textit{a priori} drop the term $\Hcal \o$ from the background redshift drift, because this observer contribution multiplies perturbations at the source position, hence it is not just a monopole term that can be removed (we shall discuss in detail the consequences of this fact in Subsect.~\ref{subsec:plots-RD}).

\subsection{General mathematical formulae for the bispectrum}
The bispectrum of the redshift drift is defined as the connected part of the three-point correlation function, \textit{i.e.}
\begin{align}
    B(z_1, z_2, z_3, \hat{\mathbf{n}}_1, \hat{\mathbf{n}}_2, \hat{\mathbf{n}}_3) \equiv \left \langle \frac{\widetilde{\Delta z}}{\Delta \tau \o }(z_1,\hat{\mathbf{n}}_1)\frac{\widetilde{\Delta z}}{\Delta \tau \o }(z_2,\hat{\mathbf{n}}_2)\frac{\widetilde{\Delta z}}{\Delta \tau \o }(z_3,\hat{\mathbf{n}}_3)\right \rangle_\t{c} \, ,
\end{align}
where, following the notation of \cite{DiDio:2015bua}, the observation direction is defined as the unit vector $-\hat{\mathbf{n}}$, while  $\hat{\mathbf{n}}$ represents the photon  propagation direction in the sky.  We can then expand the redshift drift in spherical harmonics as
\begin{align}
   \frac{\widetilde{\Delta z}}{\Delta \tau \o }(z,\hat{\mathbf{n}}) = \sum_{\ell=1}^\infty \sum_{m=-\ell}^\ell a_{\ell m}(z)Y_{\ell m}(\hat{\mathbf{n}})\qquad , \qquad  a_{\ell m}(z) = \int \t{d}\Omega_{\hat{\mathbf{n}}} \, \frac{\widetilde{\Delta z}}{\Delta \tau \o }(z,\hat{\mathbf{n}}) Y^*_{\ell m}(\hat{\mathbf{n}}) \, , 
   \label{eq:harmonic-decomposition}
\end{align}
and the bispectrum as
\begin{align}
    B(z_1, z_2, z_3, \hat{\mathbf{n}}_1, \hat{\mathbf{n}}_2, \hat{\mathbf{n}}_3)  = \underset{m_1m_2m_3}{\sum_{\ell_1 \ell_2 \ell_3}} 
     B^{m_1 m_2 m_3}_{\ell_1 \ell_2 \ell_3} Y_{\ell_1 m_1}(\hat{\mathbf{n}}_1)Y_{\ell_2 m_2}(\hat{\mathbf{n}}_2)Y_{\ell_3 m_3}(\hat{\mathbf{n}}_3) \, , 
\end{align}
having defined the coefficients $B^{m_1 m_2 m_3}_{\ell_1 \ell_2 \ell_3}$ as
\begin{align}
    B^{m_1 m_2 m_3}_{\ell_1 \ell_2 \ell_3} (z_1, z_2, z_3) &\equiv \left \langle a_{\ell_1 m_1}(z_1)a_{\ell_2 m_2}(z_2)a_{\ell_3 m_3}(z_3) \right \rangle \nex
    & = \int \t{d}\Omega_1  \t{d}\Omega_2  \t{d}\Omega_3 \, B(z_1, z_2, z_3, \hat{\mathbf{n}}_1, \hat{\mathbf{n}}_2, \hat{\mathbf{n}}_3)   Y^*_{\ell_1 m_1}(\hat{\mathbf{n}}_1)  Y^*_{\ell_2 m_2}(\hat{\mathbf{n}}_2) Y^*_{\ell_3 m_3}(\hat{\mathbf{n}}_3) \, . 
\end{align}
The statistical isotropy of the background geometry enforces  the bispectrum to be of the form
\begin{align}
    B^{m_1 m_2 m_3}_{\ell_1 \ell_2 \ell_3} (z_1, z_2, z_3) = \mathcal{G}^{m_1m_2m_3}_{\ell_1 \ell_2 \ell_3} \, b_{\ell_1 \ell_2 \ell_3}(z_1,z_2,z_3) \, .
\end{align}
The quantity $\mathcal{G}^{m_1m_2m_3}_{\ell_1 \ell_2 \ell_3}$ is the so-called Gaunt integral, defined by
\begin{align}   \mathcal{G}^{m_1m_2m_3}_{\ell_1 \ell_2 \ell_3} &\equiv \int \t{d}\Omega \, Y_{\ell_1 m_1} (\hat{\mathbf{n}})Y_{\ell_2 m_2} (\hat{\mathbf{n}}) Y_{\ell_3 m_3} (\hat{\mathbf{n}})\nex
    & = \begin{pmatrix}
        \ell_1 &  \ell_2 & \ell_3 \\
        0 & 0 & 0 
    \end{pmatrix}
    \begin{pmatrix}
        \ell_1 &  \ell_2 & \ell_3 \\
        m_1 &  m_2 & m_3
    \end{pmatrix}
    \sqrt{\frac{(2\ell_1+1)(2\ell_2+1)(2\ell_3+1)}{4\pi}} \, , 
\end{align}
where we have introduced the Wigner $3j$ symbols  (see, for example, \cite{10.5555/1098650,Komatsu:2001ysk}). Note that the Gaunt integral is non-zero only for
\begin{equation}
    m_1 +m_2+m_3 =0 \quad , \quad |\ell_2 - \ell_3| \leqslant \ell_1 \leqslant \ell_2 + \ell_3 \quad , \quad \ell_1 + \ell_2 + \ell_3 = 2n \,\,  \forall n \in \mathbb{N} \, ,
    \label{eq:triangle}
\end{equation}
where the last two conditions are just the triangular inequality.
Assuming Gaussian initial conditions, namely
\begin{equation}
\left \langle \bigg (\frac{\widetilde{\Delta z}}{\Delta \tau \o }\bigg)^{(1)}(z_1,\hat{\mathbf{n}}_1)\bigg (\frac{\widetilde{\Delta z}}{\Delta \tau \o }\bigg)^{(1)}(z_2,\hat{\mathbf{n}}_2)\bigg (\frac{\widetilde{\Delta z}}{\Delta \tau \o }\bigg)^{(1)}(z_3,\hat{\mathbf{n}}_3)\right \rangle_\t{c} =0 \, , 
\label{eq:GIC}
\end{equation}
we are then interested in computing the tree-level bispectrum given by
\begin{equation}
    \left \langle \bigg (\frac{\widetilde{\Delta z}}{\Delta \tau \o }\bigg)^{(2)}(z_1,\hat{\mathbf{n}}_1)\bigg (\frac{\widetilde{\Delta z}}{\Delta \tau \o }\bigg)^{(1)}(z_2,\hat{\mathbf{n}}_2)\bigg (\frac{\widetilde{\Delta z}}{\Delta \tau \o }\bigg)^{(1)}(z_3,\hat{\mathbf{n}}_3)\right \rangle_\t{c} + \t{perms.} \, ,
    \label{eq:tree-level-bispectrum-def}
\end{equation}
that is,
\begin{align}
    B^{m_1m_2m_3}_{\ell_1 \ell_2 \ell_3}(z_1,z_2,z_3) &= \left \langle a^{(2)}_{\ell_1 m_1}(z_1)a^{(1)}_{\ell_2 m_2}(z_2) a^{(1)}_{\ell_3 m_3}(z_3)\right \rangle_\t{c} +\left \langle a^{(1)}_{\ell_1 m_1}(z_1)a^{(2)}_{\ell_2 m_2}(z_2) a^{(1)}_{\ell_3 m_3}(z_3)\right \rangle_\t{c} \nex
    & \quad +\left \langle a^{(1)}_{\ell_1 m_1}(z_1)a^{(1)}_{\ell_2 m_2}(z_2) a^{(2)}_{\ell_3 m_3}(z_3)\right \rangle_\t{c} \, . 
\end{align}

Next, following the notation of \cite{DiDio:2015bua}, we redefine $v_{||}\equiv \pa_r v$, $v$ being the velocity potential previously indicated as the time gauge mode $\xi^\tau$. Thus, because of Eqs.~\eqref{eq:leadingterms} (and dropping from now on the subscript \quotes{$z$}), Eq.~\eqref{eq:tree-level-bispectrum-def} amounts to computing the following three correlators (we omit the dependence of $B$ on the 6 variables $(z_1 , \dots \hat{\mathbf{n}}_3)$):  
\begin{align}
    B^{ v^{\pr^2} v  v} &=\left \langle \frac{\left(\pa^2_r v_{}(z_1, \hat{\mathbf{n}}_1)\right )^2}{\Hcal(z_1) \left [ \Hcal \o - \Hcal(z_1)\right ]} \frac{\Hcal^\pr(z_2) \pa_r v(z_2, \hat{\mathbf{n}}_2)}{\Hcal(z_2) \left [ \Hcal \o - \Hcal(z_2)\right ]} \frac{\Hcal^\pr(z_3) \pa_r v(z_3, \hat{\mathbf{n}}_3)}{ \Hcal(z_3) \left [ \Hcal \o - \Hcal(z_3)\right ]}\right \rangle_\t{c} + \t{perms.} \label{eq:bispectrum-RD-vvv}
    \, ,\\[1ex]
    B^{v^{\pr^2} \Phi^\pr \Phi^\pr }  &=\left \langle \frac{\left(\pa^2_r v_{}(z_1, \hat{\mathbf{n}}_1)\right )^2}{ \Hcal(z_1) \left [ \Hcal \o - \Hcal(z_1)\right ]} \frac{ \pa_r \Phi(z_2, \hat{\mathbf{n}}_2)}{  \Hcal \o - \Hcal(z_2) } \frac{ \pa_r \Phi(z_3, \hat{\mathbf{n}}_3)}{ \Hcal \o - \Hcal(z_3) }\right \rangle_\t{c} + \t{perms.}    \label{eq:bispectrum-RD-vphiphi}\, , \\[1ex]
    B^{v^{\pr^2} v \Phi^\pr }  &=-2 \left \langle \frac{\left(\pa^2_r v_{}(z_1, \hat{\mathbf{n}}_1)\right )^2}{ \Hcal(z_1) \left [ \Hcal \o - \Hcal(z_1)\right ]} \frac{\Hcal^\pr(z_2) \pa_r v(z_2, \hat{\mathbf{n}}_2)}{ \Hcal(z_2) \left [\Hcal \o - \Hcal(z_2)
\right ]} \frac{ \pa_r \Phi(z_3, \hat{\mathbf{n}}_3)}{ \Hcal \o - \Hcal(z_3) }\right \rangle_\t{c} + \t{perms.}   \, ,  \label{eq:bispectrum-RD-vphi-mixed}
\end{align}
and we then indicate the related reduced bispectra as 
\begin{equation}
    b^{v^{\pr^2} v v }_{\ell_1 \ell_2 \ell_3} \qquad , \qquad b^{v^{\pr^2} \Phi^\pr \Phi^\pr }_{\ell_1 \ell_2 \ell_3}  \qquad , \qquad b^{v^{\pr^2} v\Phi^\pr }_{\ell_1 \ell_2 \ell_3}\, .  
\end{equation}
By working in Fourier space, any function $f$ can be written as
\begin{align}
    f(z, \mathbf{x}) = \int \frac{\t{d}^3k}{(2\pi)^3} \, f(z, \mathbf{k}) e^{-i\mathbf{k}\cdot \mathbf{x}} =\int \frac{\t{d}^3k}{(2\pi)^3} \, f(z, \mathbf{k}) e^{-ir(z)\mathbf{k}\cdot \mathbf{n}} \, . 
\end{align}
Therefore, the first-order Newtonian potential and velocity potential are expressed as
\begin{align}
    \Phi(z, \hat{\mathbf{n}}) &= \int \frac{\t{d}^3k}{(2\pi)^3} \, \Phi\left (z, \mathbf{k}\right ) e^{-ir(z)\mathbf{k}\cdot \hat{\mathbf{n}}}  \qquad  , \qquad 
    v(z, \hat{\mathbf{n}}) = \int \frac{\t{d}^3k}{(2\pi)^3} \, v\left (z, \mathbf{k}\right ) e^{-ir(z)\mathbf{k}\cdot \hat{\mathbf{n}}}   \, . 
    \label{eq:velocity-fourier}
\end{align}
From now on, for brevity, we will adopt the shortcut $r(z)\equiv \eta \o -\eta (z)= r$.

To compute the quantities in Eqs.~\eqref{eq:bispectrum-RD-vvv}-\eqref{eq:bispectrum-RD-vphi-mixed}, we have to recall that, in linear perturbation theory, the power spectrum of the initial curvature perturbation is expressed as 
\begin{equation}
    \left \langle \mathcal{R}(\mathbf{k})\mathcal{R}(\mathbf{k}^\pr)\right \rangle = (2\pi)^3 \delta^{(3)}(\mathbf{k}+\mathbf{k}^\pr) P_\mathcal{R}(k) \, , 
\end{equation}
where $\delta^{(3)}$ is the 3-dimensional Dirac-delta function and $k \equiv |\mathbf{k}|$.
Furthermore, assuming adiabatic initial conditions, the transfer function $T_A (z, k)$ of any perturbation $A$ is defined through
\begin{equation}
    A(z, \mathbf{k}) = T_A (z, k) \mathcal{R}(\mathbf{k}) \, ,
\end{equation}
and the angular power spectra of two quantities $A$ and $B$ in multipole space is 
\begin{align}
    c_\ell^{AB}(z_1,z_2) \equiv 4\pi \int \frac{\t{d}k}{k} \, \mathcal{P}_\mathcal{R}(k) \Delta^A_\ell (z_1, k)\Delta^B_\ell (z_2, k) = \frac{2}{\pi} \int \t{d}k \, k^2 P_\mathcal{R}(k)\Delta^A_\ell (z_1, k)\Delta^B_\ell (z_2, k) \, .
    \label{eq:angular-spectra-def}
\end{align}
Here we have introduced the dimensionless primordial power spectrum of scalar perturbations as
\begin{align}
    \mathcal{P}_\mathcal{R}(k) \equiv \frac{k^3}{2\pi^2}P_\mathcal{R}(k) = \mathcal{A}_\mathcal{R} \left (\frac{k}{k_*} \right )^{n_\mathcal{R}-1} \, , 
    \label{eq:dimensionless-power-spectrum-def}
\end{align}
where $\mathcal{A}_\mathcal{R}$ is the amplitude, $n_\mathcal{R}$ is the spectral index and $k_*$ is the pivot scale. Moreover, $\Delta^A_\ell (z, k)$ represents the transfer function in multipole and redshift space associated with $A$. For $v_{||}$, $\pa_r \Phi$  and $\pa_r v_{||}$, these functions are written as (see also \cite{Bernardeau:2001qr})
\begin{align}
    \Delta^{v}_{\ell}(z, k)&= j^\pr_\ell (kr) T_v (z,k)  \, ,  \nex
\Delta^{\Phi^\pr}_\ell(z,k) &= \frac{k}{ \Hcal(z)}j^\pr_\ell (kr) T_\Phi (z,k) \, , \nex
    \Delta^{ v^{\pr}}_{\ell}(z, k)&=  \frac{k}{\Hcal(z)}j^{\pr\pr}_\ell (kr) T_v (z,k)\, ,
     \label{eq:trasn-funct}
\end{align}
where $j_\ell (kr)$ is the spherical Bessel function of order $\ell$, and the \quotes{prime} denotes the derivative w.r.t. its argument $kr$. In this notation, we indicate with $T_v$ the transfer function of the module of the velocity field, given by $T_v = T_\theta/k$, where $T_\theta$ is the transfer function of the divergence of the velocity, whose numerical values can be obtained from \texttt{CLASS}  \cite{Blas:2011rf}.

\subsection{Detailed computations of the bispectrum}
Let us now proceed with an in-depth evaluation of Eqs.~\eqref{eq:bispectrum-RD-vvv}-\eqref{eq:bispectrum-RD-vphi-mixed}, using that
\begin{align}
    v_{||}(z, \hat{\mathbf{n}}) &=  \int \frac{\t{d}^3 k}{(2\pi)^3} T_v (z,k) \left [ \pa_{(kr)}e^{-ir \mathbf{k}\cdot \hat{\mathbf{n}}}\right ]\mathcal{R}(\mathbf{k}) \, , \nex
    \pa_r \Phi(z, \hat{\mathbf{n}
    }) &= \int \frac{\t{d}^3 k}{(2\pi)^3} k T_\Phi (z,k)\left [ \pa_{(kr)}e^{-ir \mathbf{k}\cdot \hat{\mathbf{n}}}\right ]\mathcal{R}(\mathbf{k})\, , \nex
     \pa_r v_{||}(z, \hat{\mathbf{n}}) &=  \int \frac{\t{d}^3 k}{(2\pi)^3} k T_v (z,k) \left [ \pa^2_{(kr)}e^{-ir \mathbf{k}\cdot \hat{\mathbf{n}}}\right ]\mathcal{R}(\mathbf{k})\, .
\end{align}
In the following computations, we will make use of the harmonic expansion of the exponential, given by
\begin{align}
    e^{-ir\mathbf{k}\cdot \hat{\mathbf{n}}} &= 4\pi \sum_{\ell=0}^\infty \sum_{m=-\ell}^\ell (-i)^\ell j_\ell (kr) Y^*_{\ell m}(\hat{\mathbf{k}})Y_{\ell m}(\hat{\mathbf{n}}) \, ,
    \label{eq:exp-Yj}
\end{align}
where $Y_{\ell m}$ are the spherical harmonics.

\subsubsection{Term $b^{v^{\pr^2} v v }$ }
Defining for brevity
\begin{align}
    \mathscr{H}_{v^{\pr^2} v v}(z_1,z_2,z_3) \equiv \frac{\Hcal^\pr(z_2) \Hcal^\pr(z_3)}{\Hcal(z_1) \left [ \Hcal \o - \Hcal(z_1)\right ] \Hcal(z_2) \left [ \Hcal \o - \Hcal(z_2)\right ]\Hcal(z_3) \left [ \Hcal \o - \Hcal(z_3)\right ]} \, , 
    \label{eq:h-prefactor}
\end{align}
we can use   the Wick theorem to compute the correlation function in Eq.~\eqref{eq:bispectrum-RD-vvv} as
\begin{align}
    \langle \cdots \rangle_\t{c} &= \mathscr{H}_{v^{\pr^2} v v}(z_1,z_2,z_3) \int \frac{\t{d}^3k \, \t{d}^3 k_1 \, \t{d}^3 k_2 \, \t{d}^3 k_3}{(2\pi)^{12}} \, \left [\pa^2_{(kr_1)} e^{-ir_1\mathbf{k}\cdot \hat{\mathbf{n}}_1}\right ]\left [\pa^2_{(k_1r_1)} e^{-ir_1\mathbf{k}_1\cdot \hat{\mathbf{n}}_1}\right ]  \nex
    & \quad \times \left [\pa_{(k_2r_2)} e^{-ir_2\mathbf{k}_2\cdot \hat{\mathbf{n}}_2}\right ]\left [\pa_{(k_3r_3)} e^{-ir_3\mathbf{k}_3\cdot \hat{\mathbf{n}}_3}\right ] kk_1 T_v (z_1, k)T_v (z_1, k_1) T_v (z_2,k_2) T_v (z_3,k_3)  \nex
    & \quad \times \left \langle \mathcal{R}(\mathbf{k})\mathcal{R}(\mathbf{k}_1)\mathcal{R}(\mathbf{k}_2)\mathcal{R}(\mathbf{k}_3)\right \rangle \nex
    & =2\mathscr{H}_{v^{\pr^2} v v}(z_1,z_2,z_3) \int \frac{\t{d}^3k \, \t{d}^3 k_1 \,}{(2\pi)^{6}} \, \left [\pa^2_{(kr_1)} e^{-ir_1\mathbf{k}\cdot \hat{\mathbf{n}}_1}\right ]\left [\pa^2_{(k_1r_1)} e^{-ir_1\mathbf{k}_1\cdot \hat{\mathbf{n}}_1}\right ]  \nex
    & \quad \times \left [\pa_{(kr_2)} e^{ir_2\mathbf{k}\cdot \hat{\mathbf{n}}_2}\right ]\left [\pa_{(k_1r_3)} e^{ir_3\mathbf{k}_1\cdot \hat{\mathbf{n}}_3}\right ]kk_1 T_v(z_1, k)T_v (z_1, k_1) T_v(z_2,k) T_v(z_3,k_1) \nex
    & \quad \times P_\mathcal{R}(k)P_\mathcal{R}(k_1) \, . 
    \label{eq:comput-step1}
\end{align}
Next, using  Eq.~\eqref{eq:exp-Yj} and the orthogonality condition
\begin{equation}
    \int \t{d}\Omega_{\hat{\mathbf{k}}} \, Y_{\ell m} (\hat{\mathbf{k}}) Y^*_{\ell^\pr m^\pr} (\hat{\mathbf{k}}) =\delta_{\ell \ell^\pr}\delta_{m m^\pr} \, , 
\end{equation}
the correlator in Eq.~\eqref{eq:comput-step1} takes the form
\begin{align}
    \langle \cdots \rangle_\t{c} 
    & =  2\mathscr{H}_{v^{\pr^2} v v}(z_1,z_2,z_3) \frac{4}{\pi^2}\sum_{\ell m}\sum_{\ell^\pr m^\pr} \int  \t{d}k \, \t{d}k_1 k^3 k_1^3 T_v (z_1, k)T_v (z_1, k_1) T_v (z_2, k) T_v (z_3, k_1)  \nex
    & \quad \times P_\mathcal{R}(k)P_\mathcal{R}(k_1) j^{\pr \pr}_{\ell} (kr_1)j^{\pr}_{\ell} (kr_2)j^{\pr \pr}_{\ell^\pr} (k_1r_1)j^{\pr}_{\ell^\pr} (k_1r_3) Y_{\ell m}(\hat{\mathbf{n}}_1)Y^*_{\ell m}(\hat{\mathbf{n}}_2)Y_{\ell^\pr m^\pr}(\hat{\mathbf{n}}_1)Y^*_{\ell^\pr m^\pr}(\hat{\mathbf{n}}_3) \, , 
\end{align}
which shows that
\begin{align}
    B^{v^{\pr^2} v v}(z_1, z_2, z_3, \hat{\mathbf{n}}_1, \hat{\mathbf{n}}_2, \hat{\mathbf{n}}_3) &= \frac{4}{\pi^2} \underset{mm^\pr}{\sum_{\ell \ell^\pr}}Y_{\ell m}(\hat{\mathbf{n}}_1) Y_{\ell^\pr m^\pr}(\hat{\mathbf{n}}_1)  Y^*_{\ell m}(\hat{\mathbf{n}}_2)Y^*_{\ell^\pr m^\pr}(\hat{\mathbf{n}}_3) Z^{v^{\pr^2} v v }_{\ell \ell^\pr} (z_1, z_2, z_3)\nex
    & \quad + \t{perms.} \, , 
\end{align}
where
\begin{align}
    Z^{v^{\pr^2} v v }_{\ell \ell^\pr} (z_1, z_2, z_3) &\equiv 2\mathscr{H}_{v^{\pr^2} v v}(z_1,z_2,z_3)\int  \t{d}k \, \t{d}k_1 k^3k_1^3 T_v (z_1, k)T_v (z_1, k_1) T_v (z_2, k) \nex
    & \quad \times T_v (z_3, k_1)   P_\mathcal{R}(k)P_\mathcal{R}(k_1) j^{\pr \pr}_{\ell} (kr_1)j^{\pr}_{\ell} (kr_2)j^{\pr \pr}_{\ell^\pr} (k_1r_1)j^{\pr}_{\ell^\pr} (k_1r_3) \, .  
\end{align}
It follows that the reduced bispectrum is
\begin{align}
    b^{v^{\pr^2} v v}_{\ell_1 \ell_2 \ell_3} = \frac{4}{\pi^2}Z^{v^{\pr^2} v v }_{\ell_2 \ell_3} (z_1, z_2, z_3) + \t{perms.}  \, 
\end{align}
Finally, using Eqs.~\eqref{eq:angular-spectra-def} and \eqref{eq:trasn-funct}, we  express the final result as the product of angular power spectra:
\begin{align}
   b^{v^{\pr^2} v v }_{\ell_1 \ell_2 \ell_3}(z_1,z_2,z_3)&= 2 \bigg [\mathscr{H}_{v^{\pr^2} v v}(z_1,z_2,z_3)\Hcal(z_1)^2 \, c^{v^\pr v}_{\ell_2}(z_1,z_2)c^{v^\pr v}_{\ell_3}(z_1,z_3) \nex
   & \quad +\mathscr{H}_{v^{\pr^2} v v}(z_2,z_1,z_3)\Hcal(z_2)^2 \, c^{v^\pr v}_{\ell_1}(z_1,z_2)c^{v^\pr v}_{\ell_3}(z_2,z_3)  \nex
   & \quad +\mathscr{H}_{v^{\pr^2} v v}(z_3,z_1,z_2)\Hcal(z_3)^2 \, c^{v^\pr v}_{\ell_1}(z_1,z_3)c^{v^\pr v}_{\ell_2}(z_2,z_3)\bigg ]  \, .
   \label{eq:bispectrum:vvv-final}
\end{align}

\subsubsection{Term $b^{v^{\pr^2}\Phi^\pr \Phi^\pr}$}
Regarding the correlation function in Eq.~\eqref{eq:bispectrum-RD-vphiphi}, by following the same procedure explained in the previous subsection, and by defining
\begin{align}
    \mathscr{H}_{v^{\pr^2}\Phi^\pr \Phi^\pr}(z_1,z_2,z_3)\equiv \frac{1}{\Hcal(z_1)\left [\Hcal \o - \Hcal(z_1) \right ] \left [\Hcal \o - \Hcal(z_2) \right ] \left [\Hcal \o - \Hcal(z_3) \right ]}\, , 
\end{align}
we obtain the following reduced bispectrum:
\begin{align}
   b^{v^{\pr^2}\Phi^\pr \Phi^\pr}_{\ell_1 \ell_2 \ell_3}(z_1,z_2,z_3) &= 2 \bigg [ 
   \mathscr{H}_{v^{\pr^2}\Phi^\pr \Phi^\pr}(z_1,z_2,z_3) \Hcal(z_1)^4\, c^{v^\pr \Phi^\pr}_{\ell_2}(z_1,z_2)c^{v^\pr \Phi^\pr}_{\ell_3}(z_1,z_3)  \nex
   & \quad +  \mathscr{H}_{v^{\pr^2}\Phi^\pr \Phi^\pr}(z_2,z_1,z_3) \Hcal(z_2)^4\, c^{v^\pr \Phi^\pr}_{\ell_1}(z_1,z_2)c^{v^\pr \Phi^\pr}_{\ell_3}(z_2,z_3)   \nex
   & \quad +  \mathscr{H}_{v^{\pr^2}\Phi^\pr \Phi^\pr}(z_3,z_1,z_2) \Hcal(z_3)^4 \,  c^{v^\pr\Phi^\pr}_{\ell_1}(z_1,z_3)c^{v^\pr\Phi^\pr}_{\ell_2}(z_2,z_3)\bigg ]  \, .
   \label{eq:bispectrum:vphiphi-final}
\end{align}

\subsubsection{Term $b^{v^{\pr^2 } v\Phi^\pr}$}
Finally, by defining
\begin{align}
    \mathscr{H}_{v^{\pr^2 } v\Phi^\pr}(z_1, z_2, z_3) \equiv \frac{\Hcal^\pr(z_2)}{\Hcal (z_1)\left [\Hcal \o - \Hcal(z_1) \right ]\Hcal (z_2)\left [\Hcal \o - \Hcal(z_2) \right ]\left [\Hcal \o - \Hcal(z_3) \right ]} \, , 
\end{align}
the reduced bispectrum associated with the correlator of Eq.~\eqref{eq:bispectrum-RD-vphi-mixed} is
\begin{align}
    b^{v^{\pr^2 } v\Phi^\pr }_{\ell_1 \ell_2 \ell_3}(z_1,z_2,z_3) &= -4 \bigg [\mathscr{H}_{v^{\pr^2 } v\Phi^\pr}(z_1, z_2, z_3) \Hcal(z_1)^2 \Hcal(z_3)\, c^{v^\pr  v}_{\ell_2}(z_1,z_2)c^{v^\pr \Phi^\pr}_{\ell_3}(z_1,z_3)  \nex
   & \quad +\mathscr{H}_{v^{\pr^2 } v\Phi^\pr}(z_2, z_1, z_3) \Hcal(z_2)^2 \Hcal(z_3)\,c^{v^\pr v}_{\ell_1}(z_1,z_2)c^{ v^\pr \Phi^\pr}_{\ell_3}(z_2,z_3)   \nex
   & \quad + \mathscr{H}_{v^{\pr^2 } v\Phi^\pr}(z_3, z_1, z_2) \Hcal(z_3)^2 \Hcal(z_2)\, c^{v^\pr v}_{\ell_1}(z_1,z_3)c^{ v^\pr \Phi^\pr}_{\ell_2}(z_2,z_3)\bigg ]  \, .
   \label{eq:bispectrum:vphi-final-mixed} 
\end{align}

With these results, we are now able to provide some numerical estimations for the amplitude of the reduced bispectra presented here. This will be discussed in the following subsection.

\subsection{Numerical results}
\label{subsec:plots-RD}
Let us now focus on the numerical evaluation of the bispectrum of the redshift drift, considering the leading  contributions on sub-Hubble scales given  by $b^{v^{\pr^2}vv}_{\ell_1 \ell_2 \ell_3}$,  $b^{v^{\pr^2}\Phi^\pr \Phi^\pr}_{\ell_1 \ell_2 \ell_3}$ and  $b^{v^{\pr^2}v \Phi^\pr}_{\ell_1 \ell_2 \ell_3}$ as written in Eqs.~\eqref{eq:bispectrum:vvv-final}, \eqref{eq:bispectrum:vphiphi-final} and \eqref{eq:bispectrum:vphi-final-mixed}.  

The values of the transfer functions $\Delta^{ v} _\ell(z,k)$,  $\Delta^{\Phi^\pr}_\ell (z,k)$ and $\Delta^{v^\pr}_\ell (z,k)$, defined in Eqs. \eqref{eq:trasn-funct}, are computed with \texttt{CLASS} \cite{Blas:2011rf}, using as input the cosmological  parameters from Planck collaboration \cite{Planck:2018vyg} reported in Table \ref{tab:cosmopar}. Considering sub-Hubble scales,  the integrals in momentum space defining the coefficients  in Eq.~\eqref{eq:angular-spectra-def} are performed over the interval $k \in \left [\Hcal_0, 1 h \, \t{Mpc}^{-1}\right ]$, where $\Hcal_0$  is the value of the conformal Hubble factor at the present time.
\begin{table}
\[
\begin{array}{cc}
\toprule
    \text{Parameter} & \text{Value}  \\
 \midrule   
     A_\mathcal{R} & 2.1 \cdot 10^{-9}\\
     n_\mathcal{R}& 0.965\\
     h & 0.67 \\
     \Omega_\text{b}h^2& 0.0224\\
      \Omega_\text{cdm}h^2&0.120 \\
      \tau_\text{reio} & 0.054\\
\bottomrule
\end{array}
\]
\vspace{-10pt}
    \caption{Cosmological parameters \cite{Planck:2018vyg}  used to obtain the  figures.}
    \label{tab:cosmopar}
\end{table}

\begin{figure}
\centering

\subfloat[ $b^{v^{\pr ^2}v v}_{\ell\ell\ell}(z)$ from Eq.~\eqref{eq:bispectrum:vvv-final}\label{fig:a-bispectrum}]
{\includegraphics[width=.49\textwidth]{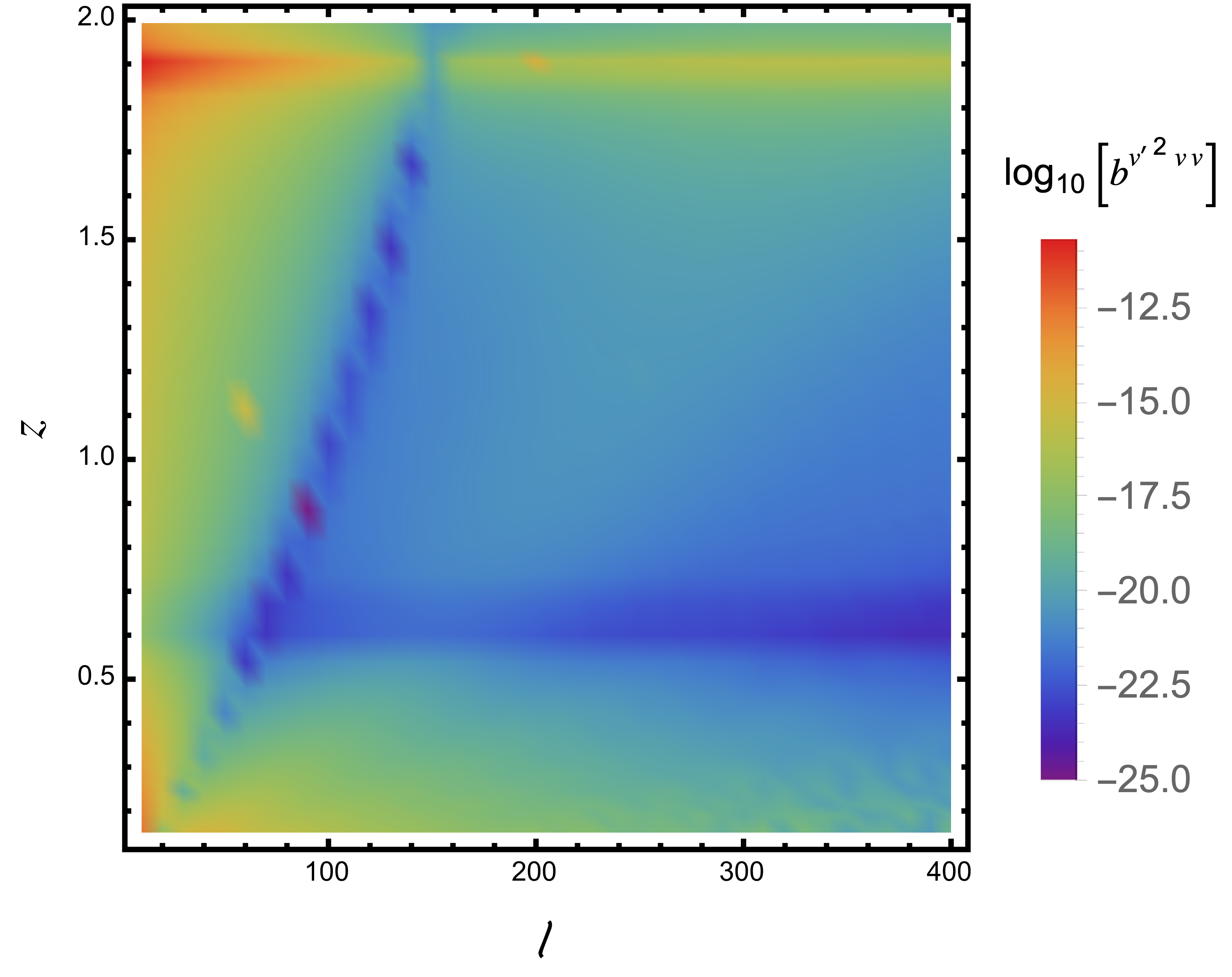}}
\hspace*{0.3em}
\subfloat[$b^{v^{\pr ^2}v \Phi^\pr}_{\ell\ell\ell }(z)$ from Eq.~\eqref{eq:bispectrum:vphi-final-mixed}\label{fig:b-bispectrum}]
{\includegraphics[width=.49\textwidth]{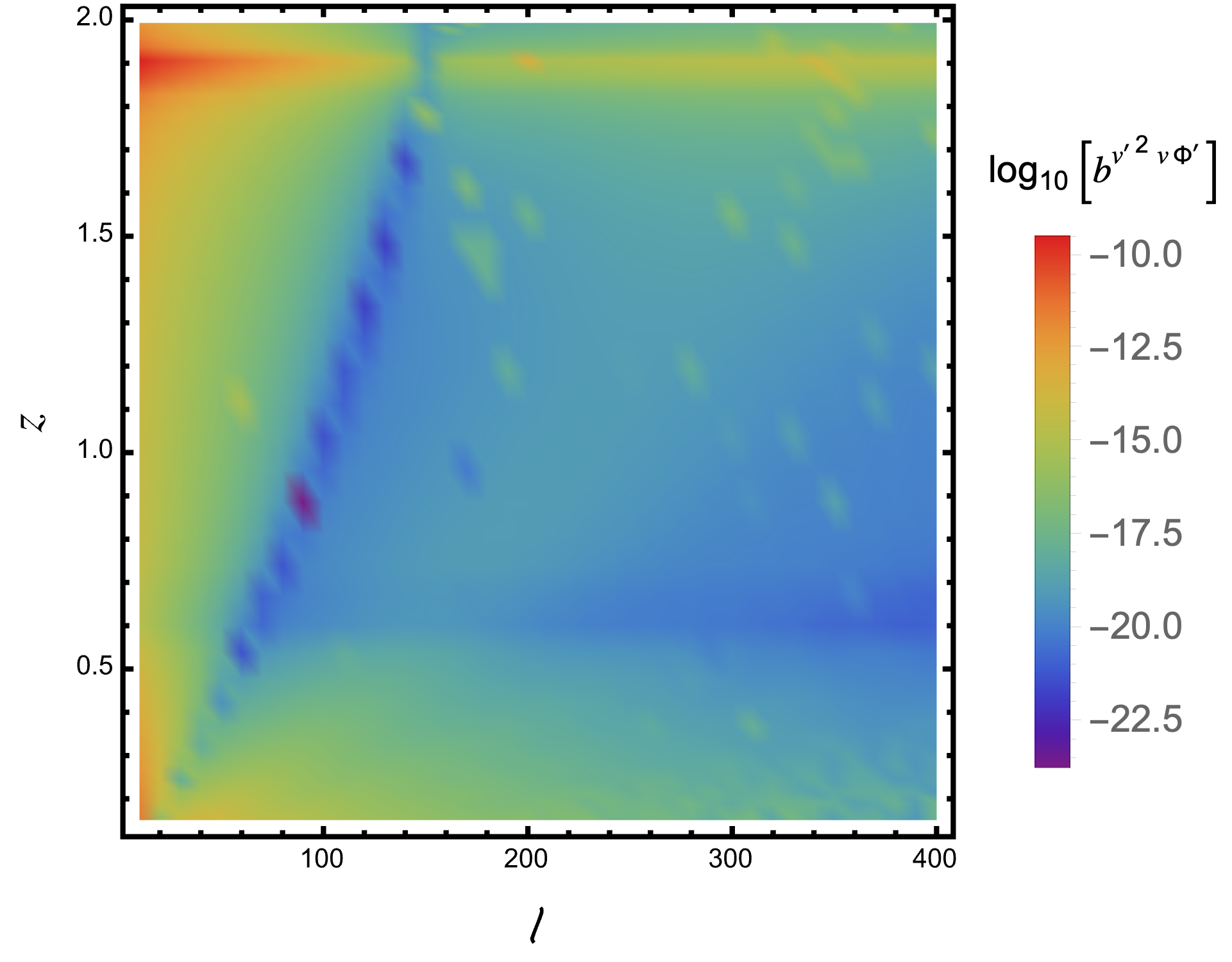}}

\medskip

\subfloat[$b^{v^{\pr ^2}\Phi^\pr \Phi^\pr}_{\ell\ell\ell}(z)$ from Eq.~\eqref{eq:bispectrum:vphiphi-final}\label{fig:c-bispectrum}]
{\includegraphics[width=.495\textwidth]{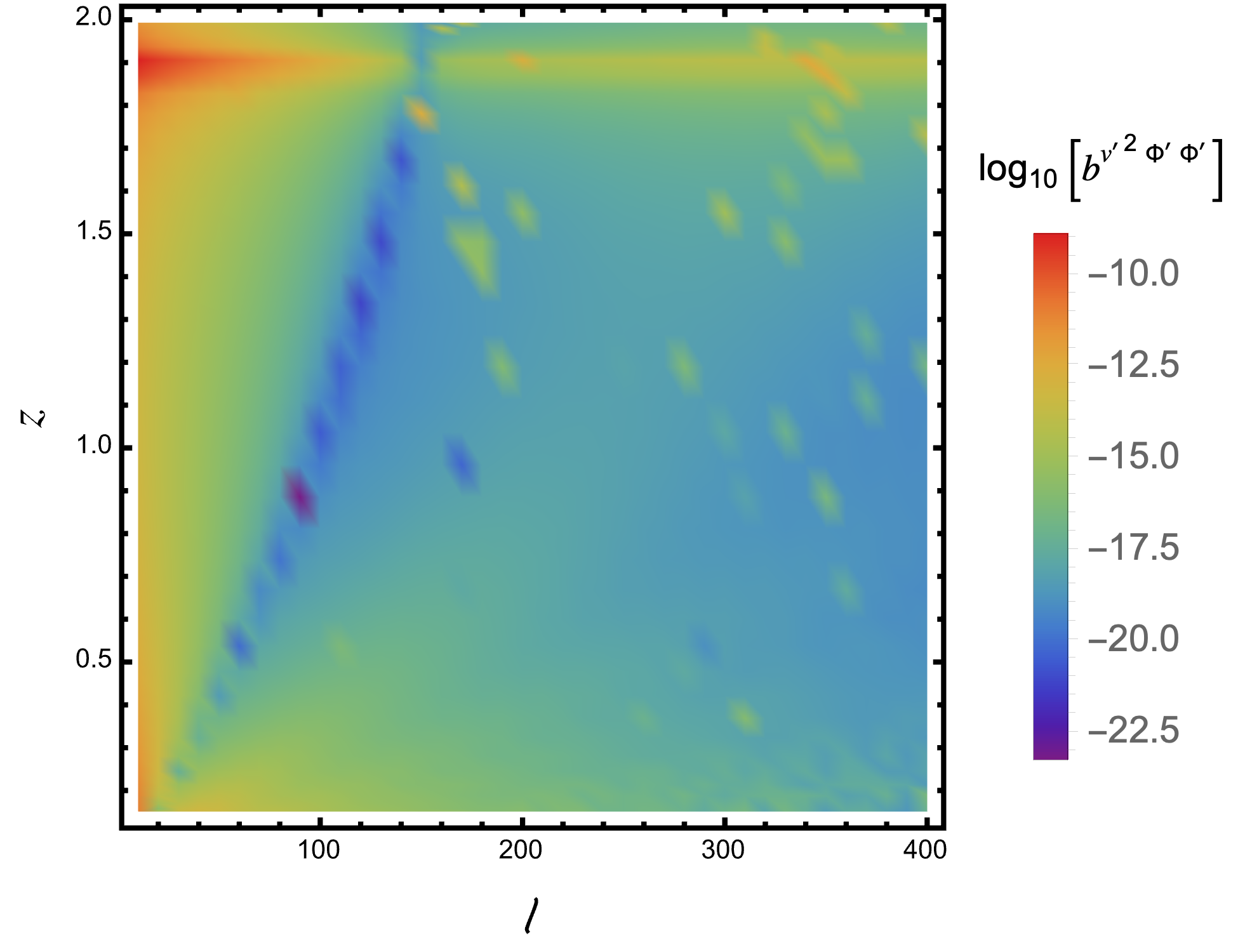}}
\hspace*{0.23em}
\subfloat[$b^{\t{tot}}_{\ell\ell\ell}(z)$  from  Eq.~\eqref{eq:btot-def} \label{fig:d-bispectrum}]
{\includegraphics[width=.47\textwidth]{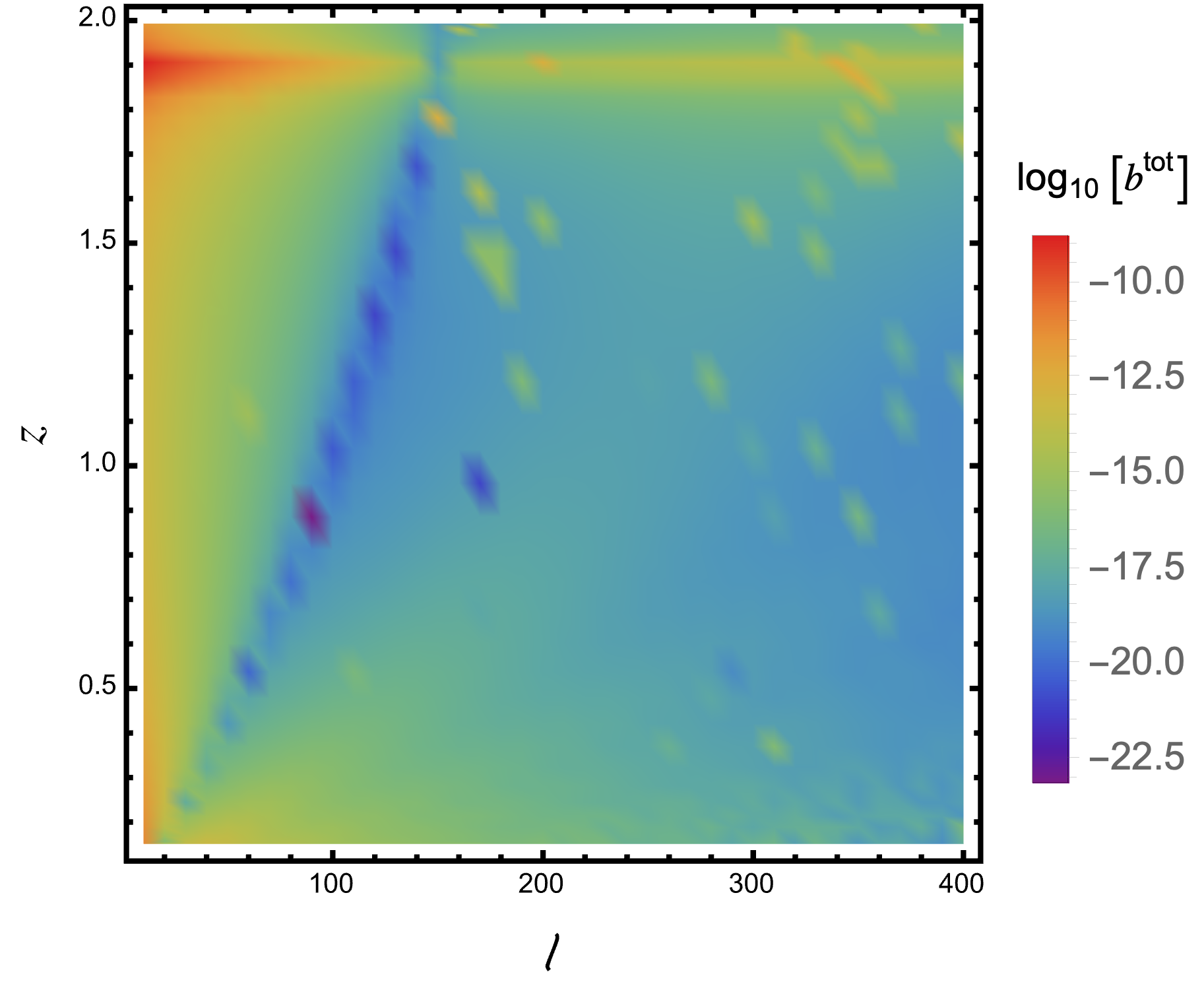}}

\caption{Amplitude of the reduced bispectra and their sum.  Both horizontal features (orange and blue) occurring in the figures are consequences of the homogeneous redshift drift in the $\Lambda$CDM model.}
\label{fig:plot-b-all}
\end{figure}

\paragraph{Plots of the bispectrum.}
In Fig.~\ref{fig:plot-b-all}, we plot each single contribution to the reduced bispectrum as well as the total reduced bispectrum given by the sum 
\begin{align}
    b^{\t{tot}}_{\ell\ell \ell}(z) \equiv b^{v^{\pr ^2}vv}_{\ell\ell\ell}(z)+b^{v^{\pr ^2}\Phi^\pr \Phi^\pr}_{\ell\ell\ell}(z)+ b^{v^{\pr ^2} v \Phi^\pr}_{\ell\ell\ell}(z) \, ,  \label{eq:btot-def}
\end{align}
 for same redshift $z_1 = z_2 = z_3 \equiv z$ and same multipole $\ell_1 = \ell_2 =\ell_3 \equiv \ell$, in the ranges $z \in [0.15,2]$ and $\ell \in [10,400]$.

We notice that all the plots show a horizontal feature   at redshift $z=z_* \sim 1.89$.  This  corresponds to the value of $z$ where 
the homogeneous redshift drift $(1+z)H_0 - H(z)$ vanishes.
In the $\Lambda$CDM model, such $z_*$ is given by \cite{Alves:2019hrg}
\begin{align}
    z_* =\frac{1-3\Omega_\t{m}^{(0)}+\sqrt{1+2\Omega_\t{m}^{(0)} -3\left (\Omega_\t{m}^{(0)}\right )^2}}{2\Omega_\t{m}^{(0)}}  \, . 
\end{align}
Since we are considering  the perturbations normalized w.r.t. their background counterpart, the magnitude of the bispectrum is enhanced around $z\sim 1.89$. 
The fact that, in the $\Lambda$CDM model, the redshift drift is expected to be  almost zero around a redshift $z \lesssim 2$ was pointed out, for example, in \cite{Loeb:1998bu,Balbi:2007fx,Liske:2008ph,Martinelli:2012vq}. 
Therefore, the background redshift drift naturally behaves as a function through which a \quotes{null test} (see \textit{e.g.} \cite{Franco:2019wbj,Bonvin:2020cxp,Bengaly:2020neu}) could be performed in order to  test models of dark energy and modified gravity against  $\Lambda$CDM predictions. 
For these reasons, differently to what done in \cite{Bessa:2023qrr}, keeping into account the term $H_0 (1+z)$ in the background value may play a relevant role in the estimators for the redshift drift.

We also note that  Figs.~\ref{fig:a-bispectrum} and \ref{fig:b-bispectrum} show a horizontal blue region around redshift $z=z_{\t{acc}} \sim 0.6$, where the magnitude of the bispectrum is significantly suppressed. This feature is due to the fact that the reduced bispectra $b^{v^{\pr^2}vv}_{\ell_1 \ell_2 \ell_3}(z)$ and $b^{v^{\pr^2}v\Phi^\pr}_{\ell_1 \ell_2 \ell_3}(z)$ in Eqs.~\eqref{eq:bispectrum:vvv-final} and \eqref{eq:bispectrum:vphi-final-mixed} respectively contain $\Hcal^\pr(z)^2$ and $\Hcal^\pr(z)$. The quantity $\Hcal^\pr(z)$ can be recast as 
\begin{align}
    \Hcal^\pr(z)   = \frac{1}{\eta^\pr(z)}\frac{\t{d}}{\t{d}z}\left (\frac{H(z)}{1+z} \right ) \,.
    \label{eq:dH}
 \end{align}
Thanks to this relation, it is easy to see that $\Hcal^\pr(z)$ changes smoothly from negative values at small redshift to positive ones at redshift bigger than $z_\t{acc}$. This is a consequence of the fact that, in the $\Lambda$CDM model, the Universe starts to accelerate at $z_\t{acc}$. 
This blue line in Fig.~\ref{fig:a-bispectrum} is more pronounced than the one in Fig.~\ref{fig:b-bispectrum}. This happens because the term $b^{v^{\pr ^2}vv}_{\ell\ell\ell}(z)$, plotted in Fig.~\ref{fig:a-bispectrum}, contains a further power of $\Hcal^\pr(z)$ w.r.t. the term $b^{v^{\pr^2}v\Phi^\pr}_{\ell\ell\ell}(z)$. We remark that this blue region does not appear in Figs.~\ref{fig:c-bispectrum} and \ref{fig:d-bispectrum}. This is because the bispectrum $b^{v^{\pr^2}\Phi^\pr \Phi^\pr}_{\ell_1 \ell_2 \ell_3}(z)$ does not contain the factor $\Hcal^\pr(z)$. Clearly, this also reflects  in the behavior of the total bispectrum.

\paragraph{Comparison between bispectrum and power spectrum.} To conclude this section, we give a numerical estimate of the reduced bispectrum of the redshift drift normalized  w.r.t. the product of the power spectra associated with the three-point function. In general, in the harmonic-redshift space, we can define the  statistic
\begin{align}
    R_{\ell_1 \ell_2 \ell_3}(z_1, z_2, z_3) &\equiv \frac{b^{\t{tot}}_{\ell_1 \ell_2 \ell_3}(z_1, z_2, z_3)}{\left [c^{\t{tot}}_{\ell_1}(z_1)c^{\t{tot}}_{\ell_2}(z_2)c^{\t{tot}}_{\ell_3}(z_3)+ 5 \, \t{perms.}\right ]^{2/3}} \, , 
\end{align}
where we have accounted for all the possible permutations of the triplet $(\ell_1, \ell_2, \ell_3)$.
More specifically, by considering only a single redshift $z_1=z_2=z_3=z$  and a single multipole $\ell_1=\ell_2=\ell_3=\ell$, such  a ratio can be explicitly written as 
\begin{align}
  R_\ell(z) &\equiv \frac{b^{\t{tot}}_{\ell\ell\ell}(z)}{6^{2/3} \left [c^{\t{tot}}_{\ell}(z) \right ]^2}  \nex
  &=6^{1/3}\frac{\left[\Hcal_0 - \Hcal(z)\right ]}{\Hcal(z)}
  \frac{\frac{\Hcal^\pr(z)^2}{\Hcal(z)^4} \left (c^{v^\pr v}_{\ell }(z) \right)^2 + \left (c^{v^\pr \Phi^\pr}_{\ell }(z) \right)^2 -2\frac{\Hcal^\pr(z)}{\Hcal(z)^2}\, c^{v^\pr v}_{\ell}(z)c^{v^\pr \Phi^\pr}_{\ell}(z) }{\left [\left (\frac{\Hcal^\pr(z)}{\Hcal(z)^2} \right )^2c^{v v}_\ell(z) + c^{\Phi^{\pr}\Phi^\pr}_{\ell }(z)-2 \frac{\Hcal^\pr(z)}{\Hcal(z)^2}\, c^{v \Phi^\pr}_{\ell }(z)  \right ]^2}\,  . 
  \label{eq:R-def-allterms}
\end{align}

As previously underlined (see 
end of Subsect.~\ref{subsec:second-order-RD}), on sub-Hubble scales this ratio is expected to be  large, since the RSD term only appears in the bispectrum, namely in the numerator of Eq.~\eqref{eq:R-def-allterms}. 
However, let us stress that, while the coefficients multiplying  the $c_\ell$'s both in the numerator and denominator of Eq.~\eqref{eq:R-def-allterms}  are all of the order of unity in the range of considered redshift,  the term $\frac{\Hcal_0 - \Hcal(z)}{\Hcal(z)}$ in front of the ratio suppresses the amplitude of the bispectrum w.r.t. the power spectrum squared.
\begin{figure}
\centering
{\includegraphics[width=.6\textwidth]{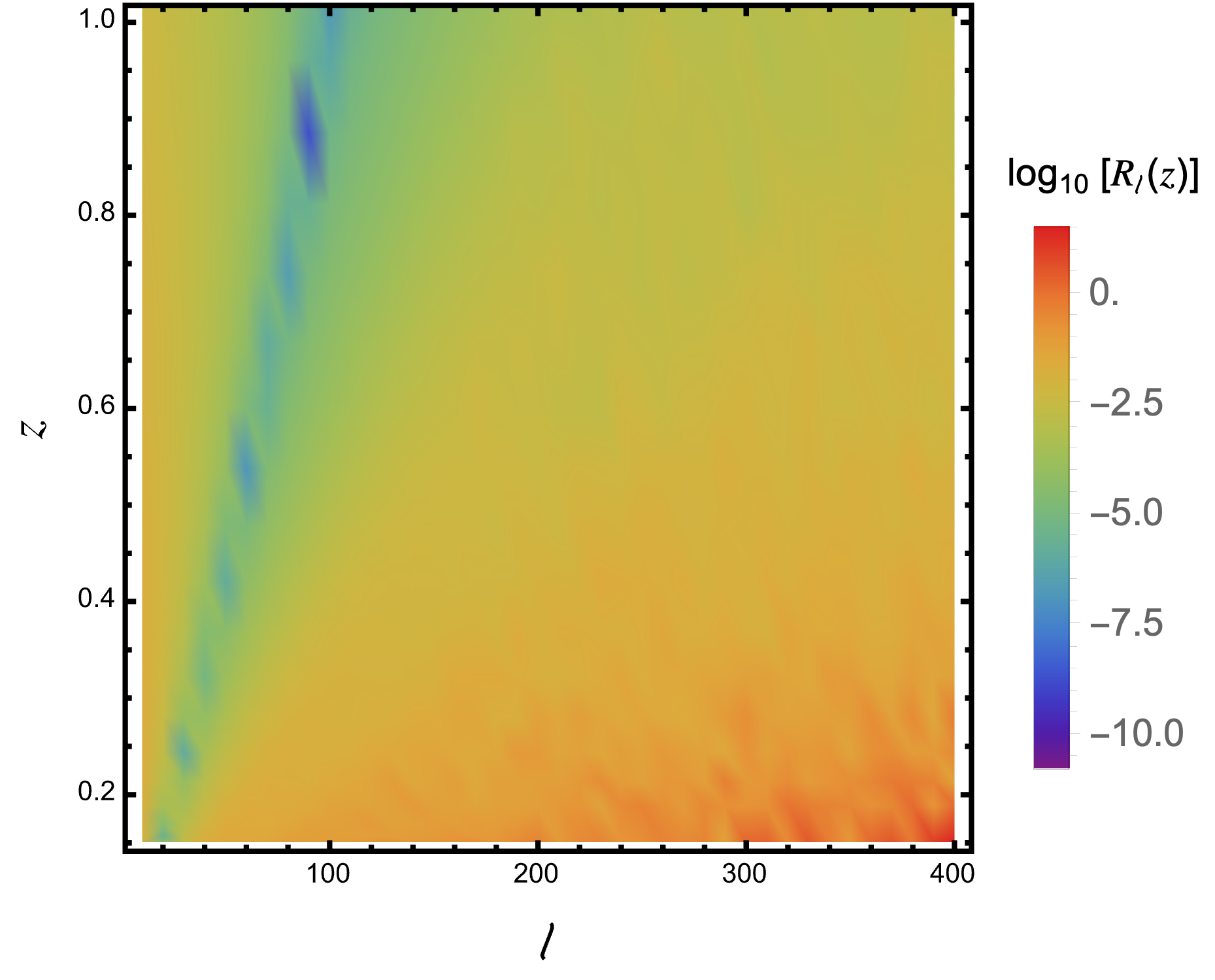}} 
\caption{Amplitude of the ratio $R_\ell(z)$ in Eq.~\eqref{eq:R-def-allterms}, for the leading perturbations  of the redshift drift. We remark that this ratio is enhanced at small redshift for high $\ell$'s. This is in line with the perturbative hierarchy explained in Subsect.~\ref{subsec:second-order-RD}. }
\label{fig:R_tot}
\end{figure}

To show these effects, in Fig.~\ref{fig:R_tot} we report a numerical evaluation of $R_\ell(z)$: we notice that there is a region in the $(\ell,z)$-plane where $R_\ell(z)\sim\mathcal{O}(10)$. This happens at very low redshift, when there has been more time for the non-linearities to grow, and on small angular scales (high $\ell$'s), since this effect becomes more evident as long as we take large momenta in Fourier space\footnote{Naively, we would expect a growth of $R_\ell\sim\ell^2$.}.
\begin{figure}
    \centering
\includegraphics[width=0.6\linewidth]{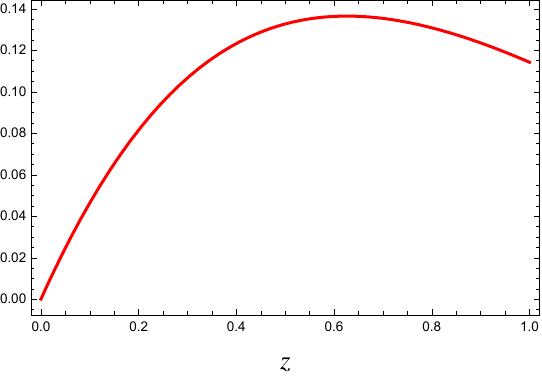}
    \caption{Evolution in redshift of the prefactor $\frac{\Hcal_0 - \Hcal(z)}{\Hcal(z)}$ in Eq.~\eqref{eq:R-def-allterms}.}
    \label{fig:H0vsHz}
\end{figure}
Meanwhile, in Fig.~\ref{fig:H0vsHz} we show that the factor $\frac{\Hcal_0 - \Hcal(z)}{\Hcal(z)}$ is roughly $\mathcal{O}(10^{-1})$ for $z \gtrsim 0.3$ and lowers down to $10^{-2}$ for $z \simeq 0.02 $, going to zero when $z$ vanishes. This shows how  this prefactor  attenuates the enhancement of the leading second-order perturbations w.r.t. the first-order ones  on sub-Hubble scales. However, from Fig.~\ref{fig:R_tot} it is clear that the enhancement generated by the second-order velocity-gradient contribution is large enough to compensate for this suppression at low redshift and large multipoles, thus allowing $R_\ell(z)$ to grow up to $\mathcal{O}(10)$ for $z \lesssim 0.2$, in accordance with what just discussed.

As an informative example, let us compare the ratio $R_\ell(z)$ against its analogous counterpart for the galaxy number count $\Delta$. To this end, we limit ourselves to the terms associated with the radial derivatives of velocity field. With this proviso, the leading first- and second-order fluctuations in the galaxy number count are \cite{DiDio:2014lka,DiDio:2015bua}
\begin{align}
    \Delta^{(1)}_{\text{leading}} = \frac{\pa_r v_{||z}}{\Hcal_z} \qquad , \qquad \Delta^{(2)}_{\text{leading}} = \frac{1}{\Hcal^2_z} \left [\left (\pa_r v_{||z} \right )^2 + v_{||z}\pa^2_r v_{||z} \right ]  \, . 
    \label{eq:leading-gnc}
\end{align}
Then, using the transfer functions written in Eqs.~\eqref{eq:trasn-funct} and defining 
\begin{align}
    \Delta^{v^{\prime \prime}}_\ell (z,k) & = \left ( \frac{k}{\Hcal(z)}\right )^2T_v(z,k)j^{\pr\pr\pr}_\ell (kr) \, , 
\end{align}
we get that the ratio of the bispectrum normalized to the power spectrum squared in harmonic space for the galaxy number count is
\begin{align}
    R^{\Delta }_\ell (z)= 6^{1/3}  \left [ 1+ \frac{c^{v^\pr v}_\ell(z)c^{v^{\pr \pr} v^\pr}_\ell(z)}{\left ( c^{ v^\pr  v^\pr}_\ell(z)\right )^2}  \right ] \, . \label{eq:RDelta}
\end{align}
Let us now compare this term to the analogous one from Eq.~\eqref{eq:R-def-allterms}, obtained by leaving aside  the terms involving $\Phi^\pr$, namely
\begin{align}
    R^{ v^\pr v}_{\ell}(z) &\equiv  6^{1/3}
    \frac{\Hcal_0 - \Hcal(z)}{\Hcal(z)}
    \frac{\Hcal(z)^4}{\Hcal^\pr(z)^2} \frac{\left ( c^{ v^\pr v}_\ell(z)\right )^2}{\left (c^{ vv}_\ell(z)\right )^2}  \,.
    \label{eq:R-rsd-doppler}
\end{align}

\begin{figure}
\centering
\subfloat[][$R^{\Delta}_\ell(z)$ from Eq.~\eqref{eq:RDelta} \label{fig:RGNC-rsd} ]
{\includegraphics[width=.485\textwidth]{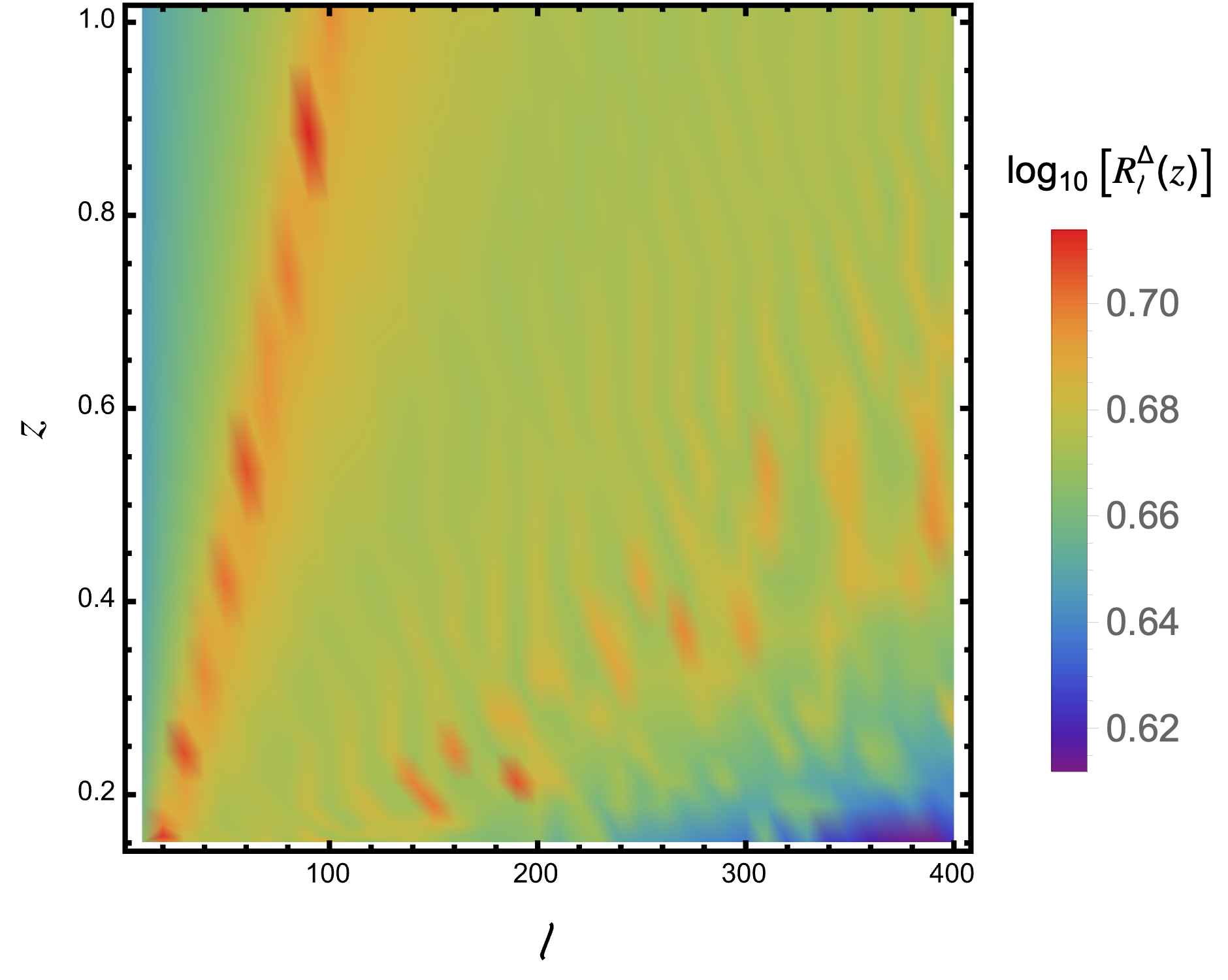}}\quad 
\subfloat[][$R^{v^\pr v}_\ell(z)$ from Eq.~\eqref{eq:R-rsd-doppler} \label{fig:Rrsdv}]
{\includegraphics[width=.49\textwidth]{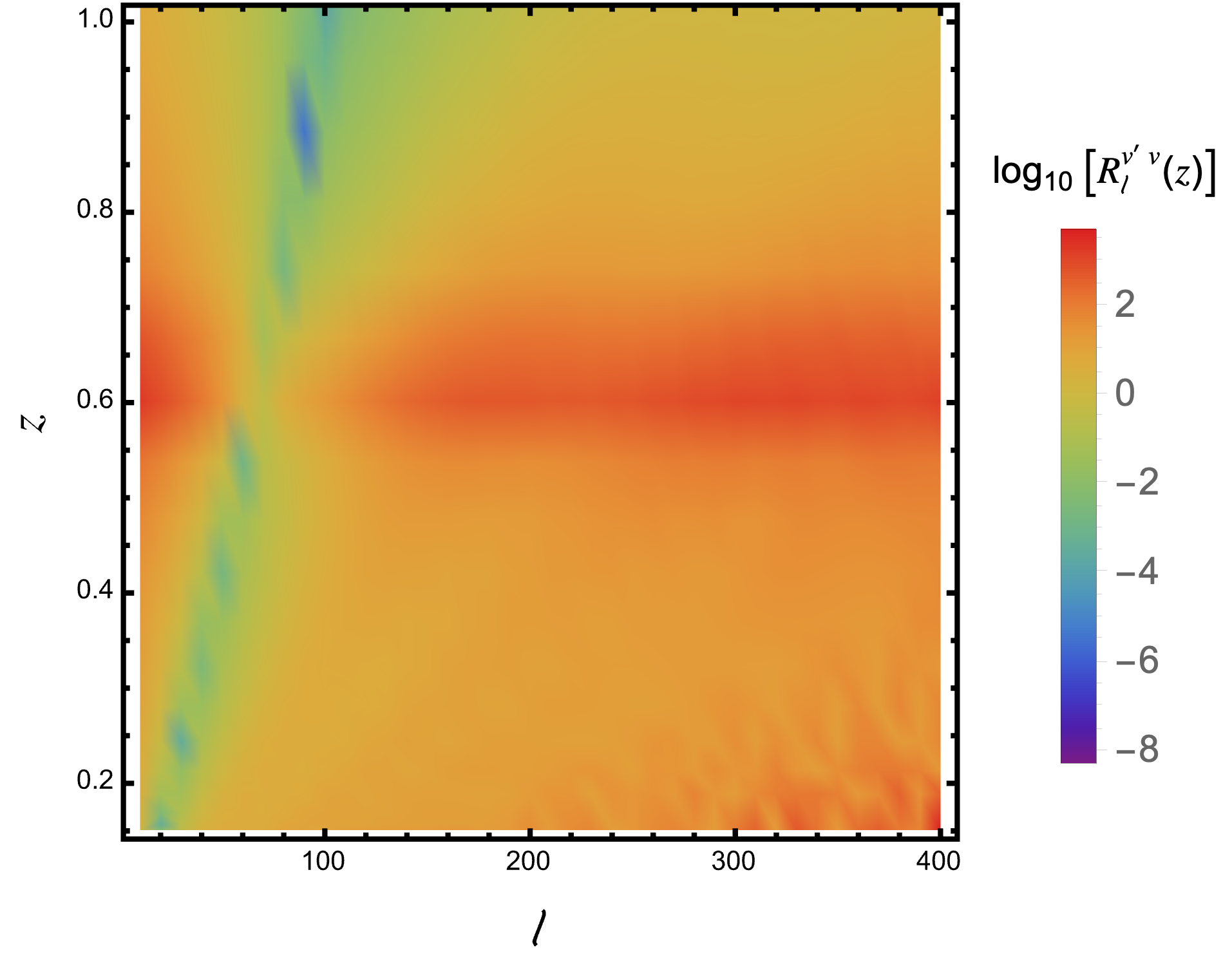}} \\
\caption{Amplitude of the ratios ${R}^{\Delta}_\ell(z)$ and $R^{v^\pr v}_\ell(z)$   for the leading perturbations involving the velocity field in the  galaxy number count  and redshift drift. }
\label{fig:R_rsdv}
\end{figure}

The values  $R^\Delta_\ell(z)$ and $R^{v^\pr v}_\ell(z)$ are reported in Fig.~\ref{fig:R_rsdv}.
In particular, Fig.~\ref{fig:RGNC-rsd} shows that the ratio $R^\Delta_\ell(z)$ for the widely studied cosmological observable as the number counts stays almost stable and $\sim\mathcal{O}(1)$ for various $\ell$'s and $z$. This is not the case for $R^{v^\pr v}_\ell(z)$ (Fig.~\ref{fig:Rrsdv}). This is because, for the galaxy number count, the leading terms in the sub-Hubble regime rescale, as expected, as $(k/\Hcal)^{2n}$ (with $n$ being the perturbative order), thus the bispectrum is intrinsically non-linear as the spectrum squared.
On the contrary, as shown in Fig.~\ref{fig:Rrsdv}, the quantity $R^{v^\pr v}_\ell(z)$  for the redshift drift ranges across three orders of magnitude, reaching  values  $\mathcal{O}(10^2)$ at large $\ell$'s and low $z$. 
Indeed, at the second perturbative order, the redshift drift is  intrinsically more sensitive to the non-linear scales w.r.t. what expected by naive perturbative arguments, as discussed at the end of Subsect.~\ref{subsec:second-order-RD}. This  is one of the most important outcomes of this manuscript.

Let us stress once again that such a different behavior can be understood by comparing Eqs.~\eqref{eq:leading-gnc} to Eqs.~\eqref{eq:leadingterms-normalized}. In the galaxy number count, the  RSD contribution scales  as  $\pa_r v_{||z}$ at first order and as $(\pa_r v_{||z})^2$ at second order. Instead, for the redshift drift, the term $\pa_r v_{||z}$ is absent among first-order perturbations, while at second order we do have a $(\pa_r v_{||z})^2$ term. This is the reason why the expected hierarchical scaling $(k/\Hcal)^{2n}$ in Fourier space is broken.

\section{Summary and Conclusions}
\label{eq:summary}
In this paper we have presented for the first time in  the literature the 
expression of the cosmological redshift drift up to the second perturbative order, including
all the physical effects and expressed as a function of the observed redshift, the observed past light-cone and observed angles.

This has been achieved by evaluating the redshift drift in the Geodesic Light-Cone (GLC) gauge, which naturally provides fully non-linear expressions for late-time cosmological observables. In this gauge, the observational angles remain constant along trajectories  connecting the source and the observer  on the same light-cone. As a consequence, the contributions to the redshift drift can be factorized into products of local perturbations evaluated at the source and observer positions. Resorting to the approach developed in \cite{Bechaz:2025ojy}, we have then promoted each perturbation to its gauge-invariant counterpart, thereby obtaining manifestly gauge-invariant formulae for the redshift drift on the light-cone.  Finally, we have 
 recast these results in terms of standard gravitational potentials, obtaining  the usual perturbative effects -- such as Shapiro time-delay,  Doppler, SW, ISW, and angular deflections -- to be manifest.

Our formulae are fully gauge invariant not only at the source but also at the observer position. Indeed, the GLC gauge admits a residual gauge freedom along the observer world-line, which we have consistently fixed following the analyses of \cite{Fanizza:2020xtv, Bechaz:2025ojy}.  This procedure  gives observables as measured by a free-falling observer and generates original  observer terms in the redshift drift already at first order, which were not properly accounted for in previous studies \cite{Bessa:2023qrr}. Also at second order, monopole and higher-multipole contributions arise at the observer position, ensuring full gauge invariance. Furthermore, since our methodology does not rely on any specific gravitational theory and allows for  non-vanishing anisotropic stress, our formulae are also applicable to (dynamical) dark energies scenarios and modified gravity theories.

In particular, from the full formulae we have isolated the leading terms on sub-Hubble scales, namely those involving the highest number of radial or angular derivatives. At first order, these leading contributions are given by the peculiar velocity of the source and local variations of the gravitational potential along the radial direction. At second order, we have found that, among hundreds of terms, the leading contribution is simply proportional to $(\pa_r v_{||z})^2$. This RSD  contribution represents a genuinely new second-order effect, since at first order terms proportional to $\pa_r v_{||z}$ cancel out. 

We have then used these leading terms to compute the correspondent contributions  to the three-point correlation function of the redshift drift in Fourier space. By working directly in the observable spherical-harmonics-redshift space, we have provided analytical expressions for the bispectrum, and showed how the leading terms of the latter can be expressed as  products of the angular    power spectrum coefficients. We have also reported some numerical evaluations of the so-called \textit{reduced} bispectrum, and compared its magnitude with that of the power spectrum squared. In particular,  the plot of the reduced bispectrum normalized to the power spectrum squared indicates that, on sub-Hubble scales, non-linear physics has more impact on the redshift drift signal than what we might naively expect considering the square of the first-order terms. This suggests that  measuring the bispectrum of the redshift drift using real data, or in $N$-body relativistic numerical simulations, may be easier  than currently thought.

In summary, we have computed the fully gauge-invariant non-linear relativistic corrections to the redshift drift, including source and observer perturbations,  as seen by a free-falling observer. We have provided analytical evaluations of  the bispectrum for the dominant terms on sub-Hubble scales, and given some numerical estimates, discussing about their potential as probes of alternative  models of Gravity and Cosmology.

It would be interesting to compare  our analytical results with  $N$-body numerical simulations, as done, for example, in \cite{Koksbang:2023tun,Bessa:2024beh,Oestreicher:2025qcs}. In particular, in \cite{Koksbang:2023tun},  it was shown that the contribution to the redshift drift
from peculiar acceleration, that would be connected with RSD terms in a perturbative approach,  is significantly suppressed w.r.t. the other terms. This reflects the fact that, at first order, the RSD  term $\pa_r v_{||z}$ cancels out on an analytical level. One could then search for the effects of $(\pa_r v_{||z})^2$ by  evaluating the bispectrum of the redshift drift through numerical $N$-body simulations, as done in \cite{Oestreicher:2025qcs} for the power spectrum. Another  follow-up of the present manuscript  could consist in  estimating the contamination that primordial non-Gaussianity evaluated through the redshift drift can 
be subject to because of the peculiar non-linearities of this observable
(similarly, for example, to what done in \cite{DiDio:2016gpd} for the galaxy number count). Furthermore, the investigation of the role of vector and tensor perturbations at first and second order will be addressed   in a future work.

\section*{Acknowledgements}
PB is   grateful to Guillem Domènech for stimulating discussions during the final stage of this work. PB and GM are supported in part by the Istituto Nazionale di Fisica Nucleare (INFN) through the Commissione Scientifica Nazionale 4 (CSN4),  under the Iniziativa Specifica (IS) Theoretical Astroparticle Physics (TAsP) and the IS Quantum Fields in Gravity, Cosmology and Black Holes (FLAG). The work of GM and MRMS  was  supported by the research grant number 2022E2J4RK \quotes{PANTHEON: Perspectives in Astroparticle and
Neutrino THEory with Old and New messengers} under the program PRIN 2022 funded by the Italian Ministero dell’Universit\`a e della Ricerca (MUR) and by the European Union – Next Generation EU. GF acknowledges the COST Action CosmoVerse, CA21136, supported by COST (European Cooperation in
Science and Technology).
GF is also member of the Gruppo Nazionale per la Fisica Matematica (GNFM) of the Istituto Nazionale di Alta Matematica (INdAM). The work of MRMS was supported by the Brazilian National Council for 
Scientific and Technological Development – CNPq, under project grant 447129/2024-4.

\appendix

\section{Light-Cone Gauge-Invariant Perturbations}
\label{app:gauge-inv-formulae}
In this appendix, we report the explicit definitions of the quantities $F^{(2)}$, $\mathscr{L}^{(2)}_{ab}$, $\mathcal{V}^{(2)}$ and $\hat{\mathcal{V}}^{(2)}$ which appear in Eqs.~\eqref{eq:xi-2nd-order}, and the definitions of $\mathbb{V}^{(2)}$, $\mathcal{D}$, $\mc{M}^{(2)}_{ab}$, $\mc{U}^{(2)}$ and $\hat{\mc{U}}^{(2)}$ arising in Eqs.~\eqref{eq:glc-gauge-inv-2nd-order}.

The quantity $F^{(2)}$ is defined as \begin{align}
F^{(2)} & = a \bigg [  \partial_\tau \big (aL^{(1)}\xi^\tau_{(1)} \big ) + a \xi^i_{(1)}\partial_i L^{(1)} - a^2 L^{(1)} \bigg ( M^{(1)}+\frac{a}{4}L^{(1)}\bigg )+   \partial_\tau \big (\xi^\mu_{(1)} \partial_\mu \xi^w_{(1)} \big )\notag \\[1ex]
& \quad + ar^2 q^{ab} \mathscr{L}^{(2)}_{ab} \bigg ]  +  2aL^{(1)}\partial_w \xi^\tau_{(1)}+2M^{(1)}\Big [a\dot{\xi}^\tau_{(1)}+ \partial_w \xi^\tau_{(1)}+2\dot{a}\xi^\tau_{(1)}\notag \\[1ex]
& \quad + a \partial_w \xi^w_{(1)} \Big ] + N^{(1)} \bigg (H\xi^\tau_{(1)}+ 2a \dot{\xi}^w_{(1)}+\frac{1}{a}\partial_w \xi^\tau_{(1)}+\partial_w \xi^w_{(1)} \bigg ) \notag \\[1ex]
& \quad +  \Big \{ 2ar^2(D_au^{(1)}+\tilde{D}_a\hat{u}^{(1)}) - 2ar^2 \partial_w (D_a \chi_{(1)}+\tilde{D}_a \hat{\chi}_{(1)})  \notag \\[1ex]
& \quad +\partial_a (\xi^\tau_{(1)}- a\xi^w_{(1)}) \Big \} \big [q^{ab}(D_b \dot{\chi}_{(1)}+\tilde{D}_b \dot{\hat{\chi}}_{(1)}) \big ] \notag \\[1ex]
& \quad + \Big \{ 2r^2 \big [ a (D_a v^{(1)}+\tilde{D}_a \hat{v}^{(1)}) +D_a u^{(1)}+\tilde{D}_a \hat{u}^{(1)}\big ] - r^2 \partial_w (D_a \chi_{(1)}+\tilde{D}_a \hat{\chi}_{(1)}) \notag \\[1ex]
& \quad + \partial_a \xi^w_{(1)}\Big \} \big [q^{ab}\partial_w (D_b \chi_{(1)}+\tilde{D}_b \hat{\chi}_{(1)}) \big ] +\frac{\big (\partial_w \xi^\tau_{(1)} \big )^2}{a^2}+2H\xi^\tau_{(1)}\dot{\xi}^\tau_{(1)}\notag \\[1ex]
& \quad + \xi^\mu_{(1)}\partial_\mu N^{(1)} -a \xi^\mu_{(1)} \partial_\mu \big (\dot{\xi}^w_{(1)}-2M^{(1)} \big ) + \frac{1}{a}\partial_w \big (\xi^\mu_{(1)}\partial_\mu \xi^\tau_{(1)} \big ) -\pa_w (\xi^\mu_{(1)} \partial_\mu \xi^w_{(1)}) \notag \\[1ex] 
& \quad + \big (\dot{\xi}^\tau_{(1)}+ \partial_w \xi^w_{(1)} \big )^2 + 4 \dot{\xi}^w_{(1)}\partial_w \xi^\tau_{(1)}-4\dot{a}\xi^\tau_{(1)}\dot{\xi}^w_{(1)}\notag \\[1ex]
& \quad + \xi^\mu_{(1)}\partial_\mu \big ( \dot{\xi}^\tau_{(1)} + \partial_w \xi^w \big )  - a\big (\dot{\xi}^\tau_{(1)}\dot{\xi}^w_{(1)}+ 3 \dot{\xi}^w_{(1)}\partial_w \xi^w_{(1)}\big ) \notag \\[1ex]
& \quad + \bigg [r^2 (D_a u^{(1)}+\tilde{D}_a \hat{u}^{(1)})+\frac{1}{a}\pa_a \xi^\tau_{(1)}-\pa_a \xi^w_{(1)}-r^2 \pa_w(D_a \chi_{(1)}+\tilde{D}_a \hat{\chi}_{(1)}) \bigg ]^2\, 
\label{eq:F2-def}
\end{align}
where $\mathscr{L}^{(2)}_{ab}$ is
\begin{align}
    \mathscr{L}^{(2)}_{ab} & \equiv \bigg ( D_a v^{(1)}+ \tilde{D}_a \hat{v}^{(1)}+\frac{1}{ar^2} D_a\xi^w_{(1)}\bigg )\bigg ( D_b v^{(1)}+ \tilde{D}_b \hat{v}^{(1)}-\frac{1}{ar^2}D_b \xi^w_{(1)}\bigg ) \, .   
\end{align}
Moreover, the perturbations $\mathcal{V}^{(2)}$ and $\hat{\mathcal{V}}^{(2)}$ are 
\begin{align}
    \mc{V}^{(2)}&\equiv -2Hr^2 q^{ab}D_a \xi^\tau_{(1)} (D_b v^{(1)} + \tilde{D}_b \hat{v}^{(1)}) -2Hr^2 \xi^\tau_{(1)}D^2 v^{(1)} \nex
    & \quad -D_a \xi^\mu_{(1)} \pa_\mu \big [r^2 q^{ab}(D_b v^{(1)} + \tilde{D}_b \hat{v}^{(1)}) \big ]-\xi^\mu_{(1)}\pa_\mu (r^2 D^2 v^{(1)})\nex
    & \quad -r^2 q^{ab}(D_a \chi_{(1)} + \tilde{D}_a \hat{\chi}_{(1)})(D_b v^{(1)} + \tilde{D}_b \hat{v}^{(1)})-q^{ab}D_a L^{(1)} D_b \xi^\tau_{(1)} -q^{ab}D_a M^{(1)}D_b \xi^w_{(1)} \nex
    & \quad - r^2 q^{ab}q^{cd} D_a (D_c v^{(1)} + \tilde{D}_c \hat{v}^{(1)})D_b (D_d \chi_{(1)} + \tilde{D}_d \hat{\chi}_{(1)}) - L^{(1)}D^2 \xi^\tau_{(1)}-M^{(1)} D^2 \xi^w_{(1)}\nex
    & \quad -r^2q^{ab} (D_a v^{(1)} + \tilde{D}_a \hat{v}^{(1)})D^2(D_b \chi_{(1)} + \tilde{D}_b \hat{\chi}_{(1)})-r^2\dot{\xi}^\tau_{(1)}D^2 v^{(1)}-r^2\dot{\xi}^w_{(1)}D^2 u^{(1)}\nex
    & \quad -2r^2q^{ab}q^{cd}D_a (q_{bc}\nu^{(1)}+D_{bc}\mu^{(1)}+\tilde{D}_{bc}\hat{\mu}^{(1)})(D_d \dot{\chi}_{(1)}+\tilde{D}_d 
    \dot{\hat{\chi}}_{(1)})\nex
    & \quad -r^2q^{ab}(D_a v^{(1)} + \tilde{D}_a \hat{v}^{(1)})D_b \dot{\xi}^\tau_{(1)}-r^2q^{ab}(D_a u^{(1)} + \tilde{D}_a \hat{u}^{(1)})D_b \dot{\xi}^w_{(1)}\nex
    & \quad -2r^2 q^{ab}q^{cd}(q_{ac}\nu^{(1)}+D_{ac}\mu^{(1)}+\tilde{D}_{ac}\hat{\mu}^{(1)})D_b (D_d \dot{\chi}_{(1)}+\tilde{D}_d 
    \dot{\hat{\chi}}_{(1)})\nex
    & \quad +2Hr^2 q^{ab}D_a \xi^\tau_{(1)} (D_b \dot{\chi}_{(1)}+\tilde{D}_b 
    \dot{\hat{\chi}}_{(1)}) +2Hr^2 \xi^\tau_{(1)}D^2 \dot{\chi}_{(1)}\nex
    & \quad +2q^{ab}\bigg (-\frac{r}{a}D_a \xi^\tau_{(1)}+rD_a \xi^w_{(1)} \bigg )(D_b \dot{\chi}_{(1)} + \tilde{D}_b \dot{\hat{\chi}}_{(1)})+2\bigg (-\frac{r}{a}\xi^\tau_{(1)}+r\xi^w_{(1)} \bigg )D^2 \dot{\chi}_{(1)}\nex
    & \quad -\frac{H}{a}q^{ab}D_a \xi^\tau_{(1)}D_b \xi^w_{(1)}-\frac{H}{a}\xi^\tau_{(1)}D^2\xi^w_{(1)}-\frac{1}{a}(q^{ab}D_a \dot{\xi}^\tau_{(1)}D_b \xi^w_{(1)}+\dot{\xi}^\tau_{(1)}D^2 \xi^w_{(1)})\nex
    & \quad -\frac{1}{a}(q^{ab}D_a \dot{\xi}^w_{(1)}D_b \xi^\tau_{(1)}+\dot{\xi}^w_{(1)}D^2 \xi^\tau_{(1)})+q^{ab}D_a \dot{\xi}^w_{(1)}D_b \xi^w_{(1)}+\dot{\xi}^w_{(1)}D^2 \xi^w_{(1)}\nex
    & \quad +r^2 q^{ab}D_a (D_c \dot{\chi}_{(1)}+\tilde{D}_c \dot{\hat{\chi}}_{(1)})\pa_b \big [q^{cd}(D_d \chi_{(1)}+\tilde{D}_d \hat{\chi}_{(1)}) \big ]  \nex
    & \quad +r^2q^{ab}(D_a \dot{\chi}_{(1)}+\tilde{D}_a \dot{\hat{\chi}}_{(1)})D^2(D_b \chi_{(1)}+\tilde{D}_b\hat{\chi}_{(1)})\nex
    & \quad +\frac{a^2r^2}{2} \pa_\tau \Big [D_a \xi^\mu_{(1)} \pa_\mu \big [q^{ab}(D_b \chi_{(1)}+\tilde{D}_b \hat{\chi}_{(1)}) \big ]+\xi^\mu_{(1)} \pa_\mu (D^2 \chi_{(1)}) \Big ]\nex
    & \quad +a^2r^2 q^{ab}(D_a \chi_{(1)}+\tilde{D}_a \hat{\chi})(D_b \dot{\chi}_{(1)}+\tilde{D}_b \dot{\hat{\chi}}) \, , \nex
    \hat{\mc{V}}^{(2)}&\equiv -2Hr^2 \xi^\tau_{(1)}D^2 \hat{v}^{(1)}- r^2 \tilde{D}_a \big [q^{bc}(D_c \chi_{(1)}+\tilde{D}_c \hat{\chi}_{(1)}) \big ]\pa_b \big [q^{ad} (D_d v^{(1)}+\tilde{D}_d \hat{v}^{(1)})  \big ]\nex
    & \quad - \xi^\mu_{(1)}\pa_\mu (r^2 D^2 \hat{v}^{(1)}) +2r^2 \varepsilon^{ab}(D_a \chi_{(1)}+\tilde{D}_a \hat{\chi}_{(1)}) (D_b v_{(1)}+\tilde{D}_b \hat{v}_{(1)})\nex
    & \quad -r^2 \dot{\xi}^\tau_{(1)}D^2 \hat{v}_{(1)} -r^2 \dot{\xi}^w_{(1)}D^2 \hat{u}_{(1)}\nex
    & \quad -2r^2 q^{ab}q^{cd}\tilde{D}_a (q_{bc}\nu^{(1)}+D_{bc}\mu^{(1)}+\tilde{D}_{bc}\hat{\mu}^{(1)})(D_d \dot{\chi}_{(1)}+\tilde{D}_d \dot{\hat{\chi}}_{(1)})\nex
    & \quad -2r^2 q^{ab}q^{cd}(q_{ac}\nu^{(1)}+D_{ac}\mu^{(1)}+\tilde{D}_{ac}\hat{\mu}^{(1)})\tilde{D}_b (D_d \chi_{(1)}+\tilde{D}_d \hat{\chi}_{(1)}) + 2Hr^2 \xi^\tau_{(1)} D^2 \dot{\hat{\chi}}_{(1)}\nex
    & \quad +2 \bigg ( -\frac{r}{a}\xi^\tau_{(1)}+r\xi^w_{(1)}\bigg ) D^2 \dot{\hat{\chi}}_{(1)}+r^2 q^{ab}\tilde{D}_a (D_c \dot{\chi}_{(1)}+\tilde{D}_c \dot{\hat{\chi}}_{(1)})\pa_b \big [q^{cd}(D_d \chi_{(1)}+\tilde{D}_d \hat{\chi}_{(1)}) \big ]\nex
    & \quad + a^2 r^2  \pa_\tau \Big [q^{cd}\tilde{D}_a (D_d \chi_{(1)}+\tilde{D}_d \hat{\chi}_{(1)}) \pa_c \big [q^{ab}(D_b \chi_{(1)}+\tilde{D}_b \hat{\chi}_{(1)}) \big ]+\xi^\mu_{(1)}\pa_\mu (D^2 \hat{\chi}_{(1)}) \Big ]\,  .
\end{align}
Next, $\mathbb{V}^{(2)}$ is 
\begin{align}
\mathbb{V}^{(2)} &\equiv -\frac{1}{a^2r^2}\xi^\tau_{(1)} \partial_\tau (a^2r^2\nu^{(1)})  -\frac{1}{r^2}\xi^w_{(1)} \partial_w (r^2 \nu^{(1)})+ \frac{1}{2}(\dot{H}+2H^2)\big (\xi^\tau_{(1)} \big )^2 \notag \\[1ex]
& \quad + \frac{H}{2}\xi^\mu_{(1)}\partial_\mu \xi^\tau_{(1)}+ 2 H\xi^\tau_{(1)} \bigg (-\frac{\xi^\tau_{(1)}}{ar}+\frac{\xi^w_{(1)}}{r} \bigg ) + \frac{1}{4r^2} q^{ab} \mc{V}^{(2)}_{ab} \, ,
\end{align}
where 
\begin{align}
\mc{V}^{(2)}_{ab} & \equiv  \Big \{ -4r^2 [q_{c(a}\nu^{(1)}+D_{c(a}\mu^{(1)}+\tilde{D}_{c(a}\hat{\mu}^{(1)}]\pa_{b)} -2r^2 \big [(\pa_c(q_{ab}\nu^{(1)})+\pa_c (D_{ab}\mu^{(1)})\nex
& \quad + \pa_c (\tilde{D}_{ab}\hat{\mu}^{(1)})\big ] + 4Hr^2\xi^\tau_{(1)}q_{c(a}D_{b)} +2\xi^\mu_{(1)}\partial_\mu [r^2 q_{c(a}\partial_{b)}]\notag \\[1ex]
& \quad + r^2 q_{c(a}\partial_{b)}  (\xi^\mu_{(1)}\partial_\mu) \Big \} \big [ q^{cd} (D_d \chi_{(1)}+\tilde{D}_d \hat{\chi}_{(1)})\big ] + \frac{1}{2}\mathbb{G}^{(2)}_{ab}-\frac{2}{a} \partial_{(a}\xi^\tau_{(1)}\partial_{b)}\xi^w_{(1)}+ \pa_a \xi^w_{(1)}\pa_b\xi^w_{(1)}\notag \\[1ex]
& \quad  + r^2q_{cd}\Big \{ \partial_{(a} \big [q^{ce}(D_e \chi_{(1)}+\tilde{D}_e\hat{\chi}_{(1)}) \big ] \partial_{b)} \big [q^{df}(D_f \chi_{(1)}+\tilde{D}_f\hat{\chi}_{(1)}) \big ] \Big \} \notag \\[1ex]
& \quad -2r^2 \Big \{[D_{(a}v^{(1)}+\tilde{D}_{(a}\hat{v}^{(1)}]\partial_{b)} \xi^\tau_{(1)} +[D_{(a}u^{(1)}+\tilde{D}_{(a}\hat{u}^{(1)}]\partial_{b)} \xi^w_{(1)} \Big \} \, ,
\label{eq:mathscrN}
\end{align}
and \begin{align}
    \mathbb{G}^{(2)}_{ab} &\equiv \xi^\rho_{(1)}\xi^\sigma_{(1)}\pa_\rho \pa_\sigma \bar{\ga}_{ab}+ \xi^\rho_{(1)}\pa_\rho \xi^\sigma_{(1)}\pa_\sigma \bar{\ga}_{ab}\notag \\ 
    & =(\xi^\tau_{(1)})^2 \ddot{\bar{\ga}}_{ab} + 2\xi^\tau_{(1)}\xi^w_{(1)} \pa_w \dot{\bar{\ga}}_{ab} + 2\xi^\tau_{(1)}\xi^c_{(1)}\pa_c \dot{\bar{\ga}}_{ab} + (\xi^w_{(1)})^2 \pa^2_w \bar{\ga}_{ab}  + 2\xi^w_{(1)}\xi^c_{(1)}\pa_w \pa_c \bar{\ga}_{ab}\notag \\[1ex]
    & \quad + \xi^c_{(1)}\xi^d_{(1)}\pa_c \pa_d \bar{\ga}_{ab} + \xi^\tau_{(1)}\dot{\xi}^\tau_{(1)}\dot{\bar{\ga}}_{ab} + \xi^w_{(1)}\pa_w \xi^\tau_{(1)}\dot{\bar{\ga}}_{ab} + \xi^c_{(1)}\pa_c \xi^\tau_{(1)}\dot{\bar{\ga}}_{ab} + \xi^\tau_{(1)}\dot{\xi}^w_{(1)}\pa_w \bar{\ga}_{ab} \notag \\[1ex]
    & \quad + \xi^w_{(1)}\pa_w \xi^w_{(1)}\pa_w \bar{\ga}_{ab} + \xi^c_{(1)}\pa_c \xi^w_{(1)}\pa_w \bar{\ga}_{ab} 
    + \xi^\tau_{(1)}\dot{\xi}^{c}_{(1)} \pa_c \bar{\ga}_{ab} + \xi^w_{(1)}\pa_w \xi^c_{(1)} \pa_c \bar{\ga}_{ab} + \xi^d_{(1)}\pa_d \xi^c_{(1)}\pa_c \bar{\ga}_{ab} \, .
    \label{eq:G2-ab}
\end{align}
Furthermore, $\mc{M}^{(a)}_{ab}$ is given by
\begin{align}
 \mc{M}^{(2)}_{ab} & \equiv \Big \{ -4r^2 [q_{c(a}\nu^{(1)}+D_{c(a}\mu^{(1)}+\tilde{D}_{c(a}\hat{\mu}^{(1)}]D_{b)}   + Hr^2 \xi^\tau_{(1)}q_{c(a}D_{b)} +\xi^\mu_{(1)}\partial_\mu  [r^2 q_{c(a}\partial_{b)}] \notag \\[1ex]
& \quad + r^2 q_{c(a}\partial_{b)} (\xi^\mu_{(1)}\partial_\mu) \Big \} \big [ q^{cd} (D_b \chi_{(1)}+\tilde{D}_b \hat{\chi}_{(1)})\big ] + \mathbb{G}^{(2)}_{ab}- \frac{4}{a}\partial_{(a}\xi^\tau_{(1)}\partial_{b)}\xi^w_{(1)}  \notag \\[1ex]
& \quad  + r^2 \Big \{ \partial_{(a} \big [q^{cd}(D_d \chi_{(1)}+\tilde{D}_d\hat{\chi}_{(1)}) \big ] \partial_{b)} \big [q^e_c(D_e \chi_{(1)}+\tilde{D}_e\hat{\chi}_{(1)}) \big ] \Big \} \notag \\[1ex]
& \quad -4r^2 \Big \{[D_{(a}v^{(1)}+\tilde{D}_{(a}\hat{v}^{(1)}]\partial_{b)} \xi^\tau_{(1)} +[D_{(a}u^{(1)}+\tilde{D}_{(a}\hat{u}^{(1)}]\partial_{b)} \xi^w_{(1)} \Big \}  \notag \\[1ex]
& \quad +  2\partial_{(a} \xi^w_{(1)}\partial_{b)}\xi^w_{(1)}-\frac{1}{a^2}\partial_\tau \big [2a^2r^2 (q_{ab}\nu^{(2)}+D_{ab}\mu^{(2)}+\tilde{D}_{ab}\hat{\mu}^{(2)}) \big ] \notag \\[1ex]
& \quad - \partial_w \big [\xi^w_{(1)}(q_{ab}\nu^{(1)}+D_{ab}\mu^{(1)}+\tilde{D}_{ab}\hat{\mu}^{(1)}) \big ] \, .
\label{eq:Mab-2-defined}
\end{align}
Finally, $\mc{U}^{(2)}$ and $\hat{\mc{U}}^{(2)}$ are 
\begin{align}
    \mc{U}^{(2)} & \equiv -2Hr^2 q^{ab}D_a \xi^\tau_{(1)} (D_b u^{(1)}+\tilde{D}_b \hat{u}^{(1)})-2Hr^2 \xi^\tau_{(1)}D^2 u^{(1)}\nex
    & \quad - D_a \xi^\mu_{(1)}\pa_\mu \big [r^2 q^{ab}(D_b u^{(1)}+\tilde{D}_b \hat{u}^{(1)}) \big ]-\xi^\mu_{(1)}\pa_\mu (r^2 D^2 u^{(1)})\nex
    & \quad -r^2 q^{ab}(D_a \chi_{(1)}+\tilde{D}_a \hat{\chi}_{(1)}) (D_b u^{(1)}+\tilde{D}_b \hat{u}^{(1)}) - q^{ab}D_a M^{(1)}D_b \xi^\tau_{(1)}-q^{ab}D_a N^{(1)}D_b \xi^w_{(1)}\nex
    & \quad - r^2 q^{ab}q^{cd}D_a (D_c u^{(1)}+\tilde{D}_c \hat{u}^{(1)})D_b (D_d \chi_{(1)}+\tilde{D}_d \hat{\chi}_{(1)})-M^{(1)}D^2 \xi^\tau_{(1)}-N^{(1)}D^2 \xi^w_{(1)}\nex
    & \quad -r^2 q^{ab}(D_a u^{(1)}+\tilde{D}_a \hat{u}^{(1)})D^2 (D_b \chi_{(1)}+\tilde{D}_b \hat{\chi}_{(1)})-r^2 \pa_w \xi^\tau_{(1)}D^2 u^{(1)}-r^2 \pa_w \xi^w_{(1)} D^2 v^{(1)} \nex
    & \quad -2r^2 q^{ab}q^{cd}D_a (q_{bc}\nu^{(1)}+D_{bc}\mu^{(1)}+\tilde{D}_{bc}\hat{\mu}^{(1)})\pa_w (D_d \chi_{(1)}+\tilde{D}_d \hat{\chi}_{(1)})\nex
    & \quad -r^2 q^{ab}(D_a v^{(1)}+\tilde{D}_a \hat{v}^{(1)})\pa_w (D_b \xi^\tau_{(1)})-r^2 q^{ab}(D_a u^{(1)}+\tilde{D}_a \hat{u}^{(1)}) \pa_w (D_b \xi^w_{(1)})\nex
    & \quad -2r^2q^{ab}q^{cd}(q_{ac}\nu^{(1)}+D_{ac}\mu^{(1)}+\tilde{D}_{ac}\hat{\mu}^{(1)}) \pa_w (D_b D_d \chi_{(1)}+D_b \tilde{D}_d \hat{\chi}_{(1)})\nex
    & \quad + 2Hr^2q^{ab}D_a \xi^\tau_{(1)}\pa_w (D_b  \chi_{(1)}+\tilde{D}_b \hat{\chi}_{(1)})+2Hr^2 \xi^\tau_{(1)}\pa_w (D^2 \chi_{(1)})\nex
    & \quad +2^{ab}\bigg (-\frac{r}{a}D_a \xi^\tau_{(1)}+rD_a \xi^w_{(1)} \bigg )\pa_w (D_b  \chi_{(1)}+\tilde{D}_b \hat{\chi}_{(1)})+2 \bigg (-\frac{r}{a}\xi^\tau_{(1)}+r\xi^w_{(1)} \bigg )\pa_w (D^2 \chi_{(1)})\nex
    & \quad -\frac{H}{a}q^{ab}(D_a \xi^\tau_{(1)}D_b \xi^\tau_{(1)}+\xi^\tau_{(1)}D^2 \xi^\tau_{(1)})+2Hq^{ab}(D_a \xi^\tau_{(1)}D_b \xi^w_{(1)}+\xi^\tau_{(1)}D^2 \xi^w_{(1)})\nex
    & \quad -\frac{1}{a}\big [ q^{ab}\pa_w (D_a \xi^\tau_{(1)})D_b \xi^w+\pa_w \xi^\tau_{(1)}D^2 \xi^w_{(1)} \big ]-\frac{1}{a}\big [q^{ab}\pa_w (D_a \xi^w_{(1)})D_b \xi^\tau_{(1)} + \pa_w \xi^w_{(1)}D^2 \xi^\tau_{(1)} \big ]\nex
    & \quad +q^{ab}\pa_w (D_a \xi^w_{(1)})D_b \xi^w_{(1)}+\pa_w \xi^w_{(1)}D^2 \xi^w_{(1)}\nex
    & \quad +r^2 q^{ab}\pa_w (D_a D_c\chi_{(1)}+D_a\tilde{D}_c \hat{\chi}_{(1)}) \pa_b \big [q^{cd}(D_d  \chi_{(1)}+\tilde{D}_d \hat{\chi}_{(1)}) \big ]\nex
    & \quad +r^2 q^{ab}\pa_w (D_a  \chi_{(1)}+\tilde{D}_a \hat{\chi}_{(1)}) D^2 (D_b  \chi_{(1)}+\tilde{D}_b \hat{\chi}_{(1)}) \nex
    & \quad +\frac{a^2r^2}{2}\pa_w \Big [D_a \xi^\mu_{(1)}\pa_\mu  \big [q^{ab} (D_b \chi_{(1)}+\tilde{D}_b \hat{\chi}_{(1)})\big ] +\xi^\mu_{(1)}\pa_\mu (D^2 \chi_{(1)})\Big ]\nex
    & \quad +a^2r^2 q^{ab}(D_a  \chi_{(1)}+\tilde{D}_a \hat{\chi}_{(1)})\pa_w (D_b  \chi_{(1)}+\tilde{D}_b \hat{\chi}_{(1)})\, , \nex
     \hat{\mc{U}}^{(2)} & \equiv   -2Hr^2 \xi^\tau_{(1)}D^2 \hat{u}^{(1)} - r^2 \tilde{D}_a \big [q^{bc}(D_c \chi_{(1)}+\tilde{D}_c \hat{\chi}_{(1)}) \big ]\pa_b \big [q^{ad} (D_d u^{(1)}+\tilde{D}_d \hat{u}^{(1)})  \big ]\nex
     & \quad - \xi^\mu_{(1)}\pa_\mu (r^2 D^2 \hat{u}^{(1)}) +2r^2 \varepsilon^{ab}(D_a \chi_{(1)}+\tilde{D}_a \hat{\chi}_{(1)}) (D_b u^{(1)}+\tilde{D}_b \hat{u}^{(1)})\nex
     & \quad -r^2 \pa_w \xi^\tau_{(1)}D^2 \hat{u}^{(1)}-r^2 \pa_w \xi^w_{(1)}D^2 \hat{u}^{(1)}\nex
     & \quad -2r^2q^{ab}q^{cd}\tilde{D}_a (q_{bc}\nu^{(1)}+D_{bc}\mu^{(1)}+\tilde{D}_{bc}\hat{\mu}^{(1)})\pa_w (D_d \chi_{(1)}+\tilde{D}_d \hat{\chi}_{(1)})\nex
     & \quad - 2r^2 q^{ab}q^{cd}(q_{ac}\nu^{(1)}+D_{ac}\mu^{(1)}+\tilde{D}_{ac}\hat{\mu}^{(1)})\pa_w (\tilde{D}_b D_d \chi_{(1)}+\tilde{D}_b D_d \hat{\chi}_{(1)})\nex
     & \quad +2Hr^2 \xi^\tau_{(1)}\pa_w (D^2 \hat{\chi}_{(1)})+2\bigg (-\frac{r}{a}\xi^\tau_{(1)}+r\xi^w_{(1)} \bigg ) \pa_w (D^2 \hat{\chi}_{(1)})\nex
     & \quad +r^2 q^{ab}\pa_w \tilde{D}_a(D_c \chi_{(1)}+\tilde{D}_c \hat{\chi}_{(1)})\pa_b \big [q^{cd}(D_d \chi_{(1)}+\tilde{D}_d \hat{\chi}_{(1)}) \big ]\nex
     & \quad +\frac{a^2r^2}{2}\pa_w \Big [\tilde{D}_a (D_d \chi_{(1)}+\tilde{D}_d \hat{\chi}_{(1)})\pa_c  \big [q^{ab} (D_b \chi_{(1)}+\tilde{D}_b \hat{\chi}_{(1)})\big ] +\xi^\mu_{(1)}\pa_\mu (D^2 \hat{\chi}_{(1)})\Big ]\, .
\label{eq:mathbU-2-defined}
\end{align}

\bibliographystyle{JHEP}
\bibliography{Ref-RedshiftDrift}

\end{document}